\documentclass[11pt,a4paper]{article}
\usepackage{jheppub}
\usepackage[T1]{fontenc}
\usepackage[latin9]{inputenc}
\usepackage{float}
\usepackage{amsmath}
\usepackage{amssymb}
\usepackage{graphicx}
\usepackage{hyperref}
\usepackage{simpler-wick}
\usepackage{fancyhdr}
\usepackage{microtype}
\usepackage{graphicx}
\usepackage{graphbox}
\usepackage{caption}
\usepackage{subcaption}
\usepackage{mathrsfs}
\usepackage{dutchcal}
\usepackage{array,multirow,bigdelim,arydshln}
\usepackage{MnSymbol}
\usepackage{dsfont}

\DeclareMathAlphabet{\mathdutchcal}{U}{dutchcal}{m}{n}
\SetMathAlphabet{\mathdutchcal}{bold}{U}{dutchcal}{b}{n}
\DeclareMathAlphabet{\mathdutchbcal}{U}{dutchcal}{b}{n}
\DeclareMathAlphabet{\mathcal}{OMS}{cmsy}{m}{n}

\hypersetup{
    colorlinks = True,
    citecolor = magenta,
    linkcolor = cyan
}

\newcommand{\nn}{\nonumber}
\newcommand{\la}{\langle}
\newcommand{\ra}{\rangle}
\renewcommand{\d}{\text{d}}
\renewcommand{\Im}{\text{Im}}
\renewcommand{\Re}{\text{Re}}
\newcommand{\ep}{\epsilon}
\newcommand{\vep}{\varepsilon}
\newcommand{\B}{\mathcal{B}_M}

\newcommand{\disc}{\text{Disc}}
\newcommand{\pert}{{\hspace{.1em}\text{pert}}}
\newcommand{\npert}{{\hspace{.1em}\text{non-pert}}}
\renewcommand{\hom}{\text{hom}}
\newcommand{\inhom}{\text{inhom}}
\renewcommand{\vec}[1]{\boldsymbol{#1}}
\newcommand{\mat}[1]{\underline{\boldsymbol{#1}}}
\newcommand{\scale}{g_s^2 C_A}
\newcommand{\Mh}{\hat{\mathdutchcal{M}}^2}
\newcommand{\A}{\mathcal{A}_{0+2}}
\newcommand{\Ah}{ {\hat{\mathcal{A}}_{0+2}} }
\renewcommand{\O}{\mathcal{O}}
\def\<{\langle}
\def\>{\rangle}
\def\mubar{\bar{\mu}}
\def\MSbar{\overline{\rm MS}}

\title{Two-point sum-rules in three-dimensional Yang-Mills theory} 

\author[a]{Simon Caron-Huot}
\author[b]{Andrzej Pokraka}
\author[a,c]{Zahra Zahraee}

\affiliation[a]{
	Department of Physics, McGill University, 
	3600 Rue University, 
	Montr\'eal,  Canada
}
\affiliation[b]{
    Department of Physics, 
    Brown University, 
    Providence, 
    RI 02912, 
    USA
}

\affiliation[c]{
CERN, 
Theoretical Physics Department, 
CH-1211 Geneva 23, 
Switzerland}
\emailAdd{schuot@physics.mcgill.ca}
\emailAdd{andrzej\_pokraka@brown.edu}
\emailAdd{zahra.zahraee@cern.ch}

\abstract{
We compute the stress-tensor two-point function in three-dimensional Yang-Mills theory to three-loops in perturbation theory.
Using its calculable shape at high momenta, we test the notion that its Borel transform is saturated at low energies by the lowest glueball state(s).
This assumption provides relatively stable estimates for the mass of the lightest glueball that we compare with lattice simulations.
We also provide estimates for the coupling of the lightest glueball to the stress tensor.
Along the way, we comment on the extent that such estimates are non-rigorous.  
Lastly, we discuss the possibility of applying the sum-rule analysis to two-point functions of higher-spin operators and obtain a crude approximation for the glueball couplings to these operators.
}

\begin{document}
\maketitle

\section{Introduction}
\label{sec:intro}

Understanding the non-perturbative dynamics of strongly coupled systems from first principles has been a long standing problem in modern quantum field theory (QFT).  Arguably, the most direct calculation of non-perturbative effects come from lattice simulations where one computes QFT correlation functions in a discretized spacetime and then extrapolates to the continuum. While less direct, one can also obtain some non-perturbative information through dispersion relations that connect correlators at large (computable) space-like momenta and small momenta, as in the famous QCD sum-rules \cite{Weinberg1967,Shifman:1978bx,Shifman:1992xu,Shifman2009}. %
Surprisingly, the low energy contribution to these sum-rules is often found to be numerically dominated by the lightest bound states, yielding estimates of their various properties.
In light of the continuing interest in rigorous results on confining theories,
we would like to revisit these old ideas in the context
of three-dimensional Yang-Mills theory, where both perturbative calculations and lattice simulations are possible.


The central object of our study will be the stress-energy 2-point function 
\begin{align} \label{Pi intro}
 \Pi^{\mu\nu\alpha\beta} (p^2)
    = i \int \d^dx\ e^{-i p \cdot x}
    \la 0 \vert \mathsf{T}\{ T^{\mu\nu}(x) T^{\alpha\beta}(0) \} \vert 0 \ra,
\end{align}
which probes intermediate glueball states $\vert G \ra$
through its imaginary/absorptive part
\begin{equation}\begin{aligned} \label{Im Pi intro}
    2\Im\, \Pi^{\mu\nu\alpha\beta} (p^2)
    &=
     \int \d^dx\ e^{-i p \cdot x}
    \la 0 \vert T^{\mu\nu}(x) T^{\alpha\beta}(0) \vert 0 \ra
\\ &=    \int \d^dx\ e^{-i p \cdot x} \sumint_{\ G}
    \la 0 \vert T^{\mu\nu}(x) \vert G \ra
    \la G \vert T^{\alpha\beta}(0) \vert 0 \ra\,.
\end{aligned}\end{equation}
Like any two-point correlator, $\Pi^{\mu\nu\alpha\beta}$ in \eqref{Pi intro} admits a K\"{a}ll\'en-Lehmann dispersion relation that expresses it as an integral over a spectral density \eqref{Im Pi intro}.
The latter consists of two non-negative functions, corresponding to spin-0 and spin-2 exchanges.
On the one hand, at large Euclidean momenta the correlator can be calculated using perturbation theory. On the other hand, the qualitative features of the spectral density are known at low energies: we expect a sum of $\delta$-function contributions from stable glueballs followed by a continuum that possibly includes further resonances.
The goal of this work is to explore the consequences of the dispersion relation that connects these quantities.

One of our motivations is recent work on the S-matrix bootstrap
\cite{Karateev:2019ymz,Correia:2022dyp} in which scattering amplitudes of stable bound states are supplemented by form factors and two-point functions of local operators, in order to rigorously connect
short- and large-distance physics.
Here, we focus only on two-point functions and numerically explore less rigorous connections in the spirit of QCD sum-rules.  
Three-dimensional Yang-Mills is a natural model to study from this perspective since it is super-renormalizable (i.e., amenable to perturbation theory) and has interesting non-perturbative dyanmics (i.e., confinement). 
The presence of a (perturbative) mass scale in three-dimensional Yang-Mills theory is an additional simplification with respect to QCD where the mass scale is provided by non-perturbative condensates.
At the same time, three-dimensional Yang-Mills theory (especially without fermions) is readily amenable to lattice simulations and excellent data exists on its spectrum \cite{Teper1997, Teper1998, Diakonov:1999fq, Lucini:2002wg, Meyer:2002mk, Meyer:2003wx, Bringoltz2007, Buisseret:2013ch, Bursa2013, Athenodorou2016, Athenodorou:2016ebg, Lau2017, Teper2018, Conkey2019}.

Concretely, we will calculate the stress-tensor correlator \eqref{Pi intro} in pure three-dimensional Yang-Mills theory to three-loop order in perturbation theory.
Following traditional sum-rules approach, we then apply a Borel transform with respect to energy to improve convergence (see equation \eqref{eq:borel suppression} below).  
The main question is whether the Borel transform is dominated at low energies by the lightest glueball(s). 
We test this by assuming it is true and seeing whether it predicts reasonable values for the lowest glueball mass and its coupling to the stress tensor. 
The former is then compared with known lattice results, while the latter (to our knowledge) is a prediction.

In principle, this method can also be extended to higher-spin operators.
Knowing the set of couplings $\<G|\mathcal{O}^\ell|0\>$ of a given glueball to minimal-twist operators of various spins amounts to knowing its so-called lightcone
wavefunction.  
This wavefunction is closely related but distinct from parton distribution functions (which control deep inelastic scattering at high energies)
in that it controls elastic scattering at high energies  \cite{Chernyak:1977as,Lepage:1979za,Brodsky:1980ny}.
In the QCD context, such quantities have been estimated using sum-rules for higher-spin currents \cite{Chernyak:1987nt}.  
We initiate the investigation of higher-spin sum-rules for three-dimensional Yang-Mills theory.

In section \ref{sec:2ptfn}, we describe the stress-energy two-point function and provide some details of our 3-loop calculation. 
While we quote only its three-dimensional limit in the main text, the $d$-dimensional results can be found in appendix \ref{app:FFd}. 
Up to two-loops, we include cross-checks on the imaginary part using on-shell methods.  
In section \ref{sec:sum-rules}, we review Borel-transformed sum-rules for two-point functions and use simple models for the spectral density to extract the glueball masses and couplings from a $\chi^2$-fit. 
We also comment on the comparison with lattice results.  
These estimates are not rigorous and we explain in section \ref{sec:UC consistency} that essentially any low-energy spectral density can be compatible with perturbative asymptotics.
Lastly, in section \ref{sec:higher-spin}, we compute the perturbative two-point functions of more general higher-spin operators and show the existence of ``superconvergent'' sum-rules. 
This analysis leads to a crude approximation of glueball couplings to these operators.

\section{Stress-energy tensor two-point function \label{sec:2ptfn}}

In this section, we review our conventions for the YM-Lagrangian and define the stress-energy two-point functions relevant to this work. 
We compute the spin-0 and spin-2 two-point functions at one-loop in section \ref{sec:1loop}. In section \ref{sec:1and2LoopUnitarity}, we cross-check the discontinuities of one-loop two-point functions and predict the  two-loop discontinuities from unitarity cuts.
Then, we compute the full two-point functions at two-and three-loops in section \ref{sec:23loops}. 
In section \ref{sec:magic}, we identify a combination of the two-point functions with particularly good behaviour near $p^2=0$. This ``superconvergent'' combination will be central to the sum-rule analysis of section \ref{sec:sum-rules}.

The YM Lagrangian is comprised of three parts: a pure YM Lagrangian $\mathcal{L}_{\text{YM}}$, a gauge fixing condition $\mathcal{L}_{\text{gf}}$ and a ghost Lagrangian $\mathcal{L}_{\text{gh}}$. Explicitly, the total Lagrangian is (we work in mostly-plus metric signature) is
\begin{equation}\begin{aligned} \label{eq:L}
	\mathcal{L} & =-\frac{1}{4g^2_s} \left( F_{\mu\nu}^a \right)^2+\mathcal{L}_{\text{gf}}+\mathcal{L}_{\text{gh}}
\end{aligned}\end{equation}
where 
\begin{align}
	F_{\mu\nu}^{a} 
		& =\partial_{\mu}A_{\nu}^{a}
			-\partial_{\nu}A_{\mu}^{a}
			+f^{abc}A_{\mu}^{b}A_{\nu}^{c}
\end{align}
is the YM field strength. 
Since we will eventually specialize to $d=3$ spacetime dimensions rather than four, it is useful to compare the mass dimension of the coupling constant:
\begin{align}
	[g_s^2] &= 4 - d\to 
		\begin{cases}
			0 & \text{for } d=4 ,
			\\ 
			1 & \text{for } d=3.
		\end{cases}
\end{align}
Comparing, we see that the coupling constant provides a natural scale in three-dimensions but not in four-dimensions. 
This is one of the main reasons we will be interested in $d=3$ in this work: confinement and the bound state spectrum is controlled by the scale $m\sim g_s^2C_A$ instead of being an inherently non-perturbative function of the cut off $\Lambda_{\rm QCD}$ in four dimensions.

The stress-energy tensor is given by the expression 
\begin{align} \label{eq:T}
	T^{\mu\nu} & = 
		\frac{1}{g^2_s}\left(\left(F^{a}\right)^{\mu\lambda} \left(F^{a}\right)_{\ \lambda}^{\nu}
			-\frac14g^{\mu\nu}F^{2}\right).
\end{align}
Since the SU($N_c$) gauge theory admits parity and charge conjugation symmetries, the spectral decomposition \eqref{Im Pi intro} admits the group theoretic expansion
\begin{align} \label{eq:TGGT}
	\Im\, \Pi^{\mu\nu\alpha\beta} (p^2)
	\propto \sum_{J,P,C} 
	\la 0 \vert T^{\mu\nu}(p^2) \vert G_{J}^{PC} \ra
	\la G_{J}^{PC} \vert T^{\alpha\beta}(p^2) \vert 0 \ra
\end{align}
where the overlap $\la G_{J=0,2}^{++} \vert T^{\mu\nu}(p^2) \vert 0 \ra \neq 0$ is only nonvanishing for spins $J=0,2$ and $PC=++$.\footnote{
In $2+1$ spacetime dimensions, parity $P$ is a reflection $(x,y,t)\mapsto(-x,y,t)$ which anticommutes with the angular momentum of a particle. Thus any massive particle of spin $J\neq 0$ comes in a degenerate multiplet
$\{|J\>, |{-}J\>\}$.  Since any such multiplet is unitarily equivalent, 
the $P$ superscript is only meaningful (in the continuum theory) for $J=0$ states, see \cite{Teper1998} for discussion.  
}
In section \ref{sec:sum-rules}, we will try to use its two-point function to extract approximations for the masses and couplings of the lowest-lying glueball states.

The stress-energy tensor two-point function has four hanging Lorentz indices.
The Ward identities imply that a certain combination is transverse with respect to the external momentum $p$ (see \cite{Policastro:2002tn}):
\begin{equation}
 p_\mu \left( \Pi^{\mu\nu\alpha\beta} (p^2)  + g^{\nu\alpha} \< T^{\mu\beta}\>
 + g^{\nu\beta} \< T^{\mu\alpha}\> - g^{\mu\nu} \< T^{\alpha\beta}\>\right)=0\,. \label{Ward}
\end{equation}
We focus on the vacuum state, where all the above objects are constrained by Lorentz invariance.
There are only two transverse tensor structures with four Lorentz indices that are symmetric in each pair:
\begin{align}
	\label{eq:phi0}
	\phi_{0}^{\mu\nu\alpha\beta}(p) & \equiv 
	\phi^{\mu\nu}\phi^{\alpha\beta},
	\\
	\label{eq:phi2}
	\phi_{2}^{\mu\nu\alpha\beta} (p) & \equiv 
		\phi^{\mu\alpha} \phi^{\nu\beta}
		+ \phi^{\mu\beta} \phi^{\nu\alpha} 
		- c_{d} \phi^{\mu\nu} \phi^{\alpha\beta}\,,
\end{align}
where $c_{d} = \frac{2}{d-1}$ and
\begin{align}
	\label{eq:phi}
	\phi^{\mu\nu} (p) & \equiv
		\left( g^{\mu\nu} - \frac{p^\mu p^\nu}{p^2} \right).
\end{align}
Consequently, the general solution, $\Pi^{\mu\nu\alpha\beta}$, to the Ward identities \eqref{Ward} has a simple form:
\begin{equation}\begin{aligned} \label{eq:TT}
	\Pi^{\mu\nu\alpha\beta} (p^2) & =
	\frac{d_G}{512} \left[
		A_{0}(p^{2})\phi_{0}^{\mu\nu\alpha\beta}\left(p\right)
		+ A_{2}(p^{2})\phi_{2}^{\mu\nu\alpha\beta}\left(p\right)
	\right] \\ &\phantom{=}+ \left(g^{\mu\alpha}g^{\nu\beta}+g^{\mu\beta}g^{\nu\alpha}-g^{\mu\nu}g^{\alpha\beta}\right)\Lambda,
\end{aligned}\end{equation}
where we have set $\<T^{\mu\nu}\> = -\Lambda \delta^{\mu\nu}$.
For future convenience, we have absorbed a numerical factor as well as a factor of $d_G$: the dimension of the gauge group ($d_G=N_c^2-1$ for $G=SU(N_c)$).
The value of $c_{d}$ was chosen so that the spin-2 structure is traceless in each pair, which also makes it orthogonal to $\phi_0$:
\begin{align}
	\left(\phi_{2}\right)_{\ \mu}^{\mu\ \ \alpha\beta}
	=0=\left(\phi_{2}\right)_{\ \ \ \ \alpha}^{\mu\nu\alpha},\qquad
 \phi_0^{\mu\nu\alpha\beta}(\phi_2){}_{\alpha\beta\mu\nu} = 0\,.
\end{align}
Therefore, the two-point functions $A_0(p^2)$ and $A_2(p^2)$ receive contributions from only spin-0 and spin-2 intermediate states in the group theory decomposition \eqref{eq:TGGT}, respectively.

In principle, the vacuum energy density $\Lambda$ could be set to zero by a judicious choice of renormalization scheme. However, since we will perform our calculations in
a preset minimal subtraction scheme such as $\MSbar$, we do not have the freedom to set it to zero.
Namely, the vacuum energy density is proportional to the gluon condensate,
$\Lambda = \frac{d-4}{4d}\<\tfrac{1}{g_s^2}F^2\>$, which was estimated
in \cite{Hietanen:2004ew,DiRenzo:2006nh} using a combination of lattice and perturbative techniques.
Its size however is of order $\sim (\scale)^3$ which is beyond the accuracy of our calculations and thus we can effectively ignore the second line of \eqref{eq:TT}.

Beyond the decomposition into tensor structures, each two-point correlator is also decomposed into a loop-expansion
\begin{align}
	A_\bullet = \sum_{l=0} A_\bullet^{(l)}
\end{align}
where each $A_\bullet^{(l)}$ is proportional to the coupling $(\scale)^l$ and corresponds to the $L=l+1$ loop contribution to the two-point correlators. 
By dimensional analysis, the zeroth order terms come with a power $A_\bullet\propto p^d$ while each subsequent correction comes with an additional $1/p$ suppression at large momentum.
The loop-expansion of the two-point functions is described explicitly to three-loops in sections \ref{sec:1loop} and \ref{sec:23loops}.

Starting from four loops, the large-$p$ expansion ceases to be perturbatively calculable due to the appearance of non-perturbative condensates.
This can be understood by using the operator product expansion to separate calculable high-energy components from
low-energy condensates (see \cite{Novikov:1984rf}):
\begin{align} \label{OPE condensate}
    \int \d^dx\ e^{i p \cdot x} 
        \la T^{\mu\nu}(x) T^{\alpha\beta}(0) \ra_c
    \sim C_1^{\mu\nu\alpha\beta}(p)\ \la \mathds{1} \ra
    + C_{F^2}^{\mu\nu\alpha\beta}(p)\ \la \tfrac{1}{g_s^2}F^2(0) \ra
    + \cdots
    \ ,
\end{align}
where it is easy to see from tree-level Wick contractions 
that $C_{F^2}\sim (p^2)^0 (\scale)^0$ while the corresponding expectation value is $\sim (\scale)^3$.
The condensates will play no role in the present paper.

The structure of logarithms $\log(p^2)$ can be anticipated by 
applying the renormalization group equation (Callan-Symanzik equation)
to \eqref{OPE condensate} (see \cite{Zinn-Justin:2002ecy,Shifman1992}).
The theory has a single coupling $g_s^2$ whose running, by dimensional analysis, cannot be affected by perturbative quantum corrections at any order.  The stress tensor $T^{\mu\nu}(x)$ does not renormalize multiplicatively. It mixes additively with $g^{\mu\nu}\mathds{1}$ at four loops \cite{Kajantie:2002wa}, but this does not affect the connected correlator \eqref{OPE condensate}.  Thus the left-hand-side is independent of 
the $\MSbar$ scale $\mubar$ except for possible contact terms (polynomial in $p$), which can only appear at two and four loops by dimensional analysis, and can only affect specific combinations of $A_0$, $A_2$ and $\Lambda$ in accordance with \eqref{eq:TT}.  On the right-hand-side we can have mixing between the condensates, which again is only relevant starting from four loops. The most physically relevant combination, to be introduced in subsection \ref{sec:magic}, will turn out to cancel both two- and four-loop divergences.

\subsection{One-loop two-point functions \label{sec:1loop} }

In this section we present the one-loop two-point functions for $d=3$. 
The generic $d$ results can be found in appendix \ref{app:FFd}.

The two-point functions ($A_0$ and $A_2$) were computed in generic dimension $d$ using standard Feynman diagram techniques.
We extract these two-point functions from the stress-tensor correlator using the tensor decomposition \eqref{eq:TT}. 
This ensures that at each step we were working with Lorentz invariant quantities and is essential for the application of standard integration-by-parts (IBP) software such as {\tt FIRE} \cite{Smirnov2019}.

In practice, we used Feynman rules in Feynman gauge. 
While gauge invariance of the two-point functions was not checked due to this choice, two other consistency checks were performed. 
First, the conservation of $\Pi^{\mu\nu\alpha\beta}$ was checked by contracting a factor of $p$ into each hanging index of $\Pi^{\mu\nu\alpha\beta}$ while keeping the rest free. 
After applying IBP reduction, we find that the contraction of $p$ with any index of $\Pi^{\mu\nu\alpha\beta}$ vanishes. 
Secondly, we cross-check the discontinuity of $\Pi^{\mu\nu\alpha\beta}$ in $d=3$ at one- and two-loops from unitarity cuts.

\begin{figure}
	\centering
	\includegraphics[width=0.3\columnwidth]{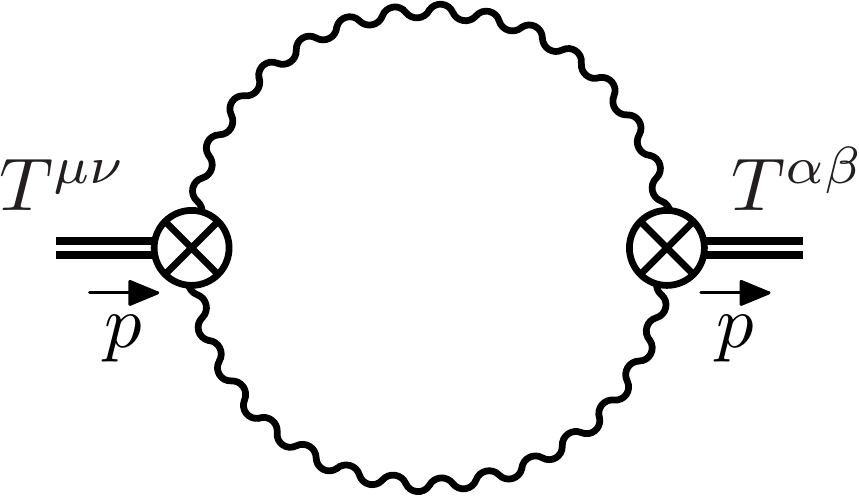}
	\caption{ \label{fig:1loopTT}
		Feynman diagram for the 1-loop 
		$TT$-correlation function.
            The circled cross denotes the vertex 
            associated to the insertion of a stress-tensor.
	}
\end{figure}

At one-loop, there is a single Feynman diagram (see fig.~\ref{fig:1loopTT}) that contributes to the one-loop correlation function $\Pi^{\mu\nu\alpha\beta}$. 
The circled cross in figure \ref{fig:1loopTT} denotes the vertex associated to the stress-tensor coupling to two gluons which can be derived via standard textbook techniques \cite{Peskin1995,Srednicki2007,Schwartz2014}. 
After integral reduction and integration, the $d=3$ two-point functions assocated to the stress-tensor two-point function \eqref{eq:TT} are
\begin{align}
\label{eq:1loop-FF}
	A^{(0)}_0(p^2) 
	&\underset{d\to3}{=} 2 (p^2)^{3/2} + \mathcal{O}(\ep),
	\\
	A^{(0)}_2(p^2)
	&\underset{d\to3}{=} (p^2)^{3/2} + \mathcal{O}(\ep).
\end{align}

\subsection{One- and two-loop two-point functions via the unitarity method \label{sec:1and2LoopUnitarity}}

In this section, we obtain the non-analytic part of the one-loop and two-loop correlation function, $\Pi^{\mu\nu\alpha\beta}$, with an independent calculation based on unitarity cuts. 

\begin{figure}[h]
\centering
\includegraphics[scale=0.3]{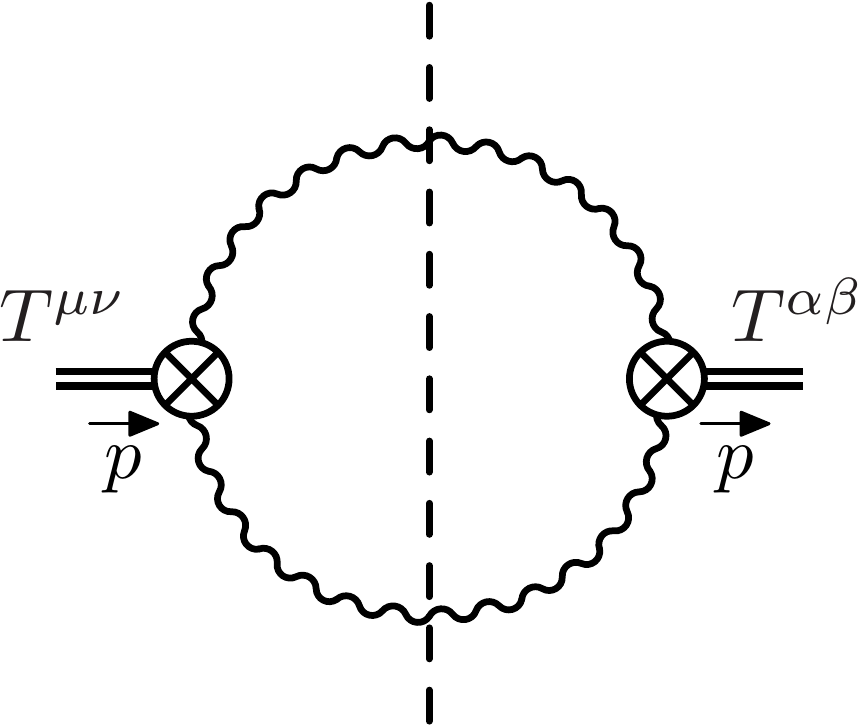}
\caption{Cut of the 1-loop 
		$TT$-correlation function is depicted. We can get each side (eq.~\eqref{cut1loop}) from basic consistency principles (Bose symmetry and conservation of stress-tensor). We can then use unitarity cut to obtain the discontinuity of the correlation function. \label{fig:1-loop-cut}}
\end{figure}

We start by writing a general ansatz for the on-shell process of creating two gluons with momentum $p_1$ and $p_2$ from a stress-tensor operator $T^{\mu\nu}(p)$. Our ansatz needs to be a symmetric in $p_1$ and $p_2$ where the coefficients are fixed by imposing the conservation of the stress-tensor operator (i.e., $\partial_{\mu}T_{\mu \nu}=0$). The form factor including the color factor is then,
\begin{equation}
\label{cut1loop}
\langle p_1^gp_2^g|T^{\mu\nu}(p)|0\rangle=\delta_{ab}(p_1^{\mu}p_2^{\nu}+p_1^{\nu}p_2^{\mu}-\delta^{\mu\nu}p_1\cdot p_2).
\end{equation}
To get the discontinuity of the one-loop correlation function depicted in figure \ref{fig:1-loop-cut}, 
we cut the diagram and glue the two sides together using the Cutckosky cutting rules. This yields,
\begin{equation}
\begin{split}
\text{Disc }\Pi^{\mu\nu\alpha\beta} (p^2) &=\frac{-id_G}{2!}\int\frac{d^2p_1}{(2\pi)^22E_1}\frac{d^2p_2}{(2\pi)^22E_2}(2\pi)^3\delta^3(p-p_1-p_2)\\
&\times \langle 0|T^{\mu\nu}(p)|p_1^gp_2^g\rangle\langle p_1^gp_2^g|T^{\alpha\beta}(p)|0\rangle,
\end{split}
\end{equation}
where 
\begin{align} 
	\disc A(p^2) 
	{=} A(p^2 {-} i \epsilon) {-} A(p^2 {+} i \epsilon) 
        .
\end{align}

Extracting the spin-0 and spin-2 two-point functions (see eq.~\eqref{cut1loop}) and going to the rest frame of $p$,
\begin{equation}
p=(\sqrt{s},0) \rightarrow \vec{p}_1+\vec{p}_2=0, \qquad E_1+E_2=\sqrt{s},
\end{equation}
the components of the correlation function (eq.~\eqref{eq:TT}) are reduced to trivial integrals over the angles. 
For example, 
\begin{equation}
\text{Disc }A_0^{(0)}(p^2)=\frac{-512i}{16\pi}\int d\theta\int\frac{dE_1}{E_1}\delta(\sqrt{s}-2E_1)\frac{(2E_1^2)^2}{4}
=\frac{4is^2}{\sqrt{s}}.
\end{equation}
Using Disc${}\sqrt{p^2}=-2i\sqrt{s}$, we can undo the cut to get the non-analytic contribution to the one-loop two-point function \eqref{eq:TT}
\begin{equation}
A_0^{(0)}(p^2)=2(p^2)^{\frac{3}{2}}, \qquad A_2^{(0)}(p^2)=(p^2)^{\frac{3}{2}}.
\end{equation}
This, of course, matches eq.~\eqref{eq:1loop-FF} when $d=3$. 

Next, we also compute the non-analytic part of the two-loop correlation function by employing unitarity cuts and on-shell form factors. 
This can then be compared with the full result including the analytic parts given in eq.~\eqref{eq:2loops-TT}. 

\begin{figure}[h]
\centering
\includegraphics[align=c,width=.3\columnwidth]{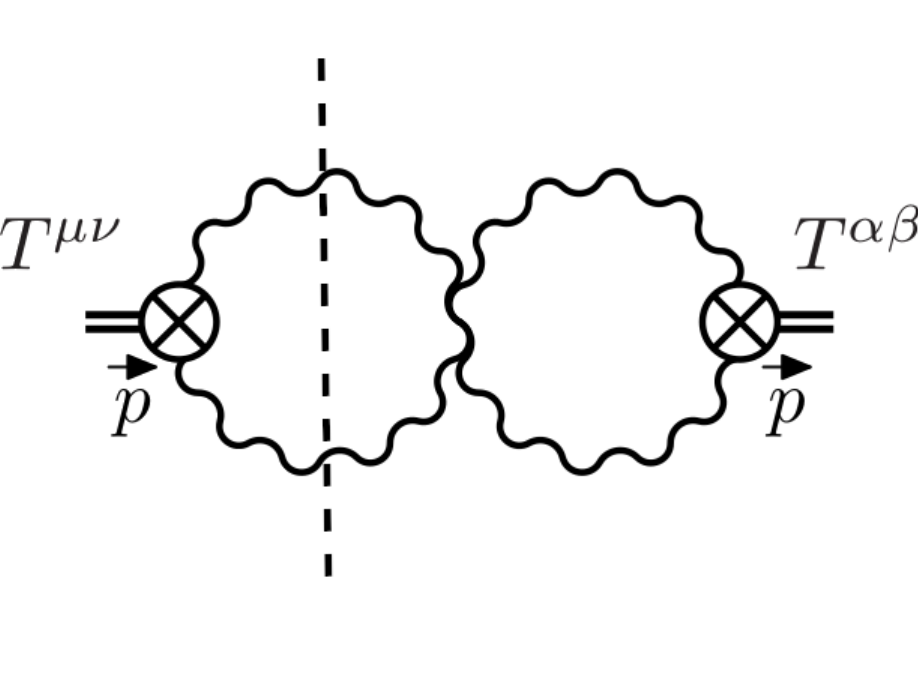}
\quad
\includegraphics[align=c,width=.3\columnwidth]{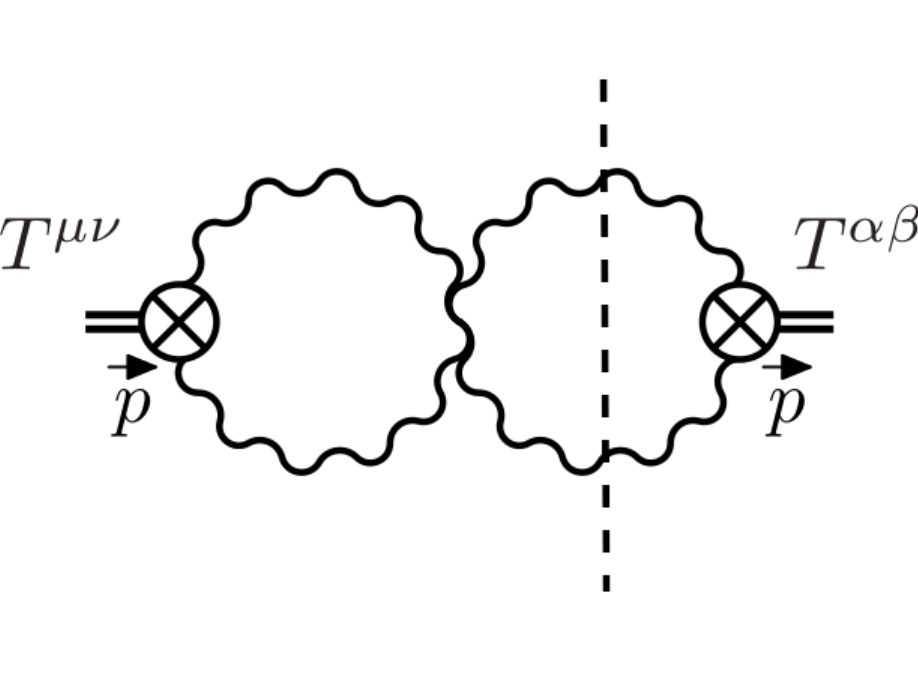}
\quad
\includegraphics[align=c,width=.23\columnwidth]{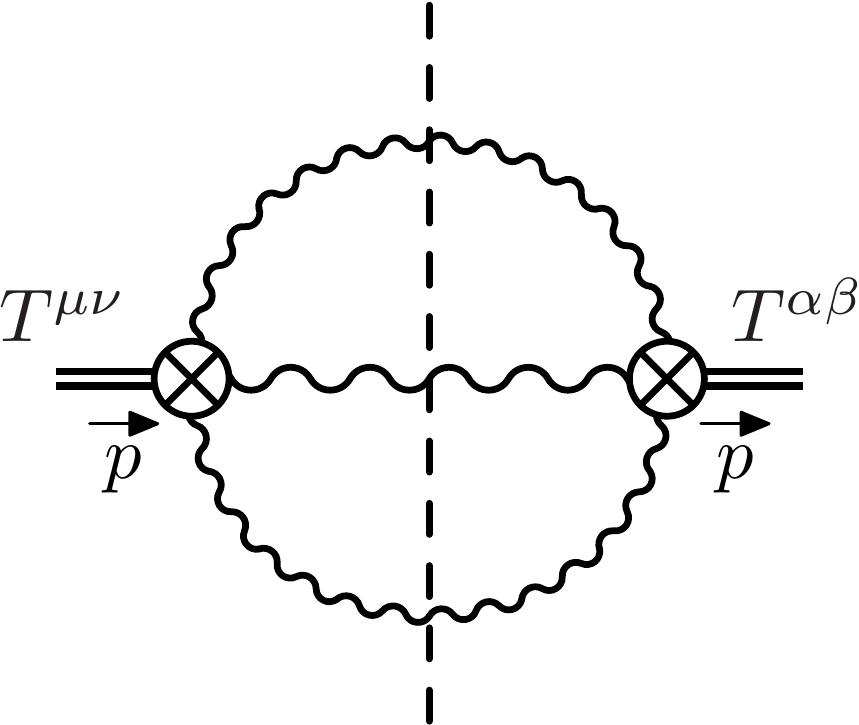}
\caption{Unitarity cuts of the diagrams contributing to the two-loop stress-tensor two-point function. The first and second diagrams are complex conjugate of each other and once added have zero discontinuity. Only the third diagram contributes to the non-analytic part. The three-gluon form factor $\langle p_1^gp_2^gp_3^g|T^{\mu\nu}|0\rangle$ in the right most figure is calculated using BCFW recursion relation. \label{fig:2-loop-cut}}
\end{figure}

We start by examining the unitarity cuts of the two-loop diagrams depicted in fig.~\ref{fig:2-loop-cut}. Importantly, the two-cuts of the double bubbles are complex conjugate to each other. 
This is because the tree-level form factor $\langle p_1^g p_2^g|T|0\rangle$ given by eq.~\eqref{cut1loop} is real and goes as $p^2$.
Moreover, the one-loop form factor, which has an additional $g_sC_A$, must scale like $(p^2)^{1/2}$ and has a discontinuity that is purely imaginary. 
Thus, the double bubbles do not contribute to the discontinuity of the correlator since their imaginary part cancels when summed.
This then means that the only contribution to the non-analytic part is contained in the right diagram in fig.~\ref{fig:2-loop-cut}.

To calculate the unitarity cut in the right most diagram in fig.~\ref{fig:2-loop-cut}, we need the on-shell three-gluon form factor $\langle p_1^gp_2^gp_3^g|T^{\mu\nu}|0\rangle$. 
This is obtained by studying the four-dimensional two gluons form factor and then using BCFW \cite{Britto:2005fq, Britto:2004ap} as explained in appendix \ref{app:bcfw}.
We can then go back to three-dimensions by using  $\epsilon^{3d}=\frac{\epsilon^++\epsilon^-}{2}$ \footnote{This is because in three-dimensions the Lorentz group is isomorphic to $SU(2)$ whereas in four-dimension it is isomorphic to $SU(2)\times SU(2)$. Accordingly, the little group for massless particles changes from $SO(2)$ to $Z_2$}.
The final result for the three-gluon form factor is,
\begin{equation}
\label{2loopcut1}
\langle p_1^gp_2^gp_3^g|T^{\mu\nu}|0\rangle=2g_{s}f^{abc}\frac{\sum_{i=1}^3(p_i^{\mu}p^{\nu}+p^{\mu}p_i^{\nu}-g^{\mu\nu}p_i\cdot p)(p_i\cdot p)-p_i^{\mu}p_i^{\nu}p^2}{\langle 12\rangle\langle 23\rangle\langle 31\rangle},
\end{equation}
where $\langle 12\rangle^2=-2p_1\cdot p_2$.
As a consistency check, we have verified that this equation correctly reproduces the two-gluon form factor, $\langle p_1^gp_2^g|T^{\mu\nu}|0\rangle$, in the soft $p_3$ limit. 

We then glue the form factors and perform the phase space integral as elucidated in appendix \ref{app:2loopcut} to obtain the non-analytic parts of the correlation function at 2-loops: 
\begin{equation} 
    \label{eq:2-loop unitarity}
    A_0^{(1)}=(g_s^2C_A)\frac{8}{3\pi}p^2\log (p^2), \qquad   A_2^{(1)}=-(g_s^2C_A)\frac{8}{3\pi}p^2\log p^2.
\end{equation}
As expected, this correctly reproduces the non-analytic part of the stress-tensor two-point functions computed using Feynman diagram methods eq.~\eqref{eq:2loops-TT} when $d=3$. 
The fact that the scale dependence cancels when we sum these two channels will be significant below.

\subsection{Two- and three-loop two-point functions \label{sec:23loops}}

As we saw in the previous section, the unitarity method gives the imaginary part of the two-loop contribution to the stress-tensor two-point function \eqref{eq:2-loop unitarity}. However, it will be useful to also have the constant part of the two-loop contribution since it contributes to the sum-rules. In fact, we do one-loop more and compute the stress-tensor two-point function to three-loops. Here, we will use Feynman diagrams because it is easier than $d$-dimensional unitarity.
Since the calculation methodology was reviewed in section \ref{sec:1loop}, we simply present the two- and three-loop two-point functions in this section 

\begin{figure}
	\centering
	
	\includegraphics[align=c,width=.3\columnwidth]{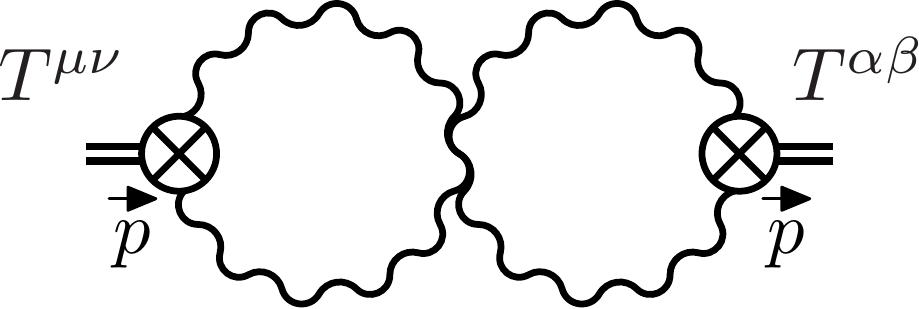}
	\quad
	\includegraphics[align=c,width=.3\columnwidth]{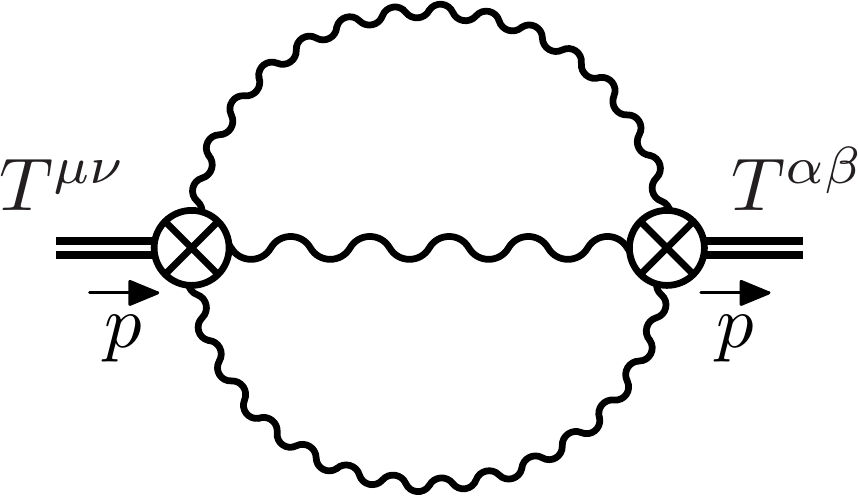}		
	\quad
	\includegraphics[align=c,width=.3\columnwidth]{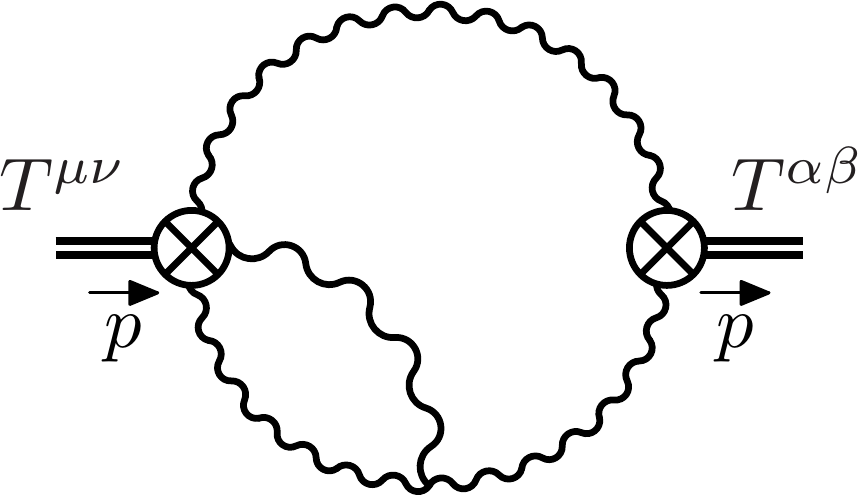}
		
	\includegraphics[align=c,width=.3\columnwidth]{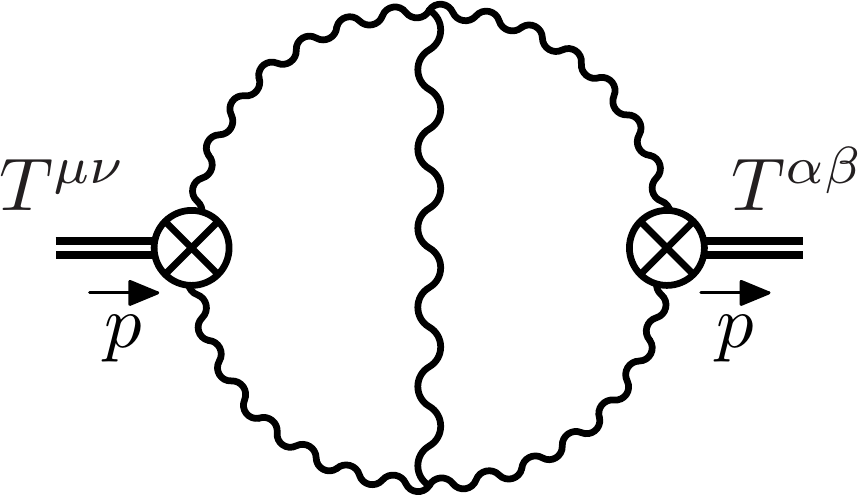}	
	\quad
	\includegraphics[align=c,width=.3\columnwidth]{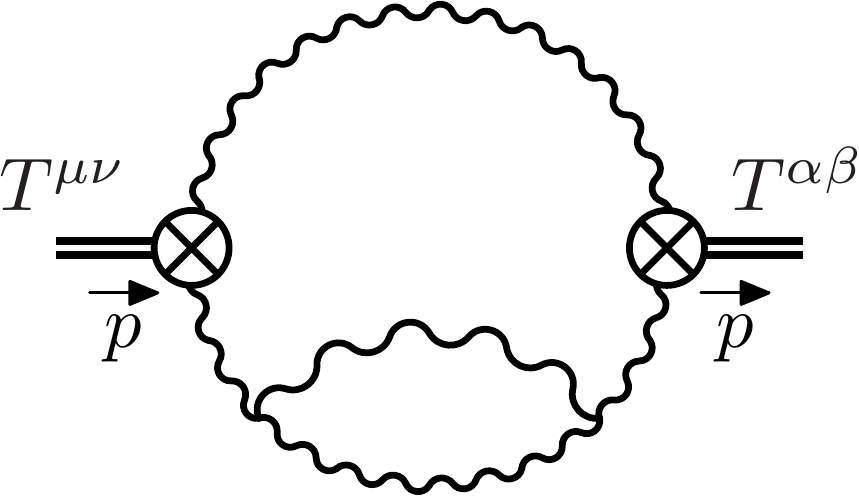}				
	\caption{ \label{fig:2loopTT}
		Feynman diagrams contributing to the 
            the two-loop stress-tensor 
            two-point function.
		Note that there is another diagram 
            not shown here
            where the gluon sub-bubble 
		in the last diagram is replaced 
            by a ghost bubble.
	}
\end{figure}

At two-loops the correlation function receives contributions form 8 diagrams including ghosts but only 7 topologies (see fig.~\ref{fig:2loopTT}). 
While there are 7 contributing topologies, there are ony two scalar master integrals at two-loops (see equation \eqref{eq:2loop masters} as well as equations \eqref{eq:2loop A0(d)} and \eqref{eq:2loop A2(d)}). 
When the dust settles, the two-loop $d=3$ two-point functions are
\begin{subequations}
\label{eq:2loops-TT}
\begin{align}
	A^{(1)}_0(p^2) 
	&\underset{d\to3}{=} 
		(g_s^2 C_A) p^2
		\bigg[ 
			{-}\frac{1}{4}
			{-}\frac{4}{3 \pi ^2 \ep }
			{+} \frac{8}{3 \pi ^2} \log\left(\frac{p^2}{\mubar^2}\right)
		\bigg]	{+} \mathcal{O}(\ep),
	\\
	A^{(1)}_2(p^2)
	&\underset{d\to3}{=} 
	 	(g_s^2 C_A) p^2
		\bigg[ 
			{-} 1
			{+} \frac{20}{3 \pi ^2}
			{+} \frac{4}{3 \pi ^2 \ep }
			{-} \frac{8}{3 \pi ^2} \log\left(\frac{p^2}{\mubar^2}\right)
		\bigg]	
		{+} \mathcal{O}(\ep),
\end{align}\end{subequations}
where $\mubar^2=4\pi e^{-\gamma_E}\mu^2$ is the $\MSbar$ renormalization scale.


\begin{figure}
	\centering
	
	\includegraphics[align=c,width=.3\columnwidth]{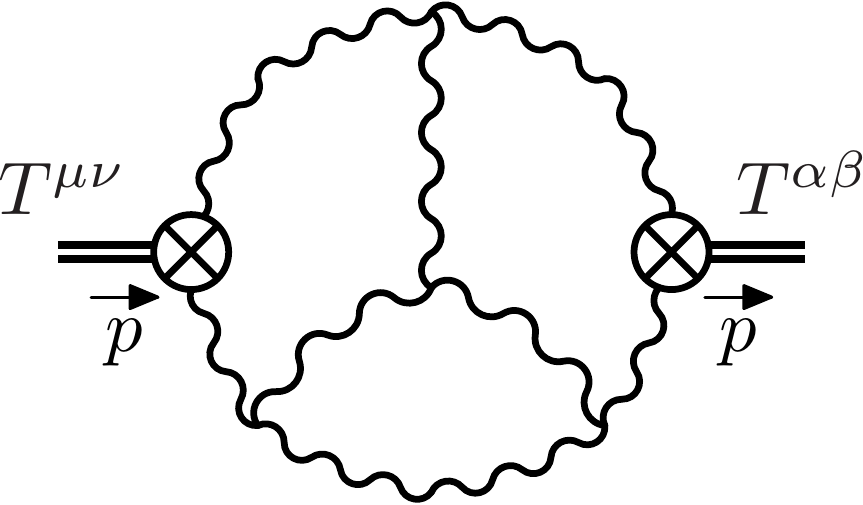}
	\quad
	\includegraphics[align=c,width=.3\columnwidth]{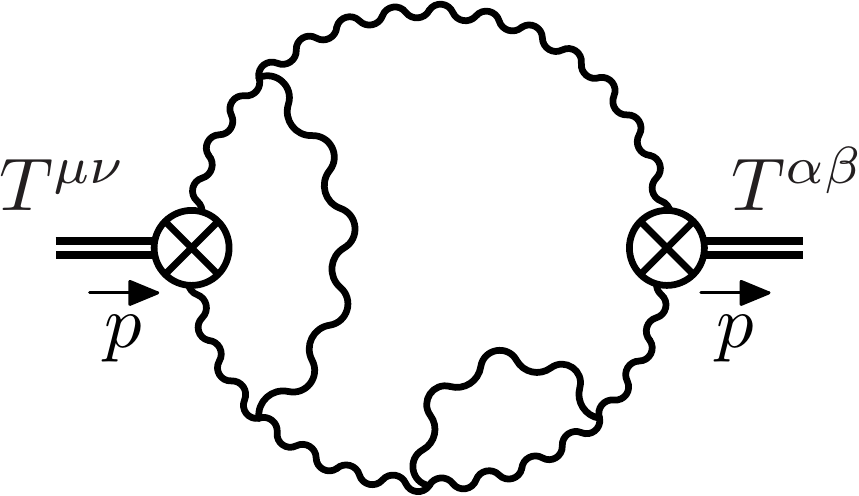}		
	\quad
	\includegraphics[align=c,width=.3\columnwidth]{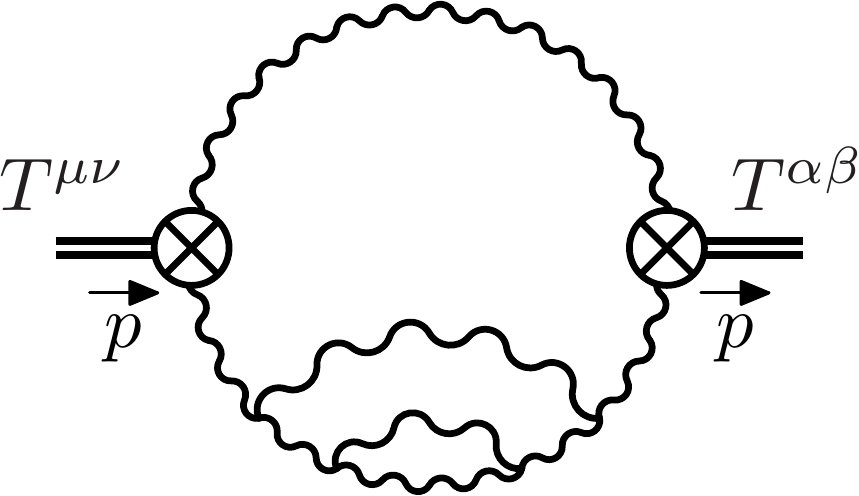}
		
	\includegraphics[align=c,width=.3\columnwidth]{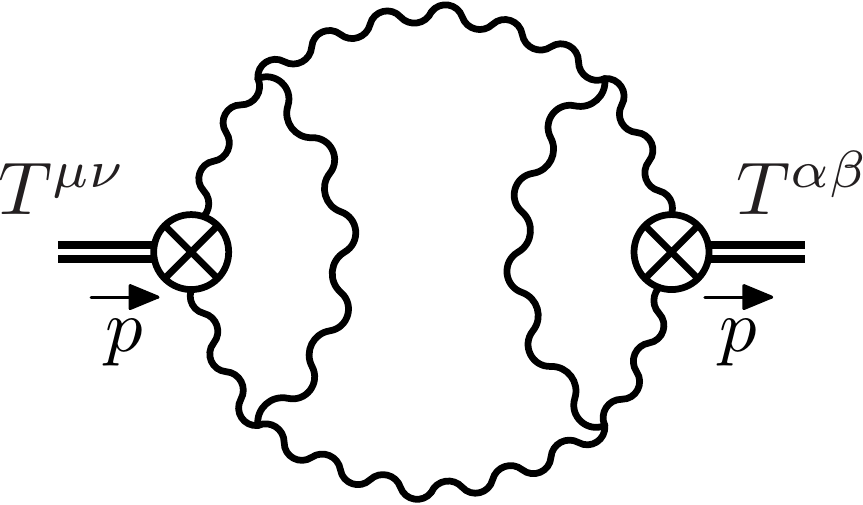}	
	\quad
	\includegraphics[align=c,width=.3\columnwidth]{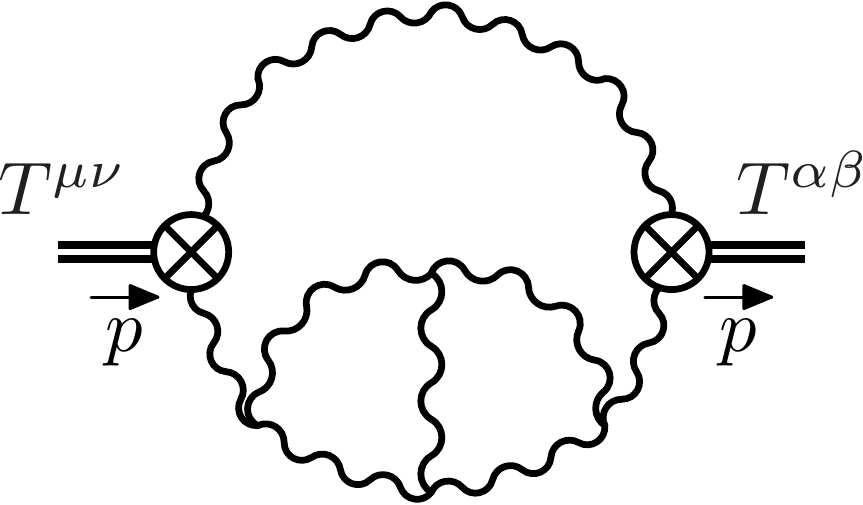}			
	\quad
	\includegraphics[align=c,width=.3\columnwidth]{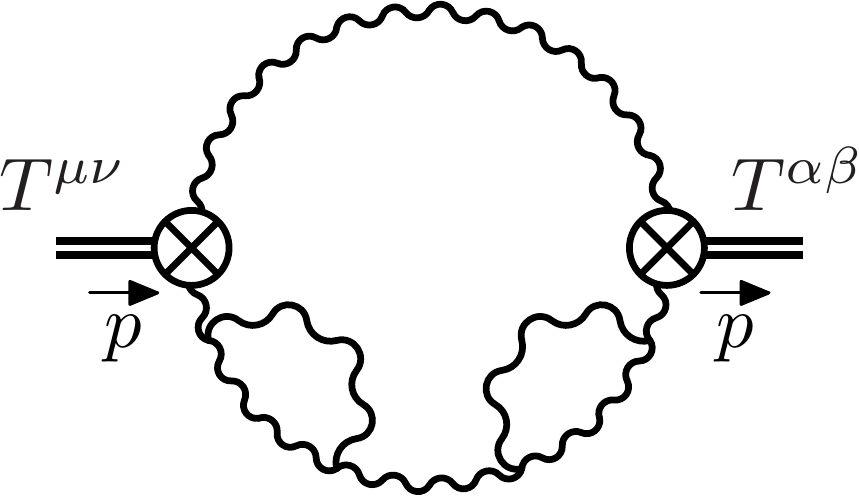}
	
	\includegraphics[align=c,width=.3\columnwidth]{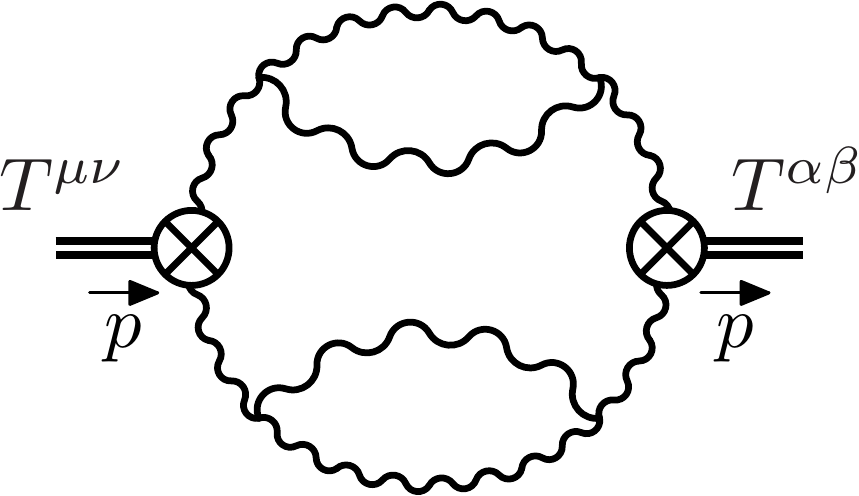}	
	\quad
	\includegraphics[align=c,width=.3\columnwidth]{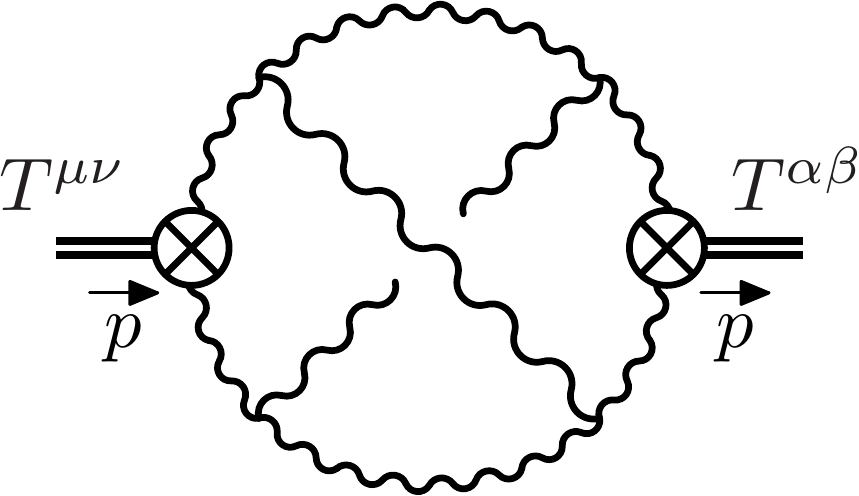}			

\caption{ \label{fig:3loopTT}
	Parent topologies for all Feynman diagrams contributing to the three-loop
	stress-tensor two-point function. All other Feynman diagrams are pinches of
	these. All ghost contributions have been suppressed.
}
\end{figure}

The three-loop correlation function receives contributions from a total of 41 different topologies where all consistent ways of distributing gluons and ghosts must be included. 
The three-loop parent topologies are listed in figure \ref{fig:3loopTT}: all other topologies can be recovered from these by pinching a subset of propagators in a parent topology to points.
Out of the 41 topologies, there are only 6 scalar master integrals (see equation \eqref{eq:3loop masters}). 
The $d$-dimensional three-loop two-point functions are presented in equations \eqref{eq:3loop A0(d)} and \eqref{eq:3loop A2(d)}. Taking the $d\to3$ limit, we find
\begin{subequations} \label{eq:3loops-TT}
\begin{align}
	A^{(2)}_0(p^2) 
	&\underset{d\to3}{=} 
		(g_s^2 C_A)^2 \sqrt{p^2}
		\bigg[ 
			\frac{155}{384}-\frac{13}{2 \pi ^2}
		\bigg]	
		{+} \mathcal{O}(\ep),
	\\
	A^{(2)}_2(p^2)
	&\underset{d\to3}{=} 
	 	(g_s^2 C_A)^2 \sqrt{p^2}
		\bigg[ 
			\frac{431}{768}-\frac{37}{9 \pi ^2}
		\bigg]	
		{+} \mathcal{O}(\ep).
\end{align}\end{subequations}
While most of the master integrals of \eqref{eq:3loop masters} are easily evaluated, $I^{(3)}_1$ and $I^{(3)}_2$ are particularly challenging.
Even though the $\ep$-expansion of these integrals is known for $d=4-2\ep$ \cite{Chetyrkin1980,Chetyrkin1981,Baikov2010}, we had to recompute the generic $d$ dependence from scratch in order to  obtain the $\ep$-expansion of these integrals in $d=3-2\ep$. 
The generic $d$ dependence of these integrals was determined using the method of dimensional recurrence and analyticity in $d$ \cite{Lee2010}.
A summary of this method along with the equations needed to recover the $d$ dependence of these integrals is presented in appendix \ref{app:dim rec and an}.


\subsection{Superconvergent combination of two-point functions \label{sec:magic}}

In this section, we introduce a ``superconvergent'' combination of two-point functions $A_0$ and $A_2$.
This combination is exceptionally well behaved as $p^2\to0$, which we can use to ameliorate the convergence of the K\"{a}ll\'en-Lehmann representation. Thus, it is ideally suited for the application of dispersive sum-rules in section \ref{sec:sum-rules}.

By expanding the tensor structure of the stress-tensor two-point funciton \eqref{eq:TT}, one finds a term with four uncontracted momenta:
\begin{align} \label{Pi magic schematic}
	\Pi^{\mu\nu\alpha\beta} (p^2)
	= \A(p^2) p^\mu p^\nu p^\alpha p^\beta + \cdots 
\end{align}
where
\begin{align}
	\A(p^2) \equiv \frac{1}{(p^2)^2}\left(A_0(p^2) + \frac{2(d-2)}{(d-1)}A_2(p^2)\right). \label{magic}
\end{align}
Here, the change of calligraphy from $A$ to $\A$ highlights that a rescaling by $1/(p^2)^2$ has been applied. 
Since other terms in \eqref{Pi magic schematic}
are proportional to $p^2$, the combination $\A(p^2)$ must be non-singular around $p^2=0$ in order for the correlator itself to be regular.
However, thanks to the denominator in \eqref{magic}, it decays faster and in fact vanishes at infinite momenta. Thus, it satisfies an unsubtracted K\"{a}ll\'en-Lehmann dispersion relation.

By combining the one-loop \eqref{eq:1loop-FF}, two-loop \eqref{eq:3loops-TT} and three-loop
\eqref{eq:3loops-TT} results, we obtain the following perturbative result for
the superconvergent two-point function $\A$ in the three-dimensional limit:
\begin{align} \label{A result}
	\A 
	=&\ \frac{a_0}{\sqrt{p^2}} 
		+ a_1 \frac{\scale}{p^2} 
		+ a_2 \frac{(\scale)^2}{(p^2)^{3/2}}
		+ a_3 \frac{(\scale)^3}{(p^2)^{2}} 
		+ \mathcal{O}\left(\frac{1}{(p^2)^{5/2}}\right).
\end{align} 
Here, 
\begin{align} \label{A result num}
    a_0 = 3,
    \qquad
    a_1 = \frac{16}{3\pi^2} -\frac{5}{4}
    \approx -0.710,
    \qquad
    a_2 = \frac{247}{256}-\frac{191}{18 \pi ^2} \approx -0.110\,,
\end{align}
and $C_A$ is the quadratic Casimir of the gauge group in the adjoint representation (i.e., $C_A = N_c$ for the gauge group $G=SU(N_c)$).

The superconvergent combination $\A$ enjoys other nice properties.
First, it is free from two-loop ultraviolet divergences, which can be checked explicitly by adding the two lines of \eqref{eq:2loops-TT}.  This is precisely as anticipated from the renormalization group argument below \eqref{OPE condensate}, since a two-loop divergence would have led to a non-polynomial term $\Pi^{\mu\nu\alpha\beta}(p)\sim \frac{p^\mu p^\nu p^\alpha p^\beta}{p^2}\log \mubar$.  Thus, all constants in \eqref{A result num} are unambiguous and scheme-independent. 

Second, even though we have not performed a four-loop calculation to determine $a_3$, we can predict that this coefficient is actually independent of the gluon condensate, which cancels out in the combination $\A$.  This can be seen from the fact that the Wick contraction of two field strengths that can give rise to the OPE coefficient
$C_{\rm F^2}^{\mu\nu\alpha\beta}$ in \eqref{OPE condensate} following \cite{Shifman1992},
cannot give rise to a $p^\mu p^\nu p^\alpha p^\beta$ term at leading order.%
\footnote{
Upon using the Ward identity \eqref{Ward} to fix all contact ambiguities and then imposing Lorentz invariance of condensates, we find specifically that
\begin{equation}
    C_{F^2}^{\mu\nu\alpha\beta}(p) = \frac{d-4}{d} \left(
     \phi_0^{\mu\nu\alpha\beta}(p) \frac{2(d-2)}{(d-1)^2}+
     \phi_2^{\mu\nu\alpha\beta}(p) \frac{1}{d-1} + \frac{g^{\mu\alpha}g^{\nu\beta}+g^{\mu\beta}g^{\nu\alpha}-g^{\mu\nu}g^{\alpha\beta}}{4}    
    \right) + \O(g_s^2/p),
\end{equation}
which is compatible with \eqref{eq:TT} and the relation between the condensate and vacuum energy.}
Therefore, the perturbative calculation of $a_3$ cannot display any infrared sensitivity and so must yield a finite, unambiguous constant.

In eq.~\eqref{A result} we have still included the term $a_3$ to parameterize our ignorance of the four-loop physics.
Given the decreasing pattern in the above coefficients, we believe that a reasonable range is for $a_3$ is
\begin{equation}
 a_3 \in \left[-\frac{1}{10},\frac{1}{10}\right]
\end{equation}
so that $|a_3| < |a_{2}|$.

Further note that the result \eqref{A result} is an asymptotic series in the large Euclidean region $p^2\gg \scale$ and its apparent singularity at $p^2=0$ is an artifact of perturbation theory since $\A$ must be regular at $p^2=0$ non-perturbatively.

\section{Sum-rules: estimating the glueball masses and couplings \label{sec:sum-rules}}

In this section, we review the dispersive sum-rules for the superconvergent combination $\A$. 
We start by constructing dispersion relations relating the $TT$-correlator in the Euclidean region to the correlator in the physical region in section \ref{sec:dis rel}. 
Then in section \ref{sec:Borel}, we describe how the Borel transform improves the convergence of the perturbative series of $\A$ in the limit $p^2\to0$. 
From the Borel transform of $\A$, we construct a function $\Mh$ that corresponds to the weighted average of the low-lying glueball masses.
Then, we compare the $\Mh$ obtained from truncating the perturbative expression of $\A$ to the $\Mh$ obtained from a non-perturbative model of $\A$. 
The Borel transform of the perturbative result for $\A$ is given in section \ref{sec:PBorel} while the Borel transform of the non-perturbative $\A$ is given in section \ref{sec:NPBorel}.
In sections \ref{sec:n=1} and \ref{sec:n=2}, we optimize the parameters of the one- and two-glueball models using a $\chi^2$ fit and extract 
estimates for the low-lying glueball masses and their couplings to the stress-tensor.
Lastly, in section \ref{sec:lattice}, we compare our values obtained from sum-rules to the lattice results.

\subsection{Dispersion relations \label{sec:dis rel}}

\begin{figure}
	\centering
	\includegraphics[scale=.25]{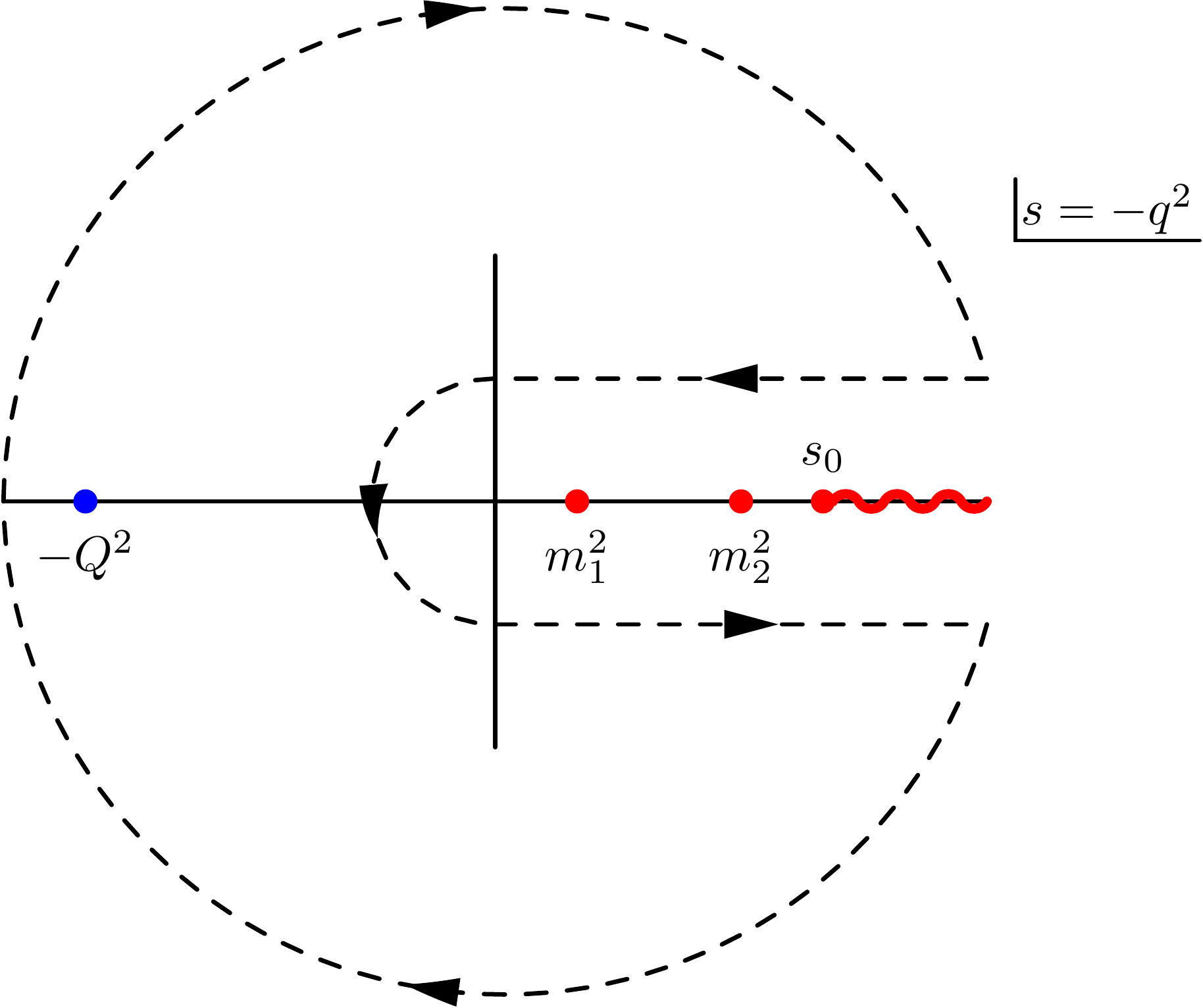}
	\caption{ \label{fig:2ptAnalyticStructure}
		Analytic structure of $\A$, which follows from 
		the K\"{a}ll\'en-Lehmann representation of $A_{0+2}$ and the fact that 
		$A_{0+2}\to0$ faster than $(p^2)^2$ as $p^2\to0$.
		The minimal assumption is that there is a pole at $q^2=-m_1^2$
		associated to the spin-0 glueball then a brach cut that 
		begins at $q^2 = s_0 \sim 4m_1^2$. 
		Here, $m_2$ is the mass of the next bound state
		(spin-2 glueball), which may or may not lie below $s_0$.
		The contour deformation used in the derivation of the dispersion
		relation is represented by the dashed line.
	}
\end{figure}

The superconvergent two-point function $\A$ inherits a K\"{a}ll\'en-Lehmann representation
\begin{align} \label{eq:KLrep}
	\A(p^2) 
	= \frac{1}{\pi} \int_{\mathbb{R}^+} \d s\ \frac{\rho(s)}{p^2+s-i\ep}
\end{align}
from the two-point functions $A_0$ and $A_2$ \eqref{eq:TT}.\footnote{In general, the Fourier transform of any 2-point function always has a K\"{a}ll\'en-Lehmann representation.} 
Here, $\rho$ is called the spectral density and is positive for timelike momenta $q^2=-s <0$.
The form of the spectral density follows directly form the spectral density of $A_0$ and $A_2$,
\begin{align} \label{eq:rho structure}
	\rho(s) 
	&= \sum_{i=1}^n \frac{2\pi g_i^2}{(m_i^2)^2}\,\delta\left(s-m_i^2\right)
		+ H(s) \, \Theta\left(s - s_0^2\right)
  ,
\end{align}
where $m_i$ is the mass of the $i^\text{th}$ bound state, 
$g_i$ is the residue of the $m_i$ pole (also the coupling constant of the $i^\text{th}$ bound state) and the continuum is assumed to start at $s_0 \sim 4m_1^2$.
Since the sum-rules are robust against perturbations in $s_0$ we set $s_0=4m_1^2$.
The $s$-dependence of $H$ is fixed by the asymptotic form of $\rho$, which is computable using perturbation theory. 
The normalization of the delta-function terms in \eqref{eq:rho structure} follows from the usual normalization of the $A_{0}$ and $A_2$ spectral densities and the fact that $\A = A_{0+2}/(p^2)^2$.
The form of the spectral density determines the analytic structure of $\A$ (see fig.~\ref{fig:2ptAnalyticStructure}). 

While the two-point functions have been computed in the limit of large spacelike momenta $q^2=Q^2>0$ (Euclidean region), we need $\A$ in the region of large timelike momenta. 
Thankfully, these regions are linked by a dispersion relation.
To see this, consider the following rewriting of $\A$ 
\begin{align}
	\A(Q^2) 
	= \oint_{q^2=Q^2} \frac{\d q^2}{2\pi i} \frac{\A(q^2)}{q^2-Q^2}
    ,
\end{align}
where the contour encircles the (spacelike) point $Q^2$.
Next, we deform the contour so that it encircles the poles along the real axis and hugs the branch cut in figure \ref{fig:2ptAnalyticStructure}
\begin{align} \label{eq:drel}
	\A(Q^2) 
	&=\int_{0}^{\infty} \frac{ \d s}{2\pi i} 
		\frac{\disc\A(-s)}{s + Q^2} 
\end{align} 
where $\disc\A$ is the discontinuity of $\A$ along the branch cut pictured in fig.~\ref{fig:2ptAnalyticStructure}
\begin{align}
	\disc\A({-}s) 
	{=} \A({-}s {-} i \epsilon) {-} \A({-}s {+} i \epsilon) 
        .
\end{align} 
Comparing \eqref{eq:KLrep} and \eqref{eq:drel} we see that the spectral density is given by the discontinuity
\begin{align} \label{eq:rho disc}
	\rho(s) 
	= \frac{\disc\A(-s)}{2i}, 
\end{align}
defined by the above contour deformation.

\subsection{Borel transformation \label{sec:Borel}}

The perturbative expansion of $\A$ is an asymptotic series and thus cannot be extended to the region of small timelike momentum $s\sim0$. 
Yet, in order to extract the mass of the low-energy bound states, we need to use the perturbative results at small $s$. 
To this end, we work with the Borel transform of $\A$, which improves the convergence of the asymptotic series and hope that the improved convergence of the perturbative result overlaps with low-energy glueball physics. 

The Borel transformation of the superconvergent two-point function is 
\begin{align} \label{eq:BT(2ptfn)}
	\Ah(M^2) 
	= \B \left[\frac{1}{\pi} \int \d s \ 
		\frac{\rho(s)}{s+Q^2} \right]
	= \frac{1}{\pi M^2} \int \d s \
		\rho(s)\ e^{-s/M^2},
\end{align}
where $M^2$ is the Borel parameter \cite{Shifman:1992xu}. 
A convenient way to implement the Borel transform of an asymptotic series in Euclidean momentum $Q^2$ is by acting with the following differential operator \cite{Shifman:1992xu}
\begin{equation}
    \label{eq:Borel op}
	\B = 
	\underset{Q^2/n=M^2}{
		\underset{Q^2\to\infty}{
			\underset{n\to\infty}{\lim}
		}
	}
	\frac{1}{(n-1)!} (Q^2)^n \left(-\frac{\d}{\d Q^2}\right)^n.
\end{equation}
In particular, all polynomials in $Q^2$ are killed by the Borel transform and the following accounts for most applications 
\begin{align}
	\B \left[  \left(\frac{1}{Q^2}\right)^n \right] 
		&= \frac{1}{\Gamma(n) (M^2)^n},
	\label{eq:borel rule 1}
	\\ 
	\B \left[ \left(\frac{1}{Q^2}\right)^n \log Q^2 \right]
		&= \frac{\log (M^2) }{\Gamma(n) (M^2)^n}
		+ \frac{\Gamma^\prime(n)}{\Gamma^2(n) (M^2)^n},
	\label{eq:borel rule 2}
\end{align}
In particular, note that the coefficient of $1/(Q^2)^n$ of the asymptotic series is suppressed by factor of $n!$ in the Borel transform
\begin{align} \label{eq:borel suppression}
	\B\left[
		\sum_{n\geq0} a_i \frac{1}{(Q^2)^n}
	\right]
	= \sum_{n\geq0} \frac{a_i}{n!} \frac{1}{(M^2)^n}.
\end{align}
The additional factors of $n!$ greatly improve the convergence of the Borel transformation for small Borel parameter $M^2$.
As a sanity check of \eqref{eq:Borel op}, one can use the above to show that 
\begin{align}
	\B\left[ \frac{1}{s+Q^2} \right] = \frac{1}{M^2} e^{-s/M^2}.
\end{align}
Then, since $\A(Q^2)$ satisfies the dispersion relation \eqref{eq:drel}, its Borel transform is exactly \eqref{eq:BT(2ptfn)}.

The Borel transform \eqref{eq:BT(2ptfn)} allows us to define a weighted average of the mass 
\begin{align} \label{eq:Mh def}
	\Mh
	\equiv \frac{\Ah^\prime(M^2)}{\Ah(M^2)}
	= \frac{\int \d s \ s\ \rho(s)\ e^{-s/M^2}
	}{
		\int \d s \ \rho(s)\ e^{-s/M^2}	
	}
\end{align}
where
\begin{align}
	\Ah^\prime(M^2)
	= -\frac{1}{M^2} \frac{\partial\left( M^2 \Ah \right)}{\partial(1/M^2)} 
	= M^2\frac{\partial\left( M^2 \Ah \right)}{\partial M^2} 
	=  \frac{1}{\pi M^2} \int \d s \
		s\ \rho(s)\ e^{-s/M^2}. 
\end{align}
Provided that the spectral density $\rho$ is dominated by the $m_1$ glueball, this quantity yields an estimate for $m_1^2$ . 
That is, at low $M^2$, $\Mh$ should have a plateau at roughly the height $m_1^2$. 
While this is indeed the case non-perturbatively, the truncated perturbative expression for $\Mh$ does not have this plateau due to the break-down of the perturbative series (as seen in  figure \ref{fig:Msqrd_pert}.

\subsection{Borel transformation of the perturbative result \label{sec:PBorel}}

\begin{figure}
	\centering
	\begin{subfigure}[b]{.49\textwidth}
		\centering
		\includegraphics[align=c,width=\textwidth]{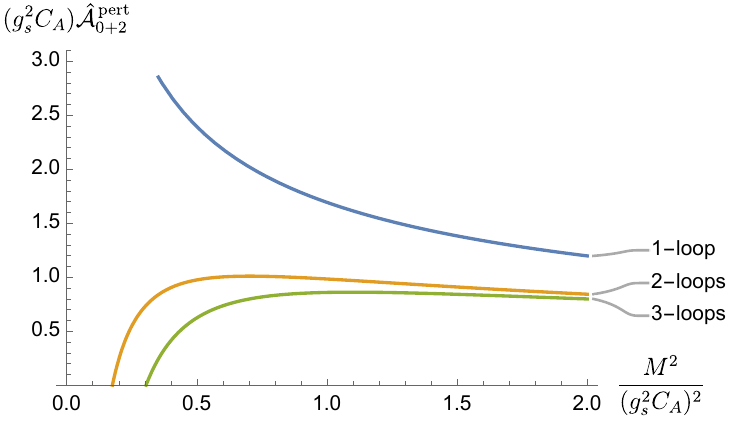}
		\caption{\label{fig:Pi0_pert}}
	\end{subfigure}
	\hfill
	\begin{subfigure}[b]{.49\textwidth}
		\centering
		\includegraphics[align=c,width=\textwidth]{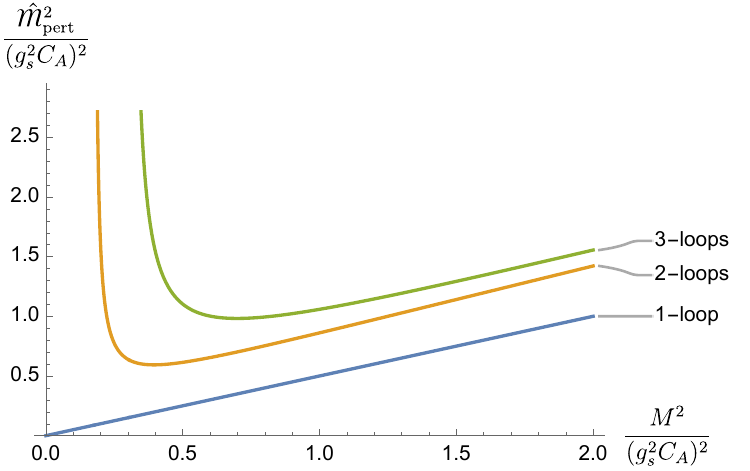}		
		\caption{\label{fig:Msqrd_pert}}
	\end{subfigure}

\caption{
	Figure \ref{fig:Pi0_pert}: 
		The Borel transform of the superconvergent combination $\Ah^\pert$ at one-, two- and three-loops. 
		The two- and three-loop curves converge quickly 
		for $M^2 \gtrsim (\scale)^2/2$.
	Figure \ref{fig:Msqrd_pert}:
		Weighted average of the mass (ratio of $\Ah^\pert$ 
		and its weighted derivative). Note that at one-loop there 
		is nothing stoping the mass from vanishing. On the other 
		hand, the two- and three-loop curves turn up producing a 
		minimum. 
}
\end{figure}

\begin{figure}
	\centering
	\begin{subfigure}[b]{.49\textwidth}
		\centering
		\includegraphics[align=c,width=\textwidth]{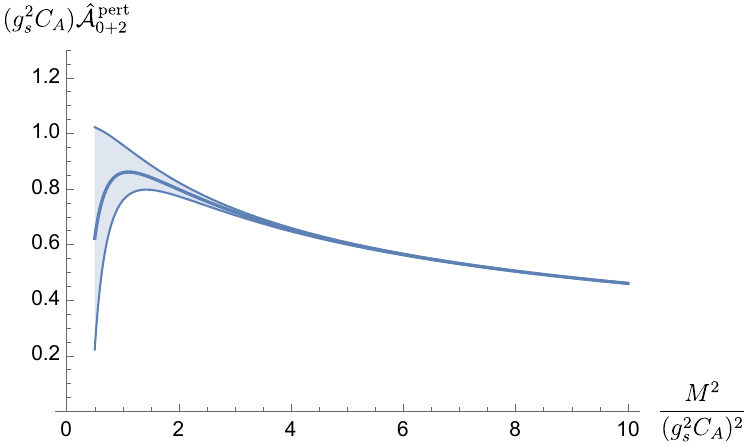}	
		\caption{\label{fig:Pi0_pert_errors}}
	\end{subfigure}
	\hfill
	\begin{subfigure}[b]{.49\textwidth}
		\centering
		\includegraphics[align=c,width=\textwidth]{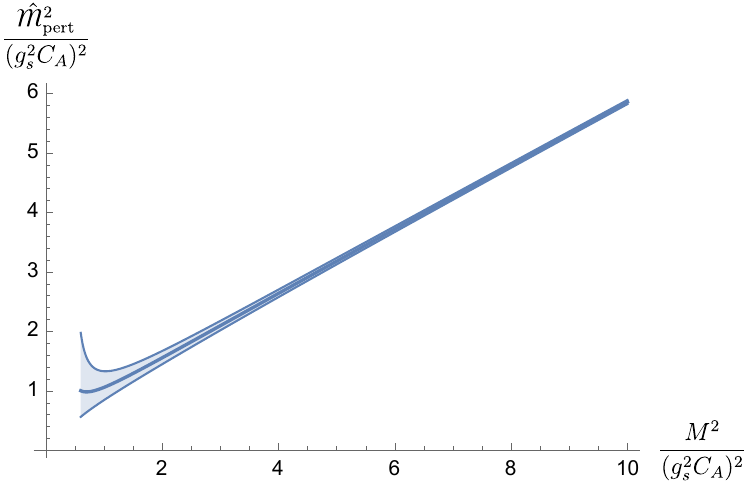}		
		\caption{\label{fig:Msqrd_pert_errors}}
	\end{subfigure}
	\caption{
		$\Ah^\pert$ and $\Mh_\pert$ where the shaded region
		represent the error in our calculations due to the unknown four-loop
		contributions. 
		The error is largest at low energy $M^2\ll1$ and shrinks to zero in the 
		high energy limit $M^2\gg1$. 
		These plots alone do not constrain the parameters of the one- and 
		two-glueball models since the parameters can always be tuned such that
		both $\Ah^\npert$ and $\Mh_\npert$ lie in the 
		corresponding shaded region. 
	}
\end{figure}

In this section, we compute the Borel transformation of the perturbative series of superconvergent two-point function. This will be  used to estimate the mass and couplings of the lightest glueball states in sections \ref{sec:n=1} and \ref{sec:n=2}.

Using equations \eqref{eq:borel rule 1} and \eqref{eq:borel rule 2}, we find that the Borel transform of the perturbative series for $\A$ is 
\begin{align} \label{eq:Ph0_pert}
	&\Ah^\pert(M^2) 
	= \B\left[ \text{three-loop truncation of } \A \right] 
	\\
	&= \frac{a_0}{\sqrt{\pi} \sqrt{M^2}}
		+ a_1 \frac{\scale}{M^2}
		+ \frac{2a_2}{\sqrt{\pi}} \frac{(\scale)^2}{(M^2)^{3/2}}
		+ a_3 
		\frac{(\scale)^3}{(M^2)^2}.
	\nn
\end{align}
The one-, two- and three-loop Borel transforms of $\Ah^\pert$ are plotted in figure \ref{fig:Pi0_pert}. 
In particular, the two- and three-loop contributions converge very quickly for $M^2 \gtrsim \scale$ signaling that the three-loop curve can be trusted for $M^2 \gtrsim \scale$.
However, it is uncertain how much we can trust the three-loop $\Ah^\pert$ for $M^2<\scale$. 

In order to try and quantify the uncertainty in $\Ah^\pert$, we have included an unknown ``four-loop'' term in $\A^\pert$.
The coefficient $a_3$ parameterizes the error in the perturbative result.
We have set the magnitude of these coefficients to be approximately the same as as the three-loop correction to $\A$: $|a_3|<\frac{1}{10}$ (see figure \ref{fig:Pi0_pert_errors}).
In particular, note that the error band shrinks as $M^2\to\infty$ where we are infinitely certain about the perturbative result but becomes very wide for small $M^2$ where we are the most uncertain of the perturbative result.

From equation \eqref{eq:Ph0_pert}, we compute the weighted mass average $\Mh_\pert$
\begin{align}
	\frac{\Mh_\pert}{(\scale)^2} = 
	\frac{
		a_0 \frac{ (M^2)^{3/2} }{ (\scale)^{3} }
		- 2 a_2 \frac{ \sqrt{M^2} }{ \scale }
		- 2 \sqrt{\pi} a_3
	}{
		2 a_0 \frac{ (M^2)^{3/2} }{(\scale)^{3}}
		+ 2 \sqrt{\pi} a_1 \frac{ M^2 }{(\scale)^2}
		+ 4 a_2 \frac{ \sqrt{M^2} }{ \scale }
		+ 2 \sqrt{\pi} a_3
        }
\end{align}
Like $\Ah^\pert$, the two- and three-loop $\Mh_\pert$ curves converge quickly for $M^2\gtrsim \scale$ (see figure \ref{fig:Msqrd_pert}). 
We can also plot a version of $\Mh_\pert$ with and error band (see figure \ref{fig:Msqrd_pert_errors}).

\subsection{Borel transformation of the non-perturbative model \label{sec:NPBorel}}

In this section, we construct an ansatz/model for the non-perturbative spectral density of the superconvergent two-point function. From this spectral density, we construct a non-perturbative Borel transform of the superconvergent two-point function $\Ah^\npert$ and the analogous weighted mass average $\Mh_\npert$. Like their perturbative cousins, these quantities will be  used to estimate the mass and couplings of the lightest glueball states in sections \ref{sec:n=1} and \ref{sec:n=2}.

We consider the following model of the non-perturbative spectral density 
\begin{align} \label{eq:rho_gen}
\rho(s) = \sum_{i=1}^{N} \frac{2\pi g_i^2}{m_i^4}\, 
	\delta\left(s-m_i^2\right)
	+ H(s) \, \Theta\left(s - 4m_1^2\right) 
\end{align}
where $H(s)$ is fixed by the asymptotic behaviour of the perturbative spectral density
\begin{align} \label{eq:H(s)}
	H(s) &\equiv \disc\A^\pert(-s)\bigg\vert_{s>4m_1^2}
	=\frac{a_0}{\sqrt{s}} 
		+ \frac{a_2(\scale)^2}{s^{3/2}} 
            + \mathcal{O}\left(\frac{1}{s^{5/2}}\right).
\end{align}
Note that this has the gross features expected non-perturbativly: a sum of delta functions for each glueball in the spectrum and a continuum that begins at the threshold of the lightest particle $s>4m_1^2$.

Using \eqref{eq:drel}, the Borel transform of $\A^\npert$ is 
\begin{align} \label{eq:A02NP}
	\Ah^\npert(M^2)
	=& \sum_{i=1}^n \frac{2g^2_i}{M^2 m_i^4} e^{-\frac{m_i^2}{M^2}}
		+ \frac{a_0}{\sqrt{\pi}\sqrt{M^2}} \text{erfc}\left(\sqrt{\frac{4m_1^2}{M^2}}\right)
		\\&
		+ \frac{\scale}{M^2}
		- \frac{a_2\scale}{\pi m_1} 
                e^{-\frac{4m_1^2}{M^2}}
		+ \frac{2a_2(\scale)^2}{\sqrt{\pi} (M^2)^{3/2}} 
			\text{erfc}\left(\sqrt{\frac{4m_1^2}{M^2}}\right)
		\nn
\end{align}
where $\text{erfc}(z) = 1-\text{erf}(z)$ is the complimentary error function. 
We can fix one parameter in our model by comparing $\Ah^\pert$ and $\Ah^\npert$ in the large $M^2$ limit where we trust perturbation theory.
Expanding in the large $M^2$ limit yields
\begin{align}
    \Ah^\npert(M^2) &=
    \frac{a_0}{\sqrt{\pi} \sqrt{M^2}}
    {+} \frac{1}{\pi M^2} \bigg[
	\sum_{i=1}^n \frac{2\pi g_i^2}{m_i^4}
	{-} 4 a_0 m_1
	{-} a_2 \frac{(\scale)^2}{m_1}
    \bigg]
    + \mathcal{O}\left(\frac{1}{(M^2)^{3/2}}\right).
\end{align}
Then, requiring 
\begin{align}
	\left[
		\Ah^\npert(M^2) - \Ah^\pert(M^2)
	\right]_{M^2\to\infty} 
	= \mathcal{O}\left(\frac{1}{(M^2)^{3/2}}\right),
\end{align}
fixes 
\begin{align} \label{eq:fixing r1}
	g_1^2
	&= a_0 \frac{2 m_1^5}{\pi }
	+ a_1 \frac{(\scale) m_1^4}{2} 
	+ a_2 \frac{(\scale)^2 m_1^3}{2\pi}
	- \sum_{i=2}^n \frac{g_i^2 m_1^4}{m_i^4}
\end{align}
and guarantees that the high energy limit of $\Ah^\npert$ matches $\Ah^\pert$.

To fix the remaining parameters of the model, we minimize
\begin{align} \label{eq:chi2}
	\chi^2 
	= \sum_{j=0}^N \left(
	\frac{
		\Mh_\pert(M^2_j)\vert_{a_3=0} - \Mh_\npert(M^2_j)
	}{
		\text{Error}(\Mh_\pert(M^2_j))
	}\right)^2
\end{align}
where $M^2_j \in R_N$ and $R_N$ is a region of Borel parameter space $R=M^2\in[R_{\min},R_{\max}]$ that has been discretized into $N+1$ points.
Since we know that the high energy limit of $\rho$ is a power law, we want to sample the low energy region of $\Mh$ more frequently (by sampling the high energy region too much the fit can be driven to a pure power law that would only be accurate at high energies). 
Thus, $R_N$ is logarithmically discretized
\begin{align}
	R_N \ni M^2_j = \exp\left(\log R_\text{min} 
		+ j\ \frac{\log R_\text{max} - \log R_\text{min}}{N}\right),
\end{align}
for $j=0,1,2,\dots,N$.

\subsection{One-glueball model ($N=1$) \label{sec:n=1}}

In this section, we minimize \eqref{eq:chi2} for a model with one-glueball ($N=1$ in \eqref{eq:rho_gen}) and extract estimates for the mass $m_1$ and coupling $g_1$. 

\begin{figure}
	\centering
	\includegraphics[align=c,scale=.7]{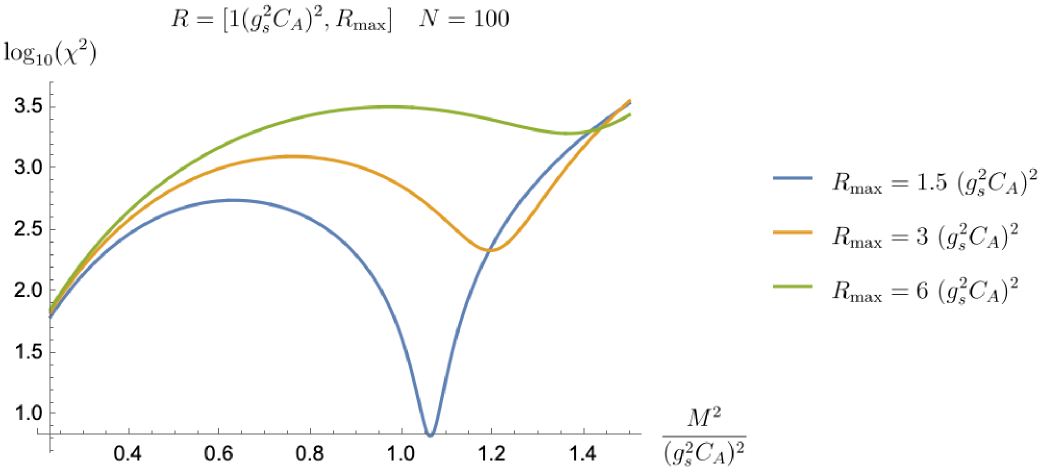}	
	\caption{ \label{fig:1Res chi}
		Plot of the 1-glueball $\log_{10}(\chi^2)$ for various 
            choices of the fitting region $R$. 
		The location of the minimum is very sensitive to the choice 
		of $R_\text{max}$. 
		For ``small'' $R_\text{max}$, the minimum is 
		roughly at $m_1/(\scale)\sim 1$. 
		After increasing $R_\text{max}$ sufficiently, 
            what was a global minimum becomes a local minimum. 
		For almost all choices of $R_\text{max}$, 
            the minimization of $\chi^2$ puts too 
            much emphasis on the high energy region. 
		This stamps out the non-linearities in  $\Mh_\npert$ 
            and drives to $g_1^2$ to zero which we deem unphysical. 
	}
\end{figure}

\begin{figure}[h]
    \centering
    \begin{subfigure}[b]{.45\textwidth}
        \centering
	\includegraphics[align=c,width=\textwidth]{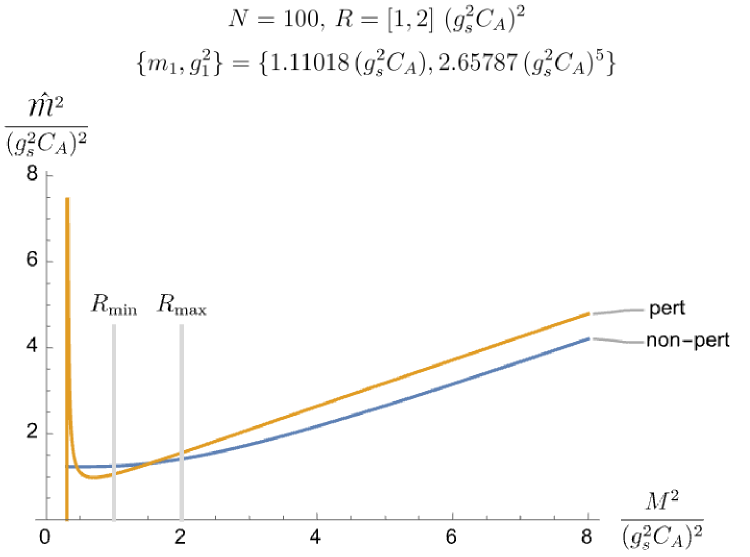}	
	\caption{\label{fig:1Res_low} }
    \end{subfigure}
    \begin{subfigure}[b]{.45\textwidth}			
	\centering
	\includegraphics[align=c,width=\textwidth]{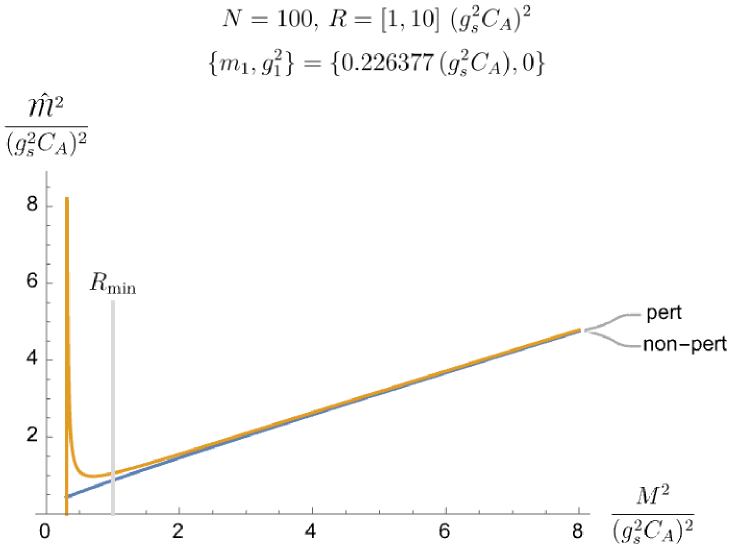}	
	\caption{\label{fig:1Res_high}}
    \end{subfigure}
    \caption{Plots of $\Mh$ (the ratio of $\Ah$ and its derivative \eqref{eq:Mh def}) 
        where the parameters have been optimized 
        over different fitting regions $R$ 
        (indicated by the region between the grey vertical lines). 
        In figure \ref{fig:1Res_low} the parameters have been optimized using 
        a small $R_\text{max}$. 
        While $m_1$ is reasonable ($m_1/\scale \sim \mathcal{O}(1)$), 
        there is still a sizable difference between 
        $\Mh_\npert$ and $\Mh_\pert$ for $M^2\geq1$. 
        In figure \ref{fig:1Res_high} the parameters have been optimized using 
        a larger $R_\text{max}$.
        Unsurprisingly, the high energy limit of $\Mh_\npert$ fits 
        $\Mh_\pert$ better. 
        By minimizing $\chi^2$ over a larger region $m_1$ 
        and $g^2_1$ are driven to their minimal values. 
        Since $\Mh_\pert$ is linear in the high energy region, 
        the minimization sets $g^2_1\to0$ so that $\Mh_\npert$
	is as close to linear as possible. 
	Given the sensitivity of the optimized parameters to the 
        range $R$, the single glueball model is perhaps not the best.
        \label{fig:n1Msqrd}
	}
\end{figure}

The simplest model for the spectral density is the single glueball model
\begin{align} \label{eq:rho1}
	\rho_1(s) 
	= \frac{2\pi g_1^2}{m_1^4} \, \delta\left(s-m_1^2\right)
	+ H(s) \, \Theta\left(s - 4m_1^2\right).
\end{align}
The spin-0 glueball residue is fixed by \eqref{eq:fixing r1}  to 
\begin{align}
	g_1^2
	&= \frac{247 (\scale)^2 m_1^3}{512\pi}
	- \frac{191 (\scale)^2 m_1^3}{36\pi^3}
	+ \frac{8 (\scale) m_1^4}{3\pi^2}
	- \frac{5 (\scale) m_1^4}{8}
	+ \frac{6 m_1^5}{\pi},
\end{align}
Moreover, since $m_1,g_1^2>0$, we obtain a lower bound on the spin-0 mass
\begin{align}
	\frac{m_1}{\scale} 
        > m_{1,\text{min}} 
        \equiv \frac{\sqrt{79744-9069 \pi ^2+225 \pi ^4}+15 \pi ^2-64}{288 \pi }
        \approx 0.226377.
\end{align}
In the one-glueball model, the minimization of $\chi^2$ is highly correlated to the selection of the region $R$. 
In figure \ref{fig:1Res chi}, $\chi^2$ is plotted as a function of the glueball mass $m_1$ for several choices of $R$. 
For small $R_\text{max}$, the fit is trying to match the low energy regions best and there is a global minimum at 
$m_1\sim 1.06 (\scale)$.
As $R_\text{max}$ is increased, what was the global minimum becomes a local minimum.   
The new global minimum forces $m_1$ to its minimal value (where $g^2_1=0$). 
By setting $g^2_1=0$, the fit wants to forget about the non-perturbative dynamics and instead match perturbation theory (figure \ref{fig:1Res_high}). 

Given the sensitivity of the optimized $m_1$ on the choice of $R_\text{max}$ and the fact that using a low energy $R_\text{max}$ leads to a significant miss-match between $\Mh_\pert$ and $\Mh_\npert$ for $M^2>1$ (figure \ref{fig:1Res_low}), we conclude that the one-glueball model does not accurately describe the non-perturbative spectral density of $\A$.

\subsection{Two-glueball model ($N=2$) \label{sec:n=2}}

Having concluded that the single glueball model does not accurately represent the non-perturbative spectral density of $\A$, we study the next simplest model containing two glueballs.
We will find that the optimized value for the masses and coupling of this model are much more stable.

Setting $N=2$ in \eqref{eq:rho_gen}, the spectral density of the two glueball model is
\begin{align}
	\rho_2(s) 
	= 2\pi\left[ 
		\frac{g_1^2}{m_1^2} \, \delta\left(s-m_1^2\right)
		+ \frac{g_2^2}{m_2^2} \, \delta\left(s-m_2^2\right)
	\right]
	+ H(s) \, \Theta\left(s - 4m_1^2\right).
\end{align}
By matching the asymptotics of $\A^\pert$ and $\A^\npert$, the coupling constant $g^2_1$ is fixed to
\begin{align} \label{eq:g1(n=2)}
	g_1^2
	&= -\frac{g_2^2 m_1^4}{m_2^4}
	+ \frac{6 m_1^5}{\pi}
	+ \left(\frac{8}{3 \pi^2} - \frac{5}{8}\right)
            m_1^4 (\scale)
	+ \left(\frac{247}{512 \pi}
	- \frac{191}{36 \pi ^3}\right) m_1^3 (\scale)^2
    .
\end{align}
Combining this with the constraints $m_i,g_i >0$ and $m_{i+1}>m_i$, we find the same constraint on $m_1$ as for one glueball model
\begin{align}
    \frac{m_1}{\scale} > m_{1,\text{min}}
    \approx 0.226377
\end{align}
as well as a constraint on the coupling $g_2^2$ 
\begin{align} 
	\frac{g_2^2}{(\scale)^5} 
	&< g_{2,\text{max}}^2
	\equiv
	\frac{6 m_1 m_2^4}{\pi (\scale)^5}
	- \frac{\left(15 \pi^2-64\right) m_2^4}{24 \pi^2 (\scale)^4}
	+ \frac{\left(2223 \pi^2 - 24448\right) m_2^4}{4608 \pi^3 (\scale)^3 m_1} 
	.
\end{align}

Minimizing the two glueball $\chi^2$, we find estimates for the model parameters. 
In particular, the minimization of the two glueball $\chi^2$ is robust to changes of the region $R$ and the discretization parameter $N$: 
\begin{center}
{\setlength{\tabcolsep}{0.5em} 
\renewcommand{\arraystretch}{1.5}
\begin{tabular}{c|c|c||c|c|c|c}
	$N$ 
        & $\frac{R_\text{min}}{(\scale)^2}$ 
        & $\frac{R_\text{max}}{(\scale)^2}$  
	& $\frac{m_1}{\scale}$ 
        & $\frac{g^2_1}{(\scale)^5}$
        & $\frac{m_2}{\scale}$  
        & $\frac{g^2_2}{(\scale)^5}$
	\\ \hline
	10-100 & 1 & 10 
        & 0.93 & 0.69 $\div$ 0.71 & 1.67 $\div$ 1.68 & 3.64 $\div$ 3.69
	\\
	100 & 1-3 & 10 
        & 0.92 $\div$ 0.93 & 0.65 $\div$ 0.70 & 1.64 $\div$ 1.67 & 3.46 $\div$ 3.65
	\\
	100 & 1 & 8-12 
        & 0.93 & 0.70 & 1.67 & 3.64 $\div$ 3.67
\end{tabular}
}
\end{center}
We caution the reader that one has to (somewhat arbitrarily) decide on a reasonable range for the input parameters ($N,R_\text{min},R_\text{max})$). 
While the estimates for $m_1$ and $g_1^2$ remain 
relatively stable, $\chi^2$ is quite flat in 
the $m_2$ and $g_2^2$ directions. 
Thus, the $m_2$ and $g_2^2$ estimates are more
sensitive to the choices of $N, R_\text{min}$ and $R_\text{max}$. 
This will be more obvious in table \ref{tab:scan a3} where we 
scan over values of the unknown four-loop contributions $a_3$.

\begin{figure}[h]
	\centering
	\begin{subfigure}[b]{.48\textwidth}
		\centering
		\includegraphics[align=c,width=\textwidth]{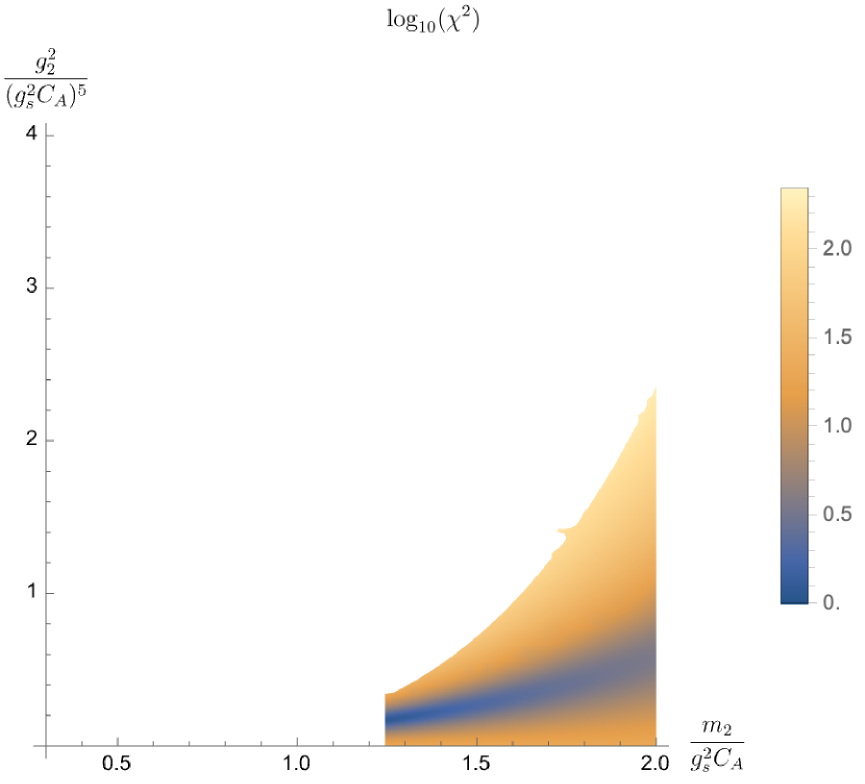}	
		\caption{\label{fig:2Res_ncompatible_min}
			Plot of $\log(\chi^2)$ at fixed
			$m_1/g_s^2 C_A = 0.30$. Here, we have chosen a value 
			of $m_1$ that is close to its minimal value which 
			places the global minimum of $\chi^2$
			outside of the allowed region. 
                Note that while it is not shown at this scale, 
                near the bottom axis the plot smoothly extends 
                all the way to the left axis at $m_1 = m_{1,\text{min}}$. 
		}
	\end{subfigure}
        \
        \begin{subfigure}[b]{.48\textwidth}
		\centering
		\includegraphics[align=c,width=\textwidth]{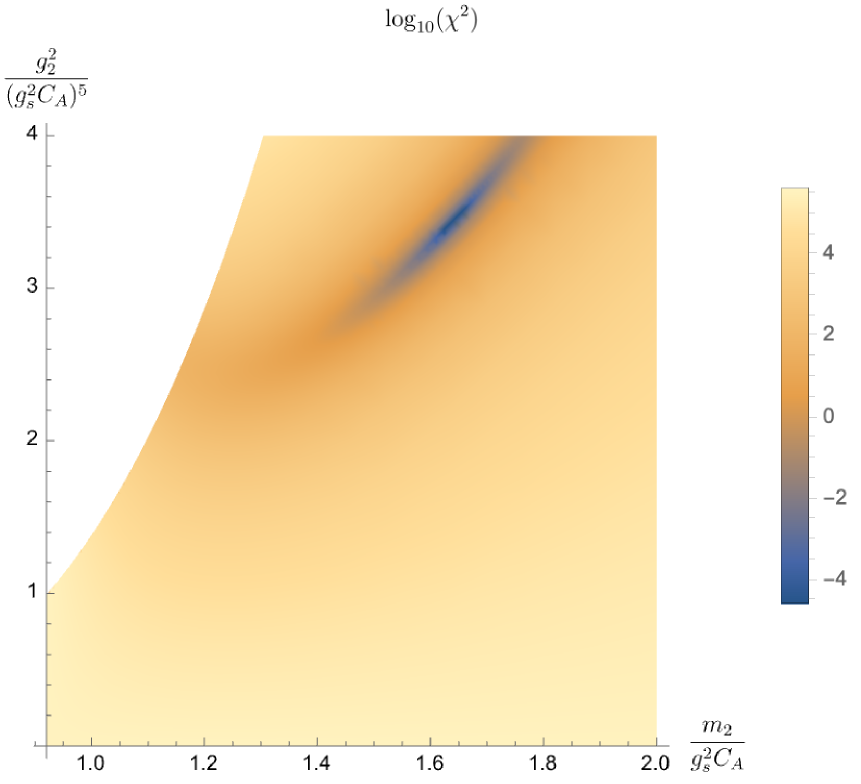}	
		\caption{\label{fig:2Res_compatible_min}
			Plot of $\log(\chi^2)$ at fixed
			$m_1/g_s^2 C_A = 0.92$. Here, we have chosen a value 
			of $m_1$ that is close to its optimized value for which the 
			global minimum of $\chi^2$ lies inside 
			the allowed region.
                \\
                \\
                \\
		}						
	\end{subfigure}
\caption{Density plots of $\log(\chi^2)$ at various values of $m_1$ in the allowed region (below equation \ref{eq:g1(n=2)}).
}
\end{figure}

Much like the one-glueball case, when $R_\text{min}$ is too large the minimization procedure puts too much emphasis on the high energy region
and drives $m_1 \to m_{1,\text{min}}$ as well as $g_1^2 \to 0$. 
In this case, the global minimum lies somewhere outside the allowed region in $(m_2,g^2_2)$-space (see figure \ref{fig:2Res_ncompatible_min}).
However, for small enough $R_\text{min}$, we get reasonable estimates for $m_1$ and $g_2^2$ where the global minimum is well inside the allowed region (see figure \ref{fig:2Res_compatible_min}).

\begin{figure}[h]
\centering
\includegraphics[align=c,scale=.6]{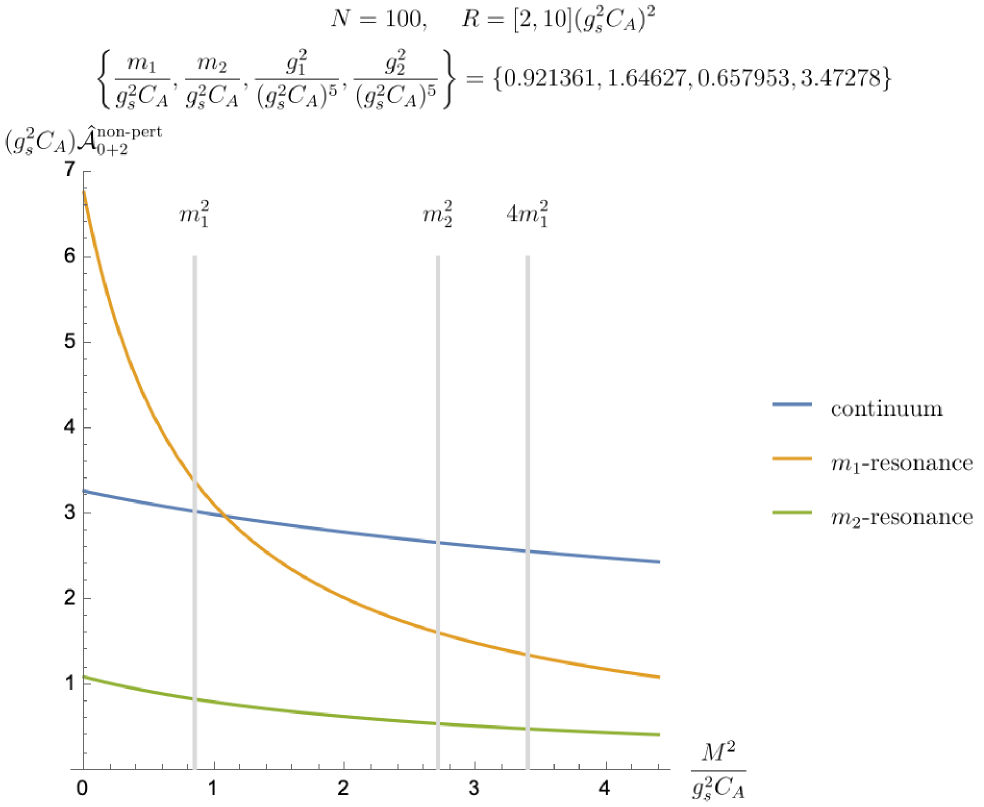}	
\caption{Plotting the individual contributions 
        to the Borel transform of the superconvergent two-point function $\Ah^\npert$.
	This plot illustrates the relative strength 
        of the threshold and glueball contributions. 
	As expected, the $m_1$ glueball dominates 
        for small  Borel parameter $M^2\sim m_1^2$ 
        while the threshold contribution dominates
        for large Borel parameter $M^2>4m_1^2$. 
	The $m_2$-glueball is always sub-dominate 
        as physically expected. 
        However, the ratio of the $m_2$ to $m_1$ contribution approaches $1/2$ asymptotically.
 \label{fig:Pi components}
}	
\end{figure}

Choosing the right $R_\text{min}$ is essential to extracting good estimates. 
This requires finding a window where one can still trust the extrapolation of perturbation theory and where the effects of the low-lying glueballs are significant. 
However, since we do not know exactly where the perturbative expansion breaks down this choice can introduce significant error into our estimates. 
As a sanity check, we compare the relative strength of the continuum and glueball contributions to the Borel transform of the superconvergent two-point function in figure \ref{fig:Pi components}. 
For the optimized values of $m_1,m_2,g_1^2$ and $g_2^2$, the relative strengths of the continuum and glueball contributions to the superconvergent combination align with physical expectations. 
Near the lowest lying glueball state, the contribution from the glueball dominates over the continuum. 
However, sometime after the first glueball but before the second glueball and threshold, the continuum starts to dominate. 
Moreover, the second glueball is subdominant in all regions and occurs below threshold. 
These properties are consistent with a physically reasonable spectral density and we are inclined to trust the optimized values of $m_1,m_2,g_1^2$ and $g_2^2$. 
We also note that the ratio of the $m_2$ to $m_1$ contribution to $\Ah^\npert$ approaches $1/2$ asymptotically. 
Since these contributions are asymptotically of the same order, this provides further evidence that the single glueball model (section \ref{sec:n=1}) misses important effects. 
Roughly speaking, this means that perturbation theory is not any more sensitive to the $m_1$ glueball compared to the $m_2$ glueball. 

On the other hand, the above analysis assumed that $a_3 = 0$ in $\A^\pert$. 
It is important to understand to what degree the unknown coefficients $a_3$ can change the optimized results. 
To get a rough idea, we repeat the above analysis for fixed $a_3$ in table \ref{tab:scan a3}. 
The only formula that changes is \eqref{eq:chi2}. 
Since we are not parameterizing the error in $a_3$, we replace $\text{Error}(\Mh_\pert(M^2_j)) \to \Mh_\pert(M^2_j)$ in \eqref{eq:chi2}. 
From table  \ref{tab:scan a3}, we see that our estimates are relatively insensitive to the unknown coefficient $a_3$.
In particular, the optimized values for the lowest-lying mass $m_1$ and its coupling to the stress-tensor $g_1$ are stable in the regions 
\begin{align}
    \frac{m_1}{\scale} \in [0.92,0.94] 
    \qquad \text{and} \qquad
    \frac{g_1^2}{(\scale)^5} \in [0.66,0.74] 
    .
\end{align}
Perhaps the insensitivity to $a_3$ can be seen from the fact that $H(s)$ (see \eqref{eq:H(s)}) does not dependent on $a_3$. This means that $a_3$ does not appear in the equations for the non-perturbative superconvergent two-point function $\Ah^\npert$ \eqref{eq:A02NP} or the lowest-lying residue \eqref{eq:fixing r1}. The only place $a_3$ appears is in the perturbative result for the weighted mass $\Mh_\npert$. Thus, $a_3$ enters into the $\chi^2$ fit in a relatively simple way. 

\begin{table}[h]
\begin{center}
\setlength{\tabcolsep}{0.5em} 
{\renewcommand{\arraystretch}{1.75}
\begin{tabular}{c|c|c|c||c|c|c|c}
		$a_3$  
		&  $N$  
		&  $\frac{R_\text{min}}{g_s^2 C_A}$  &  $\frac{R_\text{max}}{g_s^2 C_A}$ 
		&  $\frac{m_1}{g_s^2 C_A}$ & $\frac{g_1^2}{(\scale)^5}$
		& $\frac{m_2}{g_s^2 C_A}$ & $\frac{g_2^2}{(\scale)^5}$
	\\ \hline
		$\frac{1}{5}$ & 20-100 & 1-2 & 10-20 & 
			0.92 $\div$ 0.94 & 0.66 $\div$ 0.74 &
                1.64 $\div$ 1.70 & 3.48 $\div$ 3.87
	\\ \hline
		$\frac{1}{10}$ & 20-100 & 1-2 & 10-20 & 
			0.92 $\div$ 0.94 & 0.66 $\div$ 0.74 & 
                1.65 $\div$ 1.70 & 3.48 $\div$ 3.87
	\\ \hline 
		$0$ & 20-100 & 1-2 & 10-20 & 
			0.92 $\div$ 0.94 & 0.66 $\div$ 0.74 & 
                1.65 $\div$ 1.70 & 3.48 $\div$ 3.87
        \\ \hline 
		$-\frac{1}{10}$ & 20-100 & 1-2 & 10-20 & 
			0.92 $\div$ 0.94 & 0.66 $\div$ 0.74 & 
                1.65 $\div$ 1.70 & 3.48 $\div$ 3.87
        \\ \hline 
		$-\frac{1}{5}$ & 20-100 & 1-2 & 10-20 & 
			0.92 $\div$ 0.94 & 0.66 $\div$ 0.74 & 
                1.65 $\div$ 1.70 & 3.48 $\div$ 3.89
\end{tabular}
}

\end{center}
\caption{Table displaying mass estimates for 
    some compatible 
    values of the error $a_3$. 
    The above table shows that our estimates do not strongly depend on $a_3$.
    While we suspect that only $|a_3|<1/10$ are physically reasonable values, we show results  
    with $a_3$ outside this range.
    \label{tab:scan a3}
}
\end{table}

\subsection{Comparison with lattice data \label{sec:lattice}} 

In this section, we summarize the low energy spectrum of three-dimensional YM theory predicted by lattice simulations and compare with the results of section \ref{sec:n=2}. 

Roughly speaking, observables are computed in lattice simulations by directly performing the Feynman path integral over field configurations on a discretized spacetime (often done using Monte Carlo sampling).
By calculating a given observable for many different lattice spacings, one can determine a best fit for the dependence on the lattice spaceing.
Then extrapolating this fit to the limit of vanishing lattice spacing yields observables in the continuum theory. 

Fortunately, there is a lot of data from lattice simulations of three-dimensional YM \cite{Teper1997, Teper1998, Diakonov:1999fq, Lucini:2002wg, Meyer:2002mk, Meyer:2003wx, Bringoltz2007, Buisseret:2013ch, Bursa2013, Athenodorou2016, Athenodorou:2016ebg, Lau2017, Teper2018, Conkey2019}. 
In particular, the spectrum for gauge group $G = SU(N_c)$ has been computed in \cite{Teper1998, Athenodorou:2016ebg} for various $N_c$. 
However, this data must be converted from units of the string tension $\sigma$, which is the most accurate measurement on the lattice, to units of $g_s^2 N_c$.

The string tension is computed from the energy of the lowest-lying state of a static quark anti-quark pair separated by a distance $R$. If our theory has linear confinement, this energy, $E_\text{min}(R)$, provides a definition for the static quark potential as well as a definition for the string tension $\sigma$ in the large $R$ limit
\begin{align}
    E_\text{min}(R) 
    \equiv V_{q\bar{q}}(R) 
    \underset{R\to\infty}{\simeq} \sigma R
\end{align}
For large $R$, this state should be thought of as static quarks attached by a confining flux tube of length $R$. 
Reference \cite{Athenodorou:2016ebg} provides the most recent fit of the string tension in $(2{+}1)$-dimensional Yang-Mills theory
\begin{align}
    \sqrt{\sigma} 
    =\left(
        0.196573(81) 
        - \frac{0.1162(9)}{N_c^2}
    \right) g_s^2 N_c.
\end{align}
The  mass values in units of $g_s^2 N_c$ are summarized in table \ref{tab:latticeComp}.
Comparing with table \ref{tab:scan a3}, we see that the sum-rule estimates for $m_1$ are in good agreement with the lattice data with  error between $14\%$ and $19\%$ for any value of $N_c$. While the error for the $m_2$ estimates can be much larger (up to $\sim 48\%$), this comparison reveals that sum-rules capture many gross features of the low-energy non-perturbative physics. 
The discrepancy with the lattice data is likely due to the inaccuracies in our model of the spectral density that includes only one or two glueball states and a perturbative continuum. 
Despite these discrepancies, we conclude that this model is still a relative good first approximation.

\begin{table}
\centering
\begin{tabular}{c||cc|cc}
    $N_c$
    &
    $m_1/(g_s^2 N_c)$ & $J^{PC}$
    &
    $m_2/(g_s^2 N_c)$ & $J^{PC}$
    \\ \hline
     2 
     & 0.79 & $0^{++}$
     & 1.15 & $0^{++*}$
     \\
     3 
     & 0.80 & $0^{++}$
     & 1.19 & $0^{++*}$
     \\
     4
     & 0.80 & $0^{++}$
     & 1.22 & $0^{++*}$
     \\
     \vdots
     & & 
     & & 
     \\
     $\infty$
     & 0.81 & $0^{++}$
     & 1.24 & $0^{++*}$
\end{tabular}
\caption{The first two masses and states in the spectrum computed by lattice simulations \cite{Athenodorou:2016ebg}. Note that we have only included states with quantum numbers $J=C=+$ since the stress tensor has quantum numbers $J=C=+$. With this restriction the lowest lying states are are always $0^{++}$ and its excited state $0^{++*}$. 
\label{tab:latticeComp}}
\end{table}

\section{Everything is consistent with unitarity!\label{sec:UC consistency}}

In this section, we study the compatibility of the residues, $g_i^2$, of the non-perturbative superconvergent spectral density \eqref{eq:rho_gen} with the principle of unitarity  and the perturbative two-point function eq.~\eqref{A result}. 
In particular, we consider the case of a single glueball with mass $m_1$ and multi-particle threshold starting at $4m_1^2$. 
If the physical spectral density is indeed dominated by lightest the spin-0 glueball, this model would be a good approximation. 
While we have already argued against this approximation and that one should include at least two glueball states, the single glueball model is more constrained and therefore more relevant for the consistency checks.

Recalling eq.~\eqref{eq:rho1}, the spectral density for the single glueball model is
\begin{equation}
\label{eq:spec-den}
  \rho(s)
  =\frac{2\pi g_1^2}{(m_1^2)^2}\delta(s-m_1^2)+H(s) \, \Theta\left(s - s_0^2\right).
\end{equation} 
Checking unitarity of the correlation function boils down to checking the positivity of this spectral density.

To impose positivity, we consider a coordinate transformation of the $s$-plane that maps the upper half-half plane to the unit disk while moving the pole to the origin and the branch cut to the boundary of the unit disk (see \cite{Paulos:2016but, Paulos:2017fhb}):
\begin{equation}
   s\rightarrow z=\frac{\sqrt{4m_1^2-s}-\sqrt{3}m_1}{\sqrt{4m_1^2-s}+\sqrt{3}m_1}.
\end{equation}
Now, the series expansion around $z=0$ is convergent with a finite radius of convergence. Specifically, the superconvergent combination, $\mathcal{A}_{0+2}$, becomes
\begin{equation}
    \mathcal{A}_{0+2}(z)=\frac{-g_1^2}{12m_1^6z}+\sum_{n=0}^{\infty}c_nz^n,
\end{equation}
which can be truncated at some large but finite cutoff $N$.
On the other hand, the perturbative expansion for large Euclidean momenta ($s<0$) maps to an asymptotic expansion around $z=1$. Comparison with the three-loop perturbative result fixes three of the  $c_i$'s. 
Next, we impose positivity on the boundary of the unit disk in the $z$-plane
\begin{equation}
\label{eq:rho-z}
    \rho(z)
    =\frac{-\pi g_1^2}{6m_1^6}\delta(z)
    +\sum_{n=0}^Nc_n{\rm Im}(z^n)
    \geq 0, 
\end{equation}
where $z=e^{i\theta}$ and $\theta\in(0,\pi)$.
In practice, we truncate the sum \eqref{eq:rho-z} at $N=40$ and impose positivity for 2000 evenly-spaced points on the boundary of the disc.
Using simple Mathematica functions (\texttt{FindMinimum} and \texttt{FindMaximum}), we minimize/maximize the residue $g_1^2$ over the variables $c_{i=4,\dots,40}$ while enforcing the positivity condition \eqref{eq:rho-z} at the boundary points.\footnote{We have checked that the numerical solutions to the $c_i$ are stable when we change the number of boundary points.}

\begin{figure}[h]
     \centering
     \includegraphics[align=c,width=.4\columnwidth]{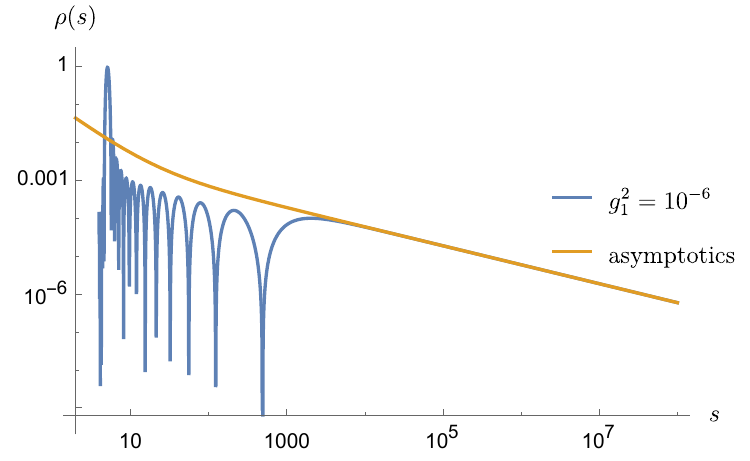}
\quad
     \includegraphics[align=c,width=0.4\columnwidth]{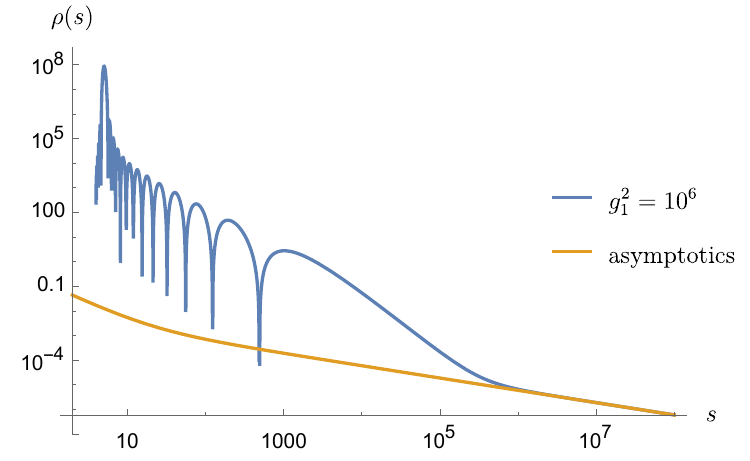}
     \caption{Log-log plot of the spectral density \eqref{eq:rho-z} at $m_1/(\scale)=1$ as well as the  asymptotic behaviour obtained from the perturbative loop expansion (equation \eqref{A result}). For the spectral density \eqref{eq:rho-z}, we set $N=40$. The left plot corresponds to a spectral density with a large value of residue ($g_1^2=10^6 (\scale)^5$) and the right plot corresponds to a one with small value of residue  ($g_1^2=10^{-6} (\scale)^5$). This exemplifies the argument that unitarity and the asymptotic behaviour of the correlation function are not strong enough to give an upper bound on the residue.}
     \label{fig:spectral-density}
 \end{figure}

Unfortunately, positivity of the physical cut alone is not enough to get a finite upper bound for the residue since both terms in \eqref{eq:rho-z} can be arbitrarily large positive numbers. 
We illustrate this in figure \ref{fig:spectral-density} (right) where we plot a positive spectral density with large residue $g_1^2=10^6 (\scale)^5$ and mass $m_1/g_s^2C_A=1$.

On the other hand, positivity only yields a trivial lower-bound for the residue. 
The minimization procedure returns negative values for the residue and adding more terms to the ansatz only increases the negativity of the residue. 
However, since the residue must be positive, we conclude that the minimal value of the residue must be zero. 
In figure \ref{fig:spectral-density} (left), we plot a positive spectral density with a small residue $g_1^2=10^{-6}(\scale)^5$ and mass $m_1/g_s^2C_A=1$ that is consistent with unitarity and the perturbative results. 

In this section, we solved the minimization/maximization problem of the residue $g_1^2$ for a wide range of mass values. 
For each mass, we find a spectral density that is compatible with unitarity and the asymptoics predicted by perturbation theory.
Hence, we conclude that one can construct a spectral density compatible with unitarity and perturbation theory for any mass and residue in the single glueball model.

\section{Higher-spin currents \label{sec:higher-spin}}

In this section, we extend our analysis to more general operators, {\it i.e.,} higher-spin currents $O^{\mu_1\ldots\mu_{\ell}}(p)$ with even spin $\ell$. 
As pointed out in the introduction, correlation functions of such operators contain important information about glueball lightcone wavefunctions, which are closely related to parton distribution functions.
However, to apply the methods of section \ref{sec:sum-rules}, we first need to identify superconvergent combinations of the higher-spin two-point functions. 


In section \ref{sec:hs-tensor}, we analyze the tensor structure of higher-spin two-point functions of interest. 
Then, we fix a basis of the higher-spin operators in section \ref{sec:hs-basis}.  
The coefficients of the tensor structures for the basis operators are then computed to two-loops. 
In section \ref{subsec:high-spin-unitarity}, we compute the imaginary part of these tensor structure coefficients using unitarity cuts at one- and two-loops.
While just the imaginary parts are enough to verify the existence of superconvergent combinations, we also compute these coefficients using Feynman diagrams (section \ref{sec:2loopsHS}) since the sum-rules are sensitive to more than just the imaginary parts. 
Lastly, in section \ref{sec:hs-superconvergent}, we explicitly show the existence of superconvergent combinations for higher-spin two-point functions and give a crude method for extracting the higher-spin residues. 

\subsection{Higher-spin Correlation Functions \label{sec:hs-tensor}}

We start our analysis of higher-spin operators by noting that higher-spin fields can be defined as traceless symmetric combinations of covariant derivatives acting on the field strength $F_{\mu\nu}^a$. In contrary to $T_{\mu\nu}$, these operators are not conserved and will have many more possible tensor structures. 

At spin-2, there is only one operator: the stress-tensor. Explicitly, 
\begin{equation}
\label{eq:spin2-lag}
    O_2^{\mu_1\mu_2}(x)=(F^a)^{\mu_1}_{\lambda}(F^a)^{\mu_2\lambda}-\frac{g^{\mu_1\mu_2}}{d                          }(F^a)_{\mu\lambda}(F^a)^{\mu\lambda}.
\end{equation}
For higher-spin operators with spin $\ell$, there are $\ell/2$ possible structures:
\begin{equation}
\label{eq:spinl-lag}
\begin{split}
    O_{0,\ell}^{\mu_1\ldots\mu_{\ell}}(x)&=\frac{1}{(\ell-2)!}[D^{(\mu_1}\ldots D^{\mu_{\frac{\ell}{2}-1}}(F^a)^{\mu_{\frac{\ell}{2}}}_{\lambda} D^{\mu_{\frac{\ell}{2}+1}}\ldots D^{\mu_{\ell}-1}(F^a)^{\mu_{\ell})\lambda}-\text{trace}],\\
    O_{i,\ell}^{\mu_1\ldots\mu_{\ell}}(x)&=D^{\mu_1}\ldots D^{\mu_i}O_{0,\ell-i}^{\mu_{i+1}\ldots\mu_{\ell}} \quad \text{for}\qquad i=2,\ldots ,\ell-2.
    \end{split}
\end{equation}
Since we are interested in even spin operators, $\ell$ and $i$ are even numbers. Any combination of these $\ell/2$ structures is a viable option for higher-spin operators. Later, we will introduce the criterion we use to select our basis of spin-$\ell$ operators.

To avoid working with indices, we utilize null vector representation for symmetric traceless tensor of spinning states, where we introduce $d+1$-dimensional null vectors $v$ with $v^2=0$, 
\begin{equation}
    f_{\mu_1\ldots \mu_{\ell}}\leftrightarrow f(v)\equiv f_{\mu_1\ldots \mu_{\ell}}v^{\mu_1}\ldots v^{\mu_{\ell}}.
\end{equation}
$f(v)$ can then be proved to be a harmonic polynomial of its $d$ variable \cite{Costa:2011mg}. Once we have the function $f(v)$, we can reconstruct $f_{\mu_1\ldots \mu_{\ell}}$ using Thomas-Todorov operator (see \cite{Costa:2011mg}):
\begin{equation}
D^{\mu}_{v}=\left(\frac{d}{2}-1+v\cdot\frac{\partial}{\partial v}\right)\frac{\partial}{\partial v_{\mu}}-\frac{1}{2}v^{\mu}\frac{\partial^2}{\partial v\cdot\partial v}.\\
\end{equation}
This differential operator imposes tracelessness directly by removing the trace.

Since our goal is to study $\langle O_{\ell}(v_1,p)O_{\ell'}(v_2,-p)\rangle$ correlation functions, we need to understand how to extract different tensor structures. 
Because the correlation function must be invariant under little group transformations (transformations that keep the momentum, $p$, fixed), $O_{\ell}(p)$ and $O_{\ell'}(-p)$ must have opposite helicity under rotation around the $p$-axis. 
This means that the correlation function can be written as a sum over expressions with fixed helicity, $j$, under these rotations. 
The helicity $j$ is an integer between 0 and $j_{\text{max}}=\text{min}(\ell,\ell')$. 
It is thus useful to define vectors that parameterize the directions perpendicular to $p$:
\begin{equation}
    (v^{\perp}_i)^{\mu}=v_i^{\mu}-\frac{v_i\cdot p}{p^2}p^{\mu}.
\end{equation}
These vectors are orthogonal to $p$ and transform nicely under the little group. 
Along with $p^{\mu}$, these vectors span all possible tensor structures. Thus, the correlation function can then be represented as,
\begin{equation}
\label{eq:proj}
\begin{split}
&\langle O_{\ell}(v_1,p)O_{\ell'}(v_2,-p)\rangle=\sum_{j=0}^{\text{min}(\ell,\ell')} [\pi_j]_{\ell,\ell'}(v,v',p)A_j^{\ell,\ell'}(p),
\\
&[\pi_j]_{\ell,\ell'}(v,v',p)=(v_1^{\perp}\cdot v_1^{\perp})^{\frac{\ell}{2}}(v_2^{\perp}\cdot v_2^{\perp})^{\frac{\ell'}{2}}T_j(\cos\theta),  
\end{split}
\end{equation}
where 
\begin{align}
    \cos\theta=\frac{v_1^{\perp}\cdot v_2^{\perp}}{|v_1^{\perp}||v_2^{\perp}|}.
\end{align}
Each helicity $j$ structure in this expansion corresponds to a channel of spin $j$ states in the K\"{a}ll\'en-Lehmann spectral decomposition of the correlator since it has the correct transformation under the little group.

\subsection{Basis for higher-spin Operators \label{sec:hs-basis}}
In this section, we define a ``nice'' basis for the operators (eqs.~\eqref{eq:spin2-lag} and ~\eqref{eq:spinl-lag}).

To find this basis, we first write the on-shell matrix elements  $\langle p_1^gp_2^g|O_{i,\ell}|0\rangle$ corresponding to these operators,
\begin{equation}
\label{eq:Ojgg1}
\begin{split}
\langle p_1^gp_2^g|O_2|0\rangle=&2(v\cdot p_1)(v\cdot p_2)\delta^{a_1a_2},\\
\langle p_1^gp_2^g| O_{i,\ell}|0\rangle=& 2\left((v\cdot p_1)+(v\cdot p_2)\right)^{i}\left((v\cdot p_1)(v\cdot p_2)\right)^{\frac{\ell-i}{2}}.
\end{split}
\end{equation}
There is a nice way of presenting these spin $\ell$ structures by introducing the angle $\phi$ via, 
\begin{equation}
\cos\phi=\frac{v^{\perp}\cdot p_1^{\perp}}{|v^{\perp}||p_1^{\perp}|}=\frac{v\cdot (p_1-p_2)}{(v\cdot p)}.
\end{equation}
In terms of this angle, the structures of eq.~\eqref{eq:Ojgg1} simplify to:
 \begin{equation}
 \langle p_1^gp_2^g|O_{i,\ell}|0\rangle=2^{i-\ell+1} (v\cdot p)^{\ell}\sin^{\ell-i}\phi.
 \end{equation}


We can then understand the decomposition of each of these two-point functions in the helicity basis $e^{ij\phi}$ using Chebyshev polynomials as a basis\footnote{This is because glueball states with spin $m$ would be states with helicity $e^{im\phi}=T_{m}(\cos\phi)+iU(\cos\phi)$ and  $e^{-im\phi}=T_{m}(\cos\phi)-iU(\cos\phi)$}:
\begin{equation}
\sin^{2n}\phi=-\frac{1}{\sqrt{\pi}}\frac{\Gamma\left(\frac{1}{2}+n\right)}{\Gamma(n+1)}+2^{1-2n}\sum_{j=0}^{n}(-1)^j\binom{2n}{n-j}T_{2j}(\cos\phi).
\end{equation}
We see that each of the form factors $ \langle p_1^gp_2^g|O_{i,\ell}|0\rangle$ have all even helicities from 0 to $\ell$. However, we can construct linear combinations of the $O_{i,\ell}$ operators such that the combinations only include helicity $\ell$, $-\ell$ and $0$:
 \begin{equation}
 \label{eq:Qgg}
\langle p_1^gp_2^g| \mathcal{Q}_{\ell}(p,v)|0\rangle=\delta^{ab}(-1)^{\frac{\ell}{2}}2^{-2}(v\cdot p)^{\ell} (T_{\ell}(\cos\phi)-1).
 \end{equation}
 From this equation it is easy to read the relations between $\mathcal{Q}_{\ell}$ and $O_{i,\ell}$,
\begin{equation}
    \mathcal{Q}_{\ell}=\sum_{m=0}^{\ell/2}\sum_{k=0}^{\frac{\ell-2m}{2}}2^{2m+2k-1}\binom{\ell}{2m}\binom{\frac{\ell-2m}{2}}{k}(-1)^{\frac{\ell+2k+2m}{2}}O_{\ell-2m-2k,\ell}+\frac{(-1)^{\frac{\ell}{2}+1}}{8}O_{\ell,\ell}.
\end{equation} 
For instance for $\mathcal{Q}_{4}$ we have,
\begin{equation}
\label{eq:Q-def}
     \mathcal{Q}_4=2^4\left(O_{0,4}-\frac{1}{4}O_{2,4}\right).
\end{equation}
This choice of basis then means that at one-loop we get:
\begin{equation}
     \langle \mathcal{Q}_{\ell}\mathcal{Q}_{\ell'}\rangle=\frac{d_G}{512}\left(\pi_{0}A_{0}^{\ell,\ell'(0)}(p^2)+\delta_{\ell\ell'}\pi_{\ell}A_{\ell}^{\ell,\ell'(0)}(p^2)\right).
\end{equation}
However, note that at higher loops, all other projection channels appear again.

 \subsection{One- and two-loop with unitarity method}
 \label{subsec:high-spin-unitarity}
 
Here, we will illustrate how to use unitarity methods to calculate the imaginary part of one- and two- loop correlation functions of spin-2 and spin 4 operators following section~\ref{sec:1and2LoopUnitarity}. 

For one-loop, the needed phase space integrals are:
\begin{equation}
\begin{split}
   \text{Disc} \left( \langle \mathcal{Q}_{\ell}\mathcal{Q}_{\ell'}\rangle \right)&= \frac{-id_G}{2!}(-1)^{\frac{\ell+\ell'}{2}}2^{-\ell-\ell'}\int\frac{d^2p_1}{(2\pi)^22E_1}\frac{d^2p_2}{(2\pi)^22E_2}(2\pi)^3\delta^3(p-p_1-p_2)\\
&\times (v_1\cdot p)^{\ell} (T_{\ell}(\cos\phi_1)-1)(v_2\cdot p)^{\ell'} (T_{\ell'}(\cos\phi_2)-1),
\end{split}
\end{equation}
where we have used eq.~\eqref{eq:Qgg}  for the form factors inside the integral. We can then perform the phase space integral as illustrated in section ~\ref{sec:1and2LoopUnitarity} to obtain the correlation functions of any spin $\ell$ and $\ell'$. The coefficients are simple to obtain for general spins:
\begin{equation}
    A_{0}^{\ell,\ell'(0)}= 2(p^2)^{\frac{\ell+\ell'-1}{2}} \quad A_{\ell}^{\ell,\ell'(0)}= (p^2)^{\frac{\ell+\ell'-1}{2}}.
\end{equation}
For example, the matrix-valued correlator of spin-2 and spin-4 operators can be written as:
  \begin{equation}
  \label{eq:Q2Q4matrix}
\left.
\begin{pmatrix}
(p^2)^{-2}\langle \mathcal{Q}_2\mathcal{Q}_2\rangle\,\, &  (p^2)^{-3}\langle\mathcal{Q}_2\mathcal{Q}_4\rangle\\
(p^2)^{-3}\langle \mathcal{Q}_4\mathcal{Q}_2\rangle & (p^2)^{-4}\langle \mathcal{Q}_4\mathcal{Q}_4\rangle
\end{pmatrix} 
\right\vert_{\text{1-loop}}=\frac{d_G}{512 \sqrt{p^2}}\Bigg( \pi_0\underbrace{\begin{pmatrix}
2&2\\
2 &2\end{pmatrix}}_{M_0^{(0)}}+\pi_2\underbrace{\begin{pmatrix}
1&0\\
0 &0\end{pmatrix}}_{M_2^{(0)}}+\pi_4\underbrace{\begin{pmatrix}
0 &0\\
0&1
\end{pmatrix}}_{M_4^{(0)}}\Bigg).
\end{equation}
This takes care of one-loop analysis. Next, we examine these correlators at two-loops.

As discussed in section~\ref{sec:1and2LoopUnitarity}, the only ingredient we need is the on-shell form factor $\langle p_1^gp_2^gp_3^g|\mathcal{Q}_{\ell}|0\rangle$ as the other two cuts in fig.~\ref{fig:2-loop-cut} cancel each other via the same argument presented in that section. To find this form factor, we use universality of the collinear and soft limit in the theory in addition to Bose symmetry. Basically, we first obtain  the universal splitting factor appearing in the collinear limit by taking $p_1$ and $p_2$ to be parallel in the stress-tensor form factor $\langle p_1^gp_2^gp_3^g|\mathcal{T}^{\mu\nu}|0\rangle$ (eq.~\eqref{2loopcut1}) to compare with $\langle p_2^gp_3^g|\mathcal{T}^{\mu\nu}|0\rangle$  in eq.~\ref{cut1loop}. This yields the following the following splitting factor:
\begin{equation}
\text{SP}=\frac{2g_{s}}{\sqrt{z(1-z)}\langle 12 \rangle}(1-z+z^2).
\end{equation}
Using this splitting factor, we obtain the three-gluon form factors:
\begin{equation}
\label{eq:ojggg}
    \begin{split}
    \langle p_1^gp_2^gp_3^g|\mathcal{Q}_2|0\rangle=&\frac{-16(p_1\cdot v)^2(p_2\cdot p_3)+8p^2(p_1\cdot v)(p_2\cdot v)+16 (p_1\cdot p_2)(v\cdot p_1)(p_1\cdot p_2)+ \text{perms}}{\langle 12\rangle\langle 23\rangle\langle 31\rangle},
    \\
    \langle p_1^gp_2^gp_3^g|\mathcal{Q}_4|0\rangle=&\frac{32\sqrt{2}}{\langle 12\rangle\langle 23\rangle\langle 31\rangle}\Big(-s_{23}(p_1\cdot v)^4+2s_{12}(p_2\cdot v) (p_1\cdot v)^3+s_{13}(p_2\cdot v) (p_1\cdot v)^3\\
    &+7s_{23}(p_2\cdot v) (p_1\cdot v)^3-4s_{12}(p_2\cdot v)^2 (p_1\cdot v)^2-7s_{13}(p_2\cdot v)^2 (p_1\cdot v)^2-s_{12}\\&\times(p_2\cdot v)(p_3\cdot v) (p_1\cdot v)^2+4s_{23}(p_2\cdot v)(p_3\cdot v) (p_1\cdot v)^2+\text{perms}\Big)
    .
    \end{split}
\end{equation}

The three-gluon form factors are used to obtain the non-analytic part of two-loop correction to the correlation function of $\mathcal{Q}_{2}$ and $\mathcal{Q}_{4}$ through phase space integral explained in appendix \ref{app:2loopcut}. We will postpone writing the explicit results of this calculation to the next section (eqs.~\eqref{eq:QQ} and ~\ref{eq:Q2Q4-mat-2loop2}) in which we do the one- and two-loop calculations using Feynman diagrams to obtain both analytic and non-analytic part.

\subsection{Two-loop higher-spin correlators \label{sec:2loopsHS}}

Now that we fixed the basis for higher-spin operators, we use the Feynman diagram approach to obtain the two-loops contributions to the correlation functions of higher-spin operators.

The calculation is almost identical to that in section \ref{sec:2loopsHS}: one replaces the vertices associated to the stress-tensor with the vertices generated by equations \eqref{eq:spin2-lag}, \eqref{eq:spinl-lag} and \eqref{eq:Q-def}. 
Unlike in section \ref{sec:2ptfn} where we only had even spin structures, these higher-spin operators couple to both even and odd spin states.

The two-loop correction for the correlation functions $\mathcal{Q}_2$ and $\mathcal{Q}_4$ in eq,~\eqref{eq:Q2Q4matrix} are:
\begin{equation}
\left.
\begin{pmatrix} \label{eq:QQ}
  (p^2)^{-2} \langle \mathcal{Q}_2\mathcal{Q}_2\rangle\,\, & \,\,(p^2)^{-3}\langle \mathcal{Q}_2\mathcal{Q}_4\rangle\\
(p^2)^{-3}\langle \mathcal{Q}_4\mathcal{Q}_2\rangle & (p^2)^{-4}\langle \mathcal{Q}_4\mathcal{Q}_4\rangle
\end{pmatrix}
\right\vert_{\text{2-loop}}=\frac{d_G g_s^2C_A}{512p^2}
\left(\frac{\mubar^2}{p^2}\right)^{2\varepsilon} \sum_{J=0}^4\pi^J M_J^{(1)}
,
\end{equation}
where 
\begin{equation}
\label{eq:Q2Q4-mat-2loop2}
\begin{split}
&M_0^{(1)}=\begin{pmatrix} -\frac{1}{4}-\frac{4}{3\pi^2}
            -\frac{4}{3\pi^2\varepsilon} &  - \frac{1}{4}
            - \frac{272}{525 \pi^2}
            -\frac{16}{15\pi^2\varepsilon}\\ - \frac{1}{4}
            - \frac{272}{525 \pi^2}
            -\frac{16}{15\pi^2\varepsilon} &   \,\,\,- \frac{1}{4}
           - \frac{14833664}{10735725 \pi ^2}
           - \frac{768}{715 \pi ^2\varepsilon}\end{pmatrix}
           ,
            \\
            &       M_1^{(1)}=\begin{pmatrix}0 &  0\\ 0 &   
           \,\,\, - \frac{6903296}{32207175 \pi ^2}
            + \frac{3584}{6435 \pi ^2\varepsilon}       \end{pmatrix}
            ,\\
       &       M_2^{(1)}=\begin{pmatrix}  - 1
            + \frac{20}{3\pi^2}
            + \frac{4}{3\pi^2\varepsilon} &  - \frac{752}{255 \pi^2}
            + \frac{16}{15\pi^2\varepsilon}\\  - \frac{752}{255 \pi^2}
            + \frac{16}{15\pi^2\varepsilon} &   
              \,\,\,- \frac{293698112}{19324305 \pi ^2} 
            -\frac{448}{1287 \pi ^2\varepsilon}
       \end{pmatrix}
       ,\\
       &M_3^{(1)}=\begin{pmatrix}  0& 0\\ 0& \,\,\,\,  
             -\frac{4632064}{5010005 \pi ^2}
            +\frac{1536}{1001 \pi ^2\varepsilon}
       \end{pmatrix}
       ,\\
         &M_4^{(1)}=\begin{pmatrix}  0& 0\\ 0& \,\,\,\,  
             - 2
            + \frac{4487877952}{225450225 \pi ^2}
            +\frac{158272 }{45045 \pi ^2\varepsilon}
       \end{pmatrix}
       .
    \end{split}
\end{equation}
We emphasize that the non-analytic part of \eqref{eq:QQ} was cross-checked by a unitary computation.

\subsection{Superconvergent Combinations \label{sec:hs-superconvergent}}
In section \ref{sec:magic} we introduced the ``superconvergent'' combination for stress-tensor 2-point function which is well-behaved non-purturbativly in the $p^2\rightarrow 0$ limit and well-suited for the application of dispersive sum-rules in section~\ref{sec:sum-rules}. In this section, we demonstrate the existence of such  combinations for spinning correlation functions. 
We focus on the spin-2 and spin-4 operators for which we use the two-loop perturbative result obtained in previous subsection inside Borel sum-rules to extract crude estimates of their coupling to the lowest-lying spin 0 particle. 
This analysis is parallel to the analysis in section~\ref{sec:n=1}.


By following the argument in section \ref{sec:magic}, we see that the superconvergent combination for the correlator $\langle O_{\ell}O_{\ell'}\rangle$ is the coefficient of $p^{\mu_1}\ldots p^{\mu_{\ell}}p^{\nu_1}\ldots p^{\nu_{\ell'}}$. From eq.~\eqref{eq:proj} it can be seen that this coefficient is given by:
\begin{equation}
\label{eq:magic-HS}
    \mathcal{A}_{\ell,\ell'}(p)=\sum_{j=0}^{\text{min }(\ell,\ell')} T_{j}(-1)\frac{A^{\ell,\ell'}_{j}(p)}{(p^2)^{\frac{\ell+\ell'}{2}}}=\sum_{j=0}^{\text{min }(\ell,\ell')}(-1)^j\frac{A_{j}^{\ell,\ell'}(p)}{(p^2)^{\frac{\ell+\ell'}{2}}}.
\end{equation}
Thus, the resulting superconvergent version of the matrix $\< \mathcal{Q}_\ell \mathcal{Q}_{\ell^\prime} \>$ to two-loops in perturbation theory is
\begin{align}
\label{eq:magic-combo-HS-FD}
    \begin{pmatrix}
        \mathcal{A}_{2,2}^\pert 
        & \mathcal{A}_{2,4}^\pert
        \\
        \mathcal{A}_{4,2}^\pert 
        & \mathcal{A}_{4,4}^\pert
    \end{pmatrix}
    &=
    \begin{pmatrix}
        3 & 2
        \\
        2 & 3
    \end{pmatrix} \frac{1}{\sqrt{p^2}}
    - 
    \begin{pmatrix}
        \frac{5}{4} - \frac{16}{3\pi^2}
        & \frac{1}{4} +\frac{1216}{315 \pi^2}
        \\ 
        \frac{1}{4} + \frac{1216}{315 \pi^2}\,\,\,
        & \frac{9}{4} - \frac{603392}{135135 \pi ^2}
    \end{pmatrix} \frac{\scale}{p^2}
    + \mathcal{O}\left(\frac{1}{(p^2)^{3/2}}\right)
    ,
    \nn\\
    &\approx 
    \begin{pmatrix}
        3 & 2
        \\
        2 & 3
    \end{pmatrix} \frac{1}{\sqrt{p^2}}
    - 
    \begin{pmatrix}
        0.710 \,\,\,& 0.641
        \\
        0.641 & 1.798
    \end{pmatrix}\frac{\scale}{p^2}
    .
\end{align}
We see that in all of these equation the logarithms vanish, as anticipated.
Like in section \ref{sec:sum-rules}, the superconvergent two-point functions inherit a K\"all\'en-Lehmann spectral representation from the K\"all\'en-Lehmann representation of the $\<\mathcal{Q}_\ell\mathcal{Q}_{\ell^\prime}\>$:
\begin{align}
    \rho_{2,2}(s) &= \eqref{eq:rho_gen}
    = \sum_{i=1}^{N} \frac{2\pi g_i^2}{m_i^4}\, 
	\delta\left(s-m_i^2\right)
	+ H(s) \, \Theta\left(s - 4m_1^2\right),
    \\
    \rho_{2,4}(s) 
    &= \sum_{i=1}^{N} \frac{2\pi g_{2,4;i}^2}{m_i^6}\, 
	\delta\left(s-m_i^2\right)
	+ H_{2,4}(s) \, \Theta\left(s - 4m_1^2\right),
    \\
    \rho_{4,4}(s) 
    &= \sum_{i=1}^{N} \frac{2\pi g_{4,4;i}^2}{m_i^8}\, 
	\delta\left(s-m_i^2\right)
	+ H_{4,4}(s) \, \Theta\left(s - 4m_1^2\right),
\end{align}
where 
\begin{align}
    H_{i,j}(s) = \left.\text{Disc}\mathcal{A}_{i,j}^\pert(-s)\right\vert_{s>4m_1^2}
    \ . 
\end{align}
Note that due to contributions from odd-spin glueballs, the spectral density $\rho_{4,4}$ is not guaranteed to be positive. This is because odd-spin glueballs contribute with the wrong sign. 

Requiring that the Borel transforms of $\mathcal{A}_{i,j}^\pert$ and $\mathcal{A}_{i,j}^\npert$ match asymptotically, 
\begin{align}
	\left[
		\mathcal{A}_{i,j}^\npert(M^2) 
            - \mathcal{A}_{i,j}^\pert(M^2)
	\right]_{M^2\to\infty} 
	= \mathcal{O}\left(\frac{1}{(M^2)^{3/2}}\right),
\end{align}
places constraints on the model parameters. Explicitly, 
\begin{align}
    0 &= 2 \sum_{i=1}^\infty 
        \frac{g_{i}^2}{m_i^4}
    - \frac{12 m_1}{\pi }
    + \left(
        \frac{15}{12} 
        - \frac{16}{3\pi ^2}
    \right) (\scale) 
    ,
    \\
    0 &=  2 \sum_{i=1}^\infty 
        \frac{g_{2,4;i}^2}{m_i^6}
    - \frac{8m_1}{\pi}
    +\left(
        \frac{1}{4}
        + \frac{1216}{315 \pi^2}
    \right) (\scale) 
    ,
    \\
    0 &= 2 \sum_{i=1}^\infty 
        \frac{g_{2,4;i}^2}{m_i^8}
    - \frac{12 m_1}{\pi}    
    + \left(
        \frac{9}{4}
        -\frac{603392}{135135 \pi ^2}
    \right) (\scale)  
    .
\end{align}
For a very crude approximation of the residue $g_1^2$, one can neglect all $g_{i>1}^2$ and solve the above equations for the coupling $m_1$-coupling. 
Such an approximation is crude because as discovered in figure \ref{fig:Pi components}, the asymptotic contributions of the $m_1$ and $m_2$ glueballs are comparable. Thus, it is questionable as to whether we can neglect the $m_2$ glueball.

\section{Conclusions}

The basic idea of QCD sum-rules is the notion that the spectral density is well-approximated by a sum of delta-function(s) and a perturbatively calculable continuum.   
In this work, we have tested this notion for three-dimensional Yang-Mills theory. 

In section \ref{sec:2ptfn}, we calculated the stress-tensor two-point function to 3-loops ($\sim \alpha_s^2$) and 
extract a perturbative approximation to the spectral density above the continuum threshold (section \ref{sec:sum-rules}) from the stress-tensor two-point function. 
Then, this is used to construct a model for the non-perturbative spectral density that, in turn, defines the non-perturbative stress-tensor two-point function.
The masses and couplings (to the stress-tensor) of the first two glueballs in the spectrum were estimated by analyzing the Borel transformations of perturbative and non-perturbative stress-tensor two-point functions. 
While our estimates are not rigorous, there exists a reasonable range of parameters in the non-perturbative model where one finds stable results that are within $14-19\%$ of the lattice data.
Here, it was important to work with the Borel transformation of the two-point functions in order to improve the convergence of the perturbative expansion. 

It was also crucial to combine the spin-0 and spin-2 parts ($A_0$ and $A_2$) of the stress-tensor two-point function into a ``superconvergent'' sum \eqref{magic}.
Otherwise, we would have had to use a subtracted dispersion relation that removes the connection between the pole (glueball) and cut (continuum) contributions; the Borel transform of such a subtracted dispersion relation kills the first term in perturbation theory increasing the sensitivity to non-perturbative condensates.
The existence of this superconvergent combination is tied to the spin of the stress tensor. 
For similar reasons, there also exists superconvergent dispersion relations for scattering amplitudes of spinning particles \cite{Kologlu:2019bco,Caron-Huot:2022ugt}.

In principle, it would be possible to extend our analysis of the stress-tensor correlator to four-loops since the non-perturbative condensate $\<0|F^2|0\>$ does not appear in the superconvergent combination. 
Even if the condensate did appear this would not necessarily be a showstopper since the $\MSbar$ condensate has been extracted from a combination of lattice and perturbative techniques \cite{Hietanen:2004ew,DiRenzo:2006nh}. 
Furthermore, one could possibly bound the lattice regularized condensate using the bootstrap techniques of \cite{Anderson:2016rcw, Kazakov:2022xuh}.

We also showed that it is mathematically possible to find positive spectral densities that display the correct asymptotic behavior at large energies that are compatible with essentially any mass spectrum and residue strength (section \ref{sec:UC consistency}).
Finally, anticipating applications to higher moments of hadron wavefunctions (form factors of higher-spin lowest-twist operators), we also verified that superconvergent combinations of higher-spin operators exist (section \ref{sec:higher-spin}). 
While we provide a crude method for approximating the higher-spin residues, we leave the analysis of higher-spin sum-rules to future work. 

Our results in three-dimensional Yang-Mills theory adds numerical evidence to the effect that the Borel transform of perturbation theory can give a reasonable approximation to continuum spectral densities at finite energy, even when using a finite number of terms. 
This is similar to what has long been observed in the QCD context.  
Of course, it has never been clear how to rigorously justify this approximation and we do not claim to have ameliorated this state of affairs.

\acknowledgments 
A.P. is grateful for support provided by the National Science and Engineering Council of Canada and the Fonds de Recherche du Qu\'ebec \textemdash \ Nature et Technologies. A.P. is also supported by the Simons Investigator Award $\#376208$ of A. Volovich.
S.C.H.'s work is supported in parts by the National Science and Engineering Council of Canada (NSERC) and by the Canada Research Chair program, reference number CRC-2022-00421.  S.C.H.'s work is additionally supported by a Simons Fellowships in Theoretical Physics and by the Simons Collaboration on the non-perturbative Bootstrap.  Z.Z. is funded by Fonds de Recherche du Qu\'ebec \textemdash \ Nature et Technologies, and the Simons Foundation through the Simons Collaboration on the non-perturbative Bootstrap. This project has received funding from the European Research Council (ERC) under the European Union's Horizon 2020 research and innovation programme (grant agreement number 949077).

\appendix

\section{$d$-dimensional form factors \label{app:FFd}}

In this appendix, we present the $d$-dimensional stress-tensor two-point functions up to three-loops. 
At three-loops, not all master integrals are know in closed form for generic $d$.
However, once $d$ is fixed these integrals can be computed via dimensional recursion \cite{Lee2010}.

\subsection{One-loop}

The one-loop two-point functions for generic dimension are
\begin{align}
	A^{(1)}_0(p^2;d) &= \frac{512}{8 (d-1)^2} \left[(d-4)^2 (d-2) p^4 I^{(1)}_1(p^2;d) 
		\right],
	\\
	A^{(1)}_2(p^2;d) &= \frac{512}{8 (d-1) (d+1)} \left[( 2d^2 - 3d - 8 ) p^4 I^{(1)}_1(p^2;d) 
		\right],
\end{align}
where $	I^{(1)}_1 $ is the scalar bubble integral and  the normalization is determined by equation \eqref{eq:TT}.
While the bubble integral is trivial, we quote it here so that our conventions are explicit
\begin{align}
	I^{(1)}_1(p^2;d) \equiv B_{1,1}(p^2;d)
\end{align}
where 
\begin{align} \label{eq:bubble integral}
    B_{a,b}(k^2;d)
    &= \int\frac{\d^{d}\ell}{i\!\left(2\pi\right)^{d}}\frac{1}{\big[\ell^{2}\big]^{a}\big[\left(\ell+k\right)^{2}\big]^{b}}
    \nn\\
    &=\frac{1}{\left(4\pi\right)^{\frac{d}{2}}}\frac{\Gamma\left(a+b-\frac{d}{2}\right)\Gamma\left(\frac{d}{2}-a\right)\Gamma\left(\frac{d}{2}-b\right)}{\Gamma\left(a\right)\Gamma\left(b\right)\Gamma\left(d-a-b\right)}\left( k^{2}\right)^{\frac{d}{2}-\left(a+b\right)}.
\end{align}
With the exception of the three-loop master integrals $I^{(3)}_1$ and $I^{(3)}_2$, all other two- and three-loop master integrals can be computed in closed form from recursive use of \eqref{eq:bubble integral}.

\subsection{Two-loops}

The two-loop contributions to the stress-tensor two-point functions in generic dimension are
\begin{align}
	\label{eq:2loop A0(d)}
	A^{(2)}_0(p^2;d) 
	&= \frac{512 (g_s^2 C_A)}{8 (d{-}1)^2}
	\bigg[
		(d{-}4) \left(d^3{-}16 d^2{+}68 d{-}88\right) p^4 I^{(2)}_1(p^2;d)
		\nn\\&\qquad
		{-} \frac{16}{3} \left(4 d^3{-}33 d^2{+}94 d{-}92\right) p^2 I^{(2)}_2(p^2;d)	
	\bigg]
	\\
	\label{eq:2loop A2(d)}
	A^{(2)}_2(p^2;d) 
	&= -\frac{512 (g_s^2 C_A)}{8 (d{-}1) (d{+}1)}
	\bigg[
		\frac{8 (d^4{-}8 d^3{+}16 d^2{+}20 d{-}68) p^4}{(d{-}4) (d{-}2)} I^{(2)}_1(p^2;d)
		\nn\\&\qquad
		{+}\frac{8 (13 d^5{-}129 d^4{+}462 d^3{-}572 d^2{-}376 d{+}1088) p^2}{3 (d{-}4)^2 (d{-}2)}	
		I^{(2)}_2(p^2;d)
	\bigg]
\end{align}
where the master integrals are 
\begin{align} \label{eq:2loop masters}
	I^{(2)}_1(p^2;d)  
	= \includegraphics[align=c,width=.2\columnwidth]{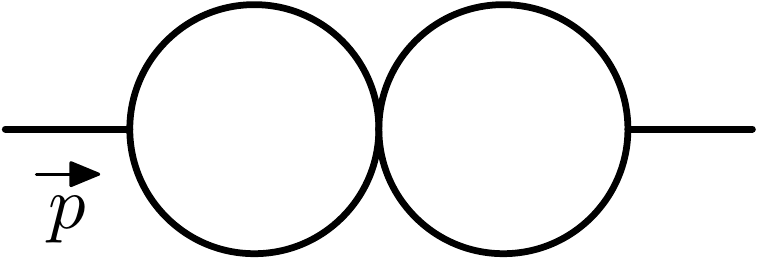} \, ,
	\qquad
	I^{(2)}_2(p^2;d)  
	= \includegraphics[align=c,width=.2\columnwidth]{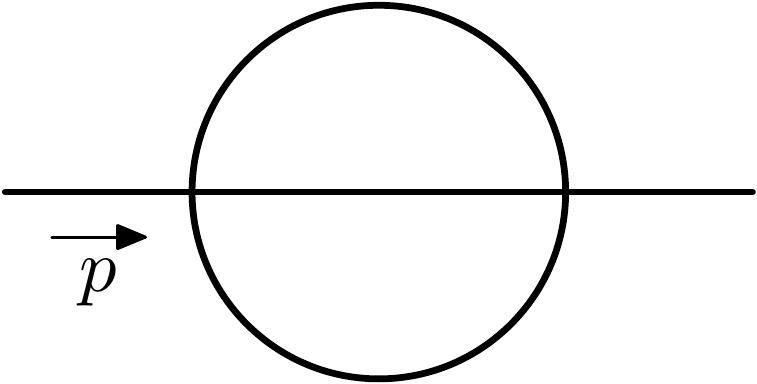} \, .
\end{align}
These master integrals are easily evaluated by repeated used of \eqref{eq:bubble integral}.

\subsection{Three-loops}

The three-loop contributions to the stress-tensor two-point functions in generic dimension are
\begin{align} \label{eq:3loop A0(d)}
	A^{(3)}_0(p^2;d)  &= \frac{512 (g_s^2 C_A)^2}{8 (d{-}1)^2}
	\bigg[
		{-}\frac{3 (d{-}4)^2 (d{-}3) (d{-}2) (3 d{-}8) p^8 I^{(3)}_1(p^2;d) }{4 (2 d{-}7) (2 d{-}5)}
	\nn\\&\
		{-}\frac{ (d^3{-}16 d^2{+}68 d{-}88)^2 p^4 I^{(3)}_3(p^2;d) }{d{-}2}
	\nn\\&\
		{+}\Big(
			657 d^7
			{-}11454 d^6
			{+}85564 d^5
			{-}354832 d^4
			{+}880176 d^3
			{-}1299616 d^2
			\nn\\&\qquad
			{+}1048384 d
			{-}350976 
		\Big)
		\frac{p^2 I^{(3)}_5(p^2;d) }{2 (d{-}4) (d{-}2) (d{-}1) (2 d{-}5)}	
	\nn\\&\
		{-}\Big(
			108 d^8
			{-}2661 d^7
			{+}28822 d^6
			{-}177546 d^5
			{+}674735 d^4
			{-}1607602 d^3
			\nn\\&\qquad
			{+}2325996 d^2
			{-}1848920 d
			{+}607968
		\Big)
		\frac{p^4 I^{(3)}_2(p^2;d) }{2 (d{-}2) (d{-}1) (2 d{-}7) (2 d{-}5)}
	\nn\\&\
		{+}\Big(
			192 d^{10}
			{-}6947 d^9
			{+}105470 d^8
			{-}907248 d^7
			{+}4958664 d^6
			{-}18113645 d^5
			\nn\\&\qquad
			{+}44930982 d^4
		 	{-}74791460 d^3
			{+}79854504 d^2
			{-}49204128 d
			\nn\\&\qquad
			{+}13194496
		 \Big)	
		 \frac{p^2 I^{(3)}_4(p^2;d) }{(d{-}4) (d{-}3) (d{-}2) (d{-}1) (2 d{-}7) (2 d{-}5)}
	\nn\\&\
		{+}\Big(
			162 d^{11}
			{-}5487 d^{10}
			{+}87553 d^9
			{-}858385 d^8
			{+}5673221 d^7
			{-}26253008 d^6
			\nn\\&\qquad
			{+}86068824 d^5
			{-}198637272 d^4
			{+}314636144 d^3	
			{-}324171296 d^2
			\nn\\&\qquad
			{+}194410240 d
			{-}50999296
		\Big)
		\frac{I^{(3)}_6(p^2;d) }{(d{-}4)^2 (d{-}3)^2 (d{-}2) (d{-}1) (2 d{-}7)}
	\bigg]
\end{align}
and
\begin{align}	\label{eq:3loop A2(d)}
	&A^{(3)}_2(p^2;d)  = \frac{512 (g_s^2 C_A)^2}{8 (d{-}1) (d{+}1)}
	\bigg[
		{-}\frac{ \left(16 d^5{-}149 d^4{+}397 d^3{+}142 d^2{-}1832 d{+}1696\right) p^8 I^{(3)}_1(p^2;d) }{4 (d{-}2) (2 d{-}7) (2 d{-}5)}
	\nn\\&\	
		{-}\frac{8  \left(4 d^8{-}62 d^7{+}371 d^6{-}939 d^5{+}128 d^4{+}4260 d^3{-}7712 d^2{+}3584 d{+}384\right) p^4 I^{(3)}_3(p^2;d) }{(d{-}4)^2 (d{-}2)^2 (d{-}1) d}
	\nn\\&\
		{+}\Big(
				1042 d^{10}
				{-}19207 d^9
				{+}147122 d^8
				{-}588708 d^7
				{+}1199632 d^6
				{-}543184 d^5
				{-}3040032 d^4
				\nn\\&\qquad
				{+}7331904 d^3
				{-}7007488 d^2
				{+}2514944 d
				{+}24576
			\Big)
			\frac{  p^2 I^{(3)}_5(p^2;d) }{2 (d{-}4)^3 (d{-}2)^2 (d{-}1) d (2 d{-}5)}
	\nn\\&\
		{-}\Big(
				1680 d^{12}
				{-}43447 d^{11}
				{+}499154 d^{10}
				{-}3324848 d^9
				{+}13961672 d^8
				{-}36985777 d^7
				\nn\\&\qquad
				{+}54553314 d^6
				{-}11375804 d^5
				{-}120445352 d^4
				{+}236351744 d^3
				{-}195105152 d^2
				\nn\\&\qquad
				{+}60414976 d
				{+}1720320
			\Big) 
			\frac{ p^2 I^{(3)}_4(p^2;d) }{(d{-}4)^3 (d{-}3) (d{-}2)^2 (d{-}1) d (2 d{-}7) (2 d{-}5)}
	\nn\\&\	
		{-}\Big(
				72 d^{12}
				{-}1812 d^{11}
				{+}24945 d^{10}
				{-}234230 d^9
				{+}1498316 d^8
				{-}6288301 d^7
				{+}16330266 d^6
				\nn\\&\qquad
				{-}22168812 d^5
				{+}1696440 d^4
				{+}41289728 d^3
				{-}55822976 d^2
				{+}25437184 d
				{-}1720320
			\Big)
			\nn\\&\qquad \times
			\frac{ p^4 I^{(3)}_2(p^2;d) }{2 (d{-}4)^2 (d{-}2)^2 (d{-}1) d (2 d{-}7) (2 d{-}5) (3 d{-}8)}
	\nn\\&\	
		{-}\Big(
				432 d^{15}
				{-}27558 d^{14}
				{+}582633 d^{13}
				{-}6158463 d^{12}
				{+}35473743 d^{11}
				{-}89675899 d^{10}
				\nn\\&\qquad
				{-}197920872 d^9
				{+}2586125488 d^8
				{-}10618482072 d^7
				{+}25226597520 d^6
				\nn\\&\qquad
				{-}36849379104 d^5
				{+}30607655680 d^4
				{-}9263259648 d^3
				{-}4995555328 d^2
				\nn\\&\qquad
				{+}3991977984 d
				{-}421134336
			\Big)
			\frac{ I^{(3)}_6(p^2;d) }{3 (d{-}4)^4 (d{-}3)^2 (d{-}2)^2 (d{-}1) d (2 d{-}7) (3 d{-}8)} 
	\bigg]
\end{align}
where
\begin{align} \label{eq:3loop masters}
	I^{(3)}_1(p^2;d) =
		\includegraphics[align=c,scale=.15]{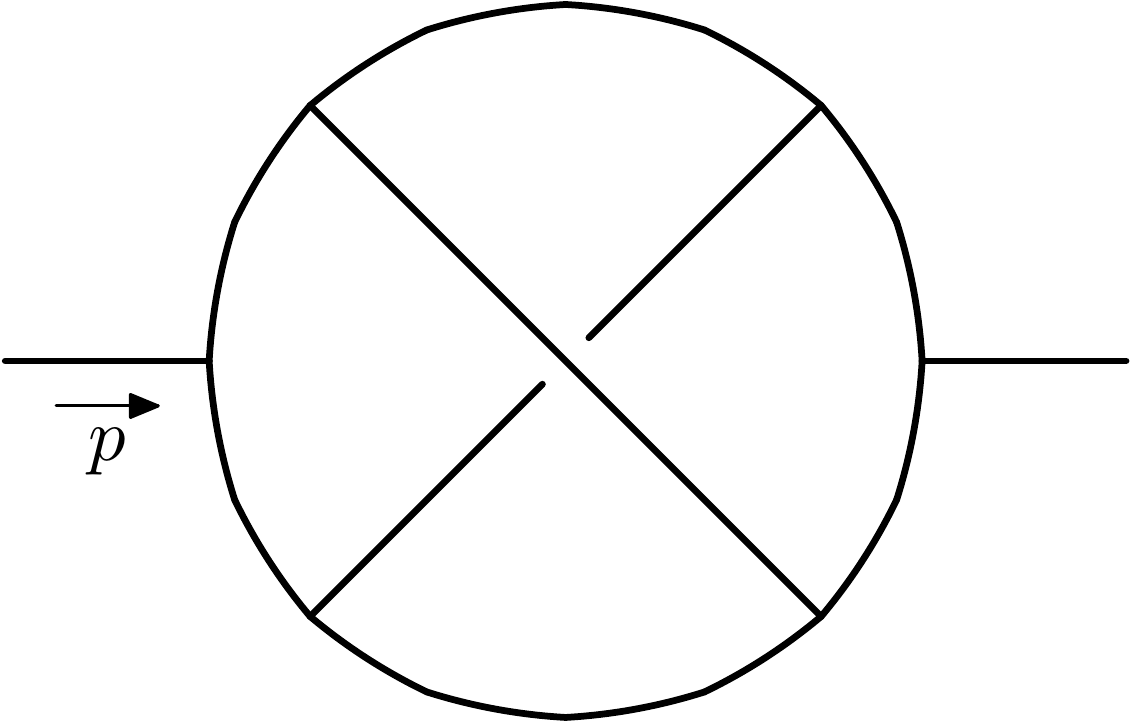} \, ,		
 	\qquad & 
	I^{(3)}_2(p^2;d) =
		\includegraphics[align=c,scale=.15]{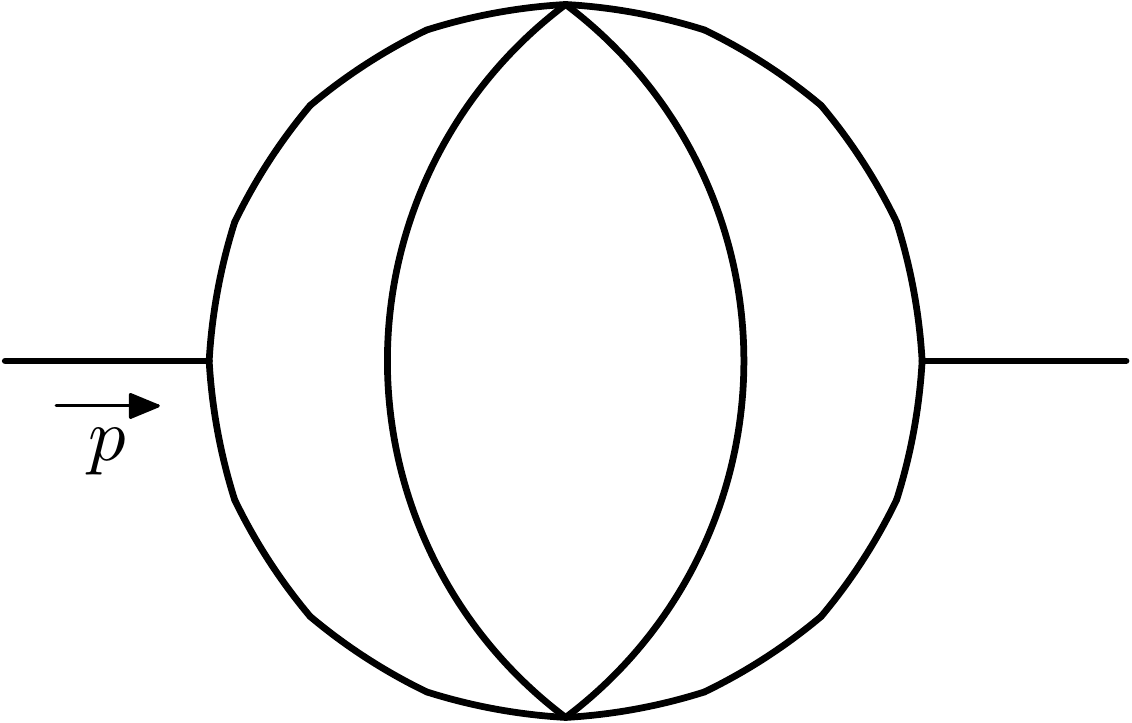}	\, ,	
	\nn\\
	I^{(3)}_3(p^2;d) =
		\includegraphics[align=c,scale=.2]{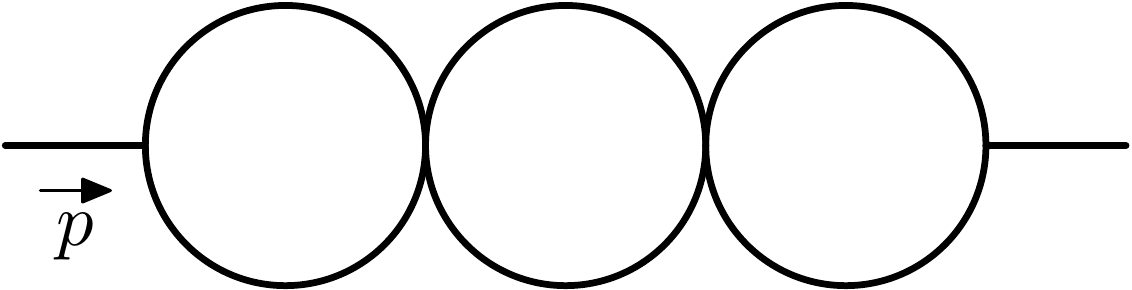} \, ,		
	\qquad & 
	I^{(3)}_4(p^2;d) =
		\includegraphics[align=c,scale=.15]{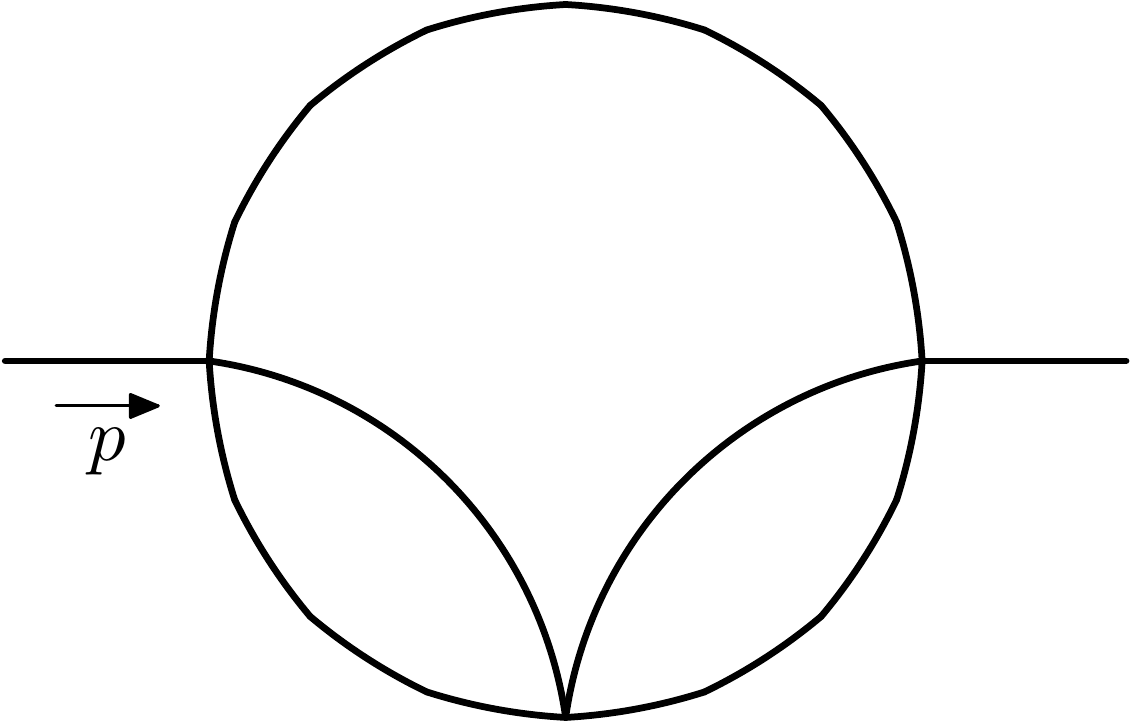} \, ,
	\nn\\
	I^{(3)}_5(p^2;d) =
		\includegraphics[align=c,scale=.2]{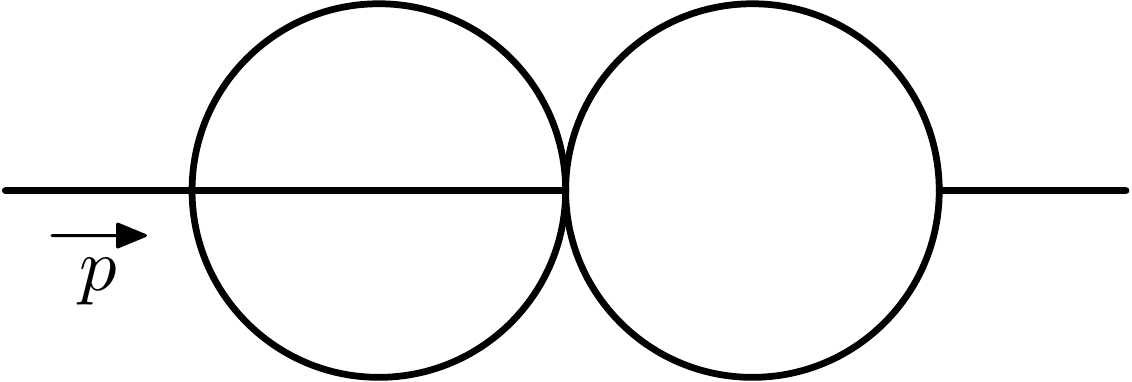} \, ,		
	\qquad & 
	I^{(3)}_6(p^2;d) =
		\includegraphics[align=c,scale=.15]{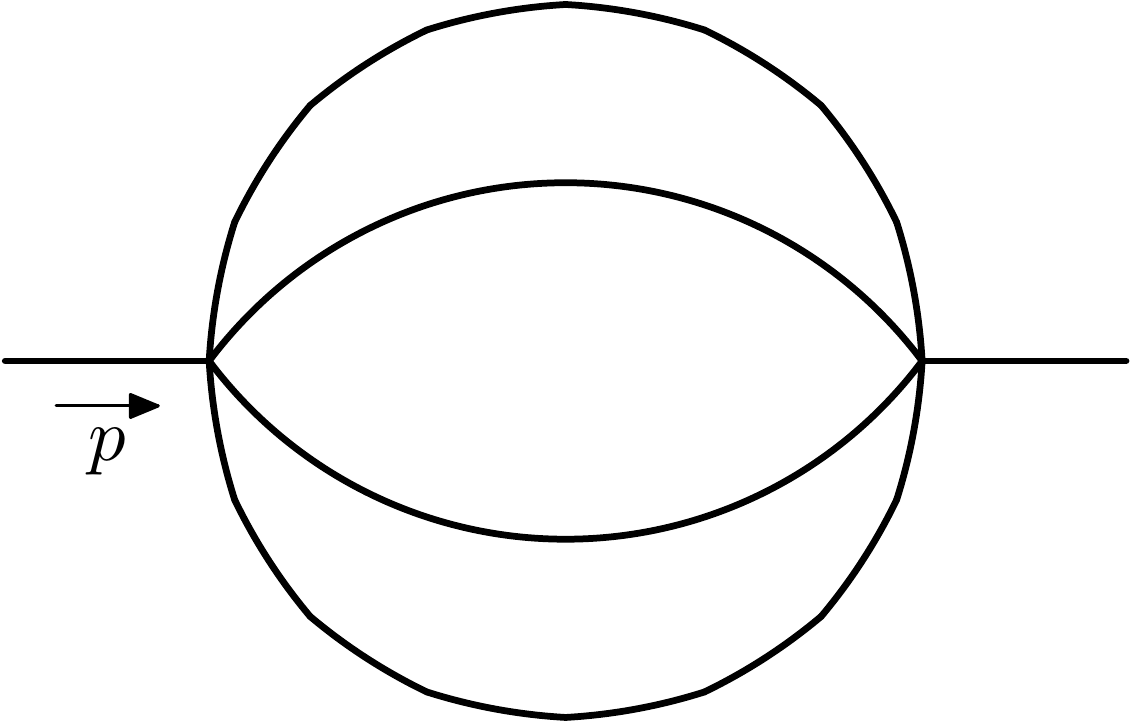}	\, ,
\end{align}
are the scalar master integrals at three-loops.
All but $I^{(3)}_1$ and $I^{(3)}_2$ are easily evaluated by repeated used of \eqref{eq:bubble integral}. 

While $\vep$-expansion of $I^{(3)}_1$ and $I^{(3)}_2$ are know near four-dimensions \cite{Chetyrkin1980, Baikov2010}, we must compute these expansons from scratch near three-dimensions since certain simplifications in four-dimensions are \emph{not} present in three-dimensions. 
We use the method of dimensional recursion \cite{Lee2010} (reviewed in appendix \ref{app:dim rec and an}) to find $d$-dimensional formulas for $I^{(3)}_1$ and $I^{(3)}_2$.

In three-dimensions, these integrals simplify to 
\begin{equation} \label{I13}
\begin{split}
	I^{(3)}_1(p^2) &\underset{d\to3}{=} 
	-\frac{(2 \pi^2 - 39)}{192 \pi^2 (p^2)^{7/2}},
	\\
	I^{(3)}_2(p^2) &\underset{d\to3}{=} 
	\frac{1}{512 (p^2)^{3/2} }.
\end{split}
\end{equation}
Since the coefficients of $I^{(3)}_1$ and $I^{(3)}_2$ in equations \eqref{eq:3loop A0(d)} and \eqref{eq:3loop A2(d)} are finite in the limit $d\to3$, the above formulas are sufficient for determining the three-loop contributions to $A_0$ and $A_2$.

\section{Computing $I^{(3)}_1$ and $I^{(3)}_2$ from dimensional reccurance \label{app:dim rec and an}}

In this appendix, we provide a short overview of the method of dimensional recurrence \ref{sec:dimensional recurrence and analyticity} and provide formulas to compute $I^{(3)}_1$ and $I^{(3)}_2$ in any dimension (section \ref{sec:computing I31 and I32}).

\subsection{Dimensional recurrence and analyticity in $d$ \label{sec:dimensional recurrence and analyticity}}

In this shot review of the method of dimensional recurrence and analyticity in $d$ \cite{Lee2010}, we keep the discussion general. We specify to the integral family relevant to $I^{(3)}_1$ and $I^{(3)}_2$ in section \ref{sec:computing I31 and I32}.

Suppose that we are given a family of Feynman integrals $\vec{I}$ that is closed under IBP relations.  
Then, this family satisfies the following dimensional recurrence relation
\begin{align} \label{eq:recurrence relation}
	\vec{I}(d+2) = \mat{R}(d) \cdot \vec{I}(d). 
\end{align}
Additionally, all Feynman integrals have the following projective parametric representation\footnote{By projective, we mean that $I$ is invariant under the rescaling of $\vec{x}$: $\vec{x}\to \lambda\vec{x}$.}
\begin{align} \label{eq:projective rep}
	I
	=\int \left(\prod_{i=1}^L\frac{\d^d\ell_i}{i (2\pi)^{d}}\right)
	\left(\prod_{j=1}^N \frac{1}{D_j^{n_j}}\right) 
	= \Gamma(\omega) \int_{(\mathbb{R}^+)^N} \d^N \vec{x}
	\left( 
		\prod_{i=1}^N
		\frac{x_i^{n_i-1}}{\Gamma(n_i)}
	\right)
	\frac{\delta(1-h(\vec{x}))}{\mathcal{U}^{\frac{d}{2}-\omega} \mathcal{F}^{\omega}}
\end{align}
where $\omega(d)=\frac{d}{2}-\vert\vec{n}\vert$ is the superficial degree of divergence, $x_i$ is the Schwinger parameter associated to the propagator $D_j$, $h(\vec{x})$ is any degree 1 homogeneous polynomial, and, $\mathcal{U}$ and $\mathcal{F}$ are the first and second Symanzik polynomials. 
Using the projective representation \eqref{eq:projective rep}, one can bound the large imaginary $d$ limit of a Feynman integral 
\begin{align} \label{eq:imd bound}
	\vert I(d) \vert 
	\lesssim \text{const.} \times \vert \Im\,d \vert^{\omega(\Re\,d)-\frac12} 
		e^{-\frac{\pi}{4}L\, \Im\,d}.
\end{align}
Then, using the above bound and provided that there exists a strip $S=\{d\in\mathbb{C} \vert d_\text{min} < \Re\,d < d_\text{max}\}$ that is known to be free from poles, the homogenous solution to the recurrence relation \eqref{eq:recurrence relation} can be constructed. 

The first step to solve the recurrence relation \eqref{eq:recurrence relation} is to define the so-called summing factors $\vec{\Sigma}(d)$ such that 
\begin{align}
	\frac{\Sigma_i(d+2)}{\Sigma_i(d)}
	= R_{ii}(d)
\end{align}
Then defining the rescaled integrals $J_i(d) = I_i(d) / \Sigma_i(d)$ and $r_i(d) = \sum_{j \neq i} R_{ij} I_j(d) / \Sigma_i(d)$ the recurrence relation \eqref{eq:recurrence relation} becomes
\begin{align} \label{eq:Jrr}
	\vec{J}(d+2) = \vec{J}(d) + \vec{r}(d). 
\end{align}
The general solution to \eqref{eq:Jrr} consists of a homogeneous and a inhomogeneous solution $\vec{J}(d) = \vec{J}_\hom(d) + \vec{J}_\inhom(d)$.

The homogeneous solution $\vec{J}_\hom(d) = \vec{f}(d)$ can be any periodic in $d$ with period 2: $\vec{f}(d+2) = \vec{f}(d)$.
Since the product $f_i(d) \Sigma_i(d)$ must obey the bound \eqref{eq:imd bound}, choosing $\Sigma_i$ such that it comes as close as possible to saturating it maximally constrains the form of $f_i$. 
In particular, it is always possible to find a $\Sigma_i$ such that \eqref{eq:imd bound} forces $\vert f(d) \vert < \vert \Im\,d \vert^\nu e^{\pi \vert \Im\,d \vert} $ for some $\nu$. 
Then, the only 2-periodic function of $d$ that satisfy this bound is $\cot$ (or $\tan$). 
Thus, $f_i$ has the following form
\begin{align}
	f_i(d) = b_{i0} + \sum_{j}^{n_{ij}} \sum_{k=1}^{L} b_{ijk} 
		\cot^k\left(\frac{\pi}{2}(d-q_{ij})\right)
\end{align}
where the $q_{ij}$ are poles that appear in $J_{i,\inhom}$, $n_{ij}$ is the number of distinct $q_{ij}$ and $L$ is the maximal order of any pole. 
Then, the $b_{ijk}$'s are fixed by requiring that $I_i$ is free from all poles in the strip $S_i$.

Sometimes, this requirement will not fix all $b_{ijk}$'s and one has to generate additional conditions. 
Additional conditions can be generated by relating $I_i$ to $\tilde{I}_i$ via an IBP relation and then requiring that $\tilde{I}_i$ is pole free in its strip $\tilde{S}_i$. 
For example, squaring all propagators defines and integral with a larger finite strip. 
Since this new new integral is related to the old integral via an IBP relation, requiring that the new integral is free from all poles in its enlarged strip may impose new constraints on the old integral. 

To obtain the inhomogeneous solution, we split $r_i(d)$ into two pieces $r_i(d) = r^+_i(d) + r_i^-(d)$ where $r_i^+(d+2k) \sim a^k$ and $r_i^-(d-2k) \sim a^k$ in the large $k$ limit with $0<a<1$. 
Then, the inhomogeneous solution $\vec{J}_\inhom(d) = \vec{g}(d)$ becomes
\begin{align} \label{eq:inhom sol}
	g_i(d) = \sum_{k=0}^\infty r_i^+(d+2k) + \sum_{k=1}^\infty r_i^-(d-2k). 
\end{align} 
Since each term in the sum is suppressed by some $a^k$ this series converges exponentially. 
While each integral in the family usually contributes only to $r^+$ or $r^-$ sometimes it is necessary to split an integral into two pieces (this will be the case for $I_1^{(3)}$). 

This method expresses integrals in terms of (nested) sums that converge rapidly. 
In practice, one computes these sums numerically to many digits and then applies the PSLQ algorithm to recover analytic results.

\subsection{Computing $I^{(3)}_{1}$ and $I^{(3)}_{2}$ \label{sec:computing I31 and I32}} 

Using the formalism outlined in the previous section, we evaluate the integrals $I^{(3)}_{1}$ and $I^{(3)}_{2}$ for $d=3$.

Before being able to apply the methods from section \ref{sec:dimensional recurrence and analyticity}, we must check if $I^{(3)}_{1}$ and $I^{(3)}_{2}$ have a strip of width at least two that is free from poles.
The integral $I^{(3)}_1$ has a strip of width two: $S_1 = \{d\in\mathbb{C}\vert\frac{10}{3}<\Re\,d<\frac{16}{3}\}$ where $d=\frac{16}{3}$ is the minimal UV divergence and $d=\frac{10}{3}$ is the maximal IR divergence. 
On the other hand, $I^{(3)}_2$ does not have a strip of width two since its minimal UV divergence is at $d=4$ and its maximal IR divergence is at $d=\frac{8}{3}$. 

In order to use the methods of the previous section, we replace $I^{(3)}$ by the related integral 
\begin{align}
	\tilde{I}^{(3)}_2 = \includegraphics[align=c,scale=.2]{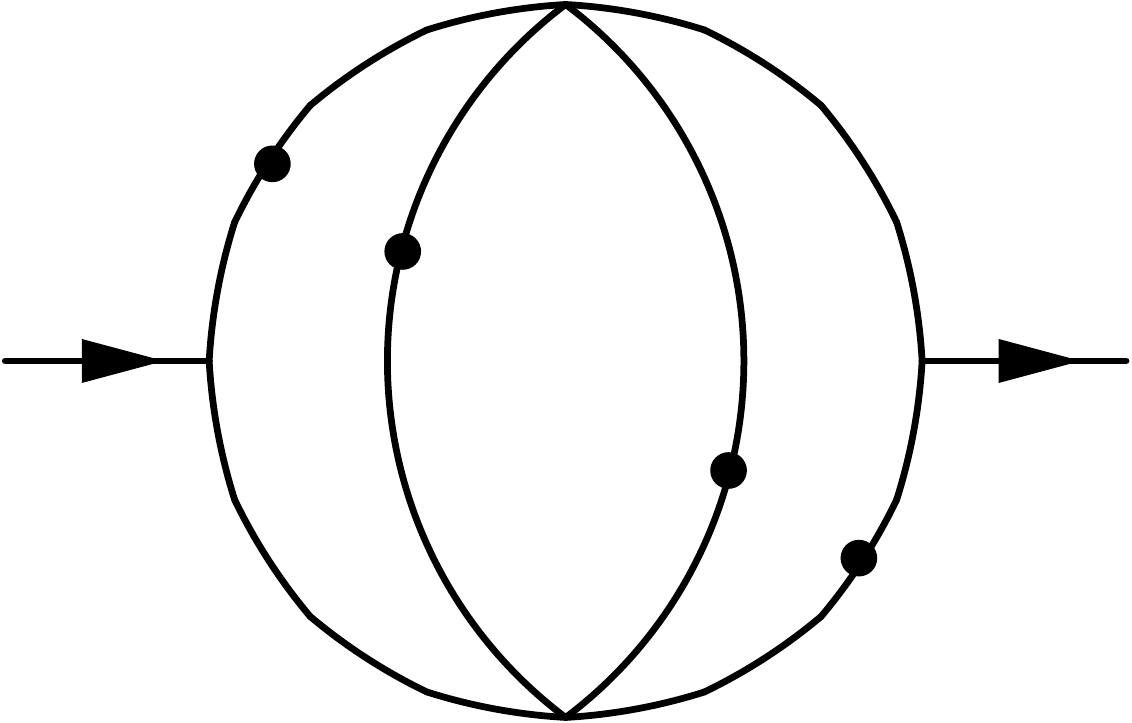}
\end{align} 
where each dotted propagator is squared.
By selectively squaring the propagators of $I^{(3)}_2$, we have enlarged the strip $S_2$ to $\tilde{S}_2=\{d\in\mathbb{C}\vert\frac{14}{3}<\Re\,d<\frac{20}{3}\}$.
Once $\tilde{I}^{(3)}_2$ is known $I^{(3)}_2$ is determined via the IBP relation 
\begin{align} \label{eq:IBPforI2}
	I^{(3)}_2(p^2;d) &= 
	\frac{
		16 (d-5) \ \left(p^2\right)^2 \tilde{I}^{(3)}_2(p^2;d)
	}{
		3 (d-3) (3 d-14) (3 d-10)\ U(d)
	}
	\nn\\&\qquad
	-\frac{
		4 (2d-5) (3d-8)\ T(d)\ \left(p^2\right)^{-1} I^{(3)}_{6}(p^2;d)
	}{
		3 (d-6)^{2} (d-5) (d-4)^{2} (3d-14) (3d-10)\ U(d)
	},
\end{align}
where
\begin{align}
	U(d) &= 3d^{2}-33d+92,
	\\
	T(d) &= 3429d^{7}-109566d^{6}+1491897d^{5}-11216508d^{4} 
	\nn\\&\qquad
		+ 50262008d^{3}-134170880d^{2}+197449040d-123506880.
\end{align}
 
Now, we can define a new family of integrals 
\begin{align} \label{eq:Iprime}
	\vec{I}^\prime
	= 
	\left(
		\frac{I^{(3)}_1}{\left(p^2\right)^{\omega_1}} ,\ 
		\frac{\tilde{I}^{(3)}_2}{\left(p^2\right)^{\tilde{\omega}_2}} ,\ 
		\frac{I^{(3)}_3}{\left(p^2\right)^{\omega_3}} ,\ 
		\dots ,\ 
		\frac{I^{(3)}_6}{\left(p^2\right)^{\omega_6}}
	\right)
\end{align}
for which the formalism of section \ref{sec:dimensional recurrence and analyticity} is applicable. 
Normalizing by $\left(p^2\right)^{-\omega_i}$ where $\omega_i$ is the superficial degree of divergence of $I^{(3)}_i$, ensures that the basis $I^\prime$ is dimensionless. 
This family of integrals satisfies the recurrence relation 
\begin{align}
	\vec{I}^\prime(d+2) = \mat{R}(d) \cdot \vec{I}^\prime(d). 
\end{align}
where $\mat{R}(d)$ is a lower triangular $6{\times}6$-matrix. 
We also define the following summing factors that (almost) saturate the bound \eqref{eq:imd bound}
\begin{align}
	\Sigma_1^\prime(d) 
	&= \frac{1}{(4\pi)^{\frac{d}{2}}}
		\left(\frac{7}{2}-d\right) 
		\Gamma\left(6-2d\right)
		\Gamma\left(\frac{d}{2}-2\right),
	\\
	\Sigma_2^\prime(d) 
	&= \frac{
			U(d)\
			\Gamma\left(\frac{3}{2}-\frac{d}{2}\right)
			\Gamma\left(\frac{8}{3}-\frac{d}{2}\right)
			\Gamma\left(\frac{10}{2}-\frac{d}{2}\right)
			\sec\left(\frac{\pi}{2}d\right)
		}{
			4^d\ 3^\frac{3d}{2}\ \pi^{\frac{d}{2}}\ (d-5)
		}.
\end{align} 
With this, the inhomogeneous solutions are given by \eqref{eq:inhom sol}. 
We remark here that $r^-_2=0$ and $r^-_1$ only receives contribution from the homogeneous solution of $I^\prime_2$. 
All other integrals contributes to $r^+_{1}$ and $r^-_2$.
 
The final piece is the homogeneous solutions
\begin{align}
	f_1^\prime(d)
	&= -\frac{16 \pi^3}{9} \bigg[
		-15 \cot^3\left(\frac{\pi}{2} (d-4)\right)
		+ 16 \cot\left(\frac{\pi}{2} (d-4)\right)
		+ 9 \cot\left(\frac{\pi}{2} (d-5)\right)
	\nn\\&\qquad\qquad\qquad
		- 2 \cot\left(\frac{\pi}{2} \left(d-\frac{10}{3}\right)\right)
		- 2 \cot\left(\frac{\pi}{2} \left(d-\frac{14}{3}\right)\right)
	\bigg],
	\\
	f_2^\prime(d) 
	&= 2187 \sqrt{3}\ \pi^{\frac32}   
		\cot\left(\frac{\pi}{2}(d-6)\right) 
		\left(1-\cot^2\left(\frac{\pi}{2}(d-6)\right)\right). 
\end{align}
While requiring $I^\prime_2$ to be free of poles in the strip $\tilde{S}_2$ fixes all the coefficients of $f^\prime_2$, requiring $I^\prime_1$ to be free from poles in $S_1$ leaves one coefficient of $f^\prime_1$ unfixed. 
The remaining coefficient was fixed by requiring the integral obtained by squaring all propagators of $I^\prime_1$, which is related to $I_1^\prime$ by IBP relations, to be free from poles in its strip.

Putting all the pieces together yields expressions for $I^\prime_1$ and $I^\prime_2$
\begin{align}
    I^\prime_{i=1,2}(d)
    = \Sigma^\prime_i(d) \left[
        f^\prime_i(d) 
        + g_i^\prime(d)
    \right].
\end{align}
For a given $d$, the infinite sum in $g_i^\prime$ can be truncated and evaluated numerically. Then, analytic expressions for $I^\prime_1$ and $I^\prime_2$ are recovered using the PSLQ algorithm.

Once $I^\prime_1$ and $I^\prime_2$ are known for a given $d$, we can determine the integrals we actually need
\begin{equation}
\begin{split}
    I^{(3)}_1(p^2;d) &=
    (p^2)^{\omega_1} I^\prime_1(p^2;d)
    ,
    \\
    I^{(3)}_2(p^2;d) &=
    \frac{
        16 (3-5) \ 
        (p^2)^{2+\tilde{\omega}_2}
        I^\prime_2(p^2;d)
    }{
	3 (3-3) (3 d-14) (3 d-10)\ U(d)
    }
    \\&\qquad
    -\frac{
	4 (2d-5) (3d-8)\ 
        T(d)\ 
        (p^2)^{\tilde{\omega}_2-1} I^{(3)}_{6}(p^2;d)
    }{
	3 (d-6)^{2} (d-5) (d-4)^{2} 
        (3d-14) (3d-10)\ U(d)
    }
    .
\end{split}
\end{equation}
Here, we have used the IBP relation \eqref{eq:IBPforI2} and the definition of the primed-basis \eqref{eq:Iprime}.
For $d=3$, we find \eqref{I13}.

\section{Ingredients for on-shell calculations}
In this appendix we presents further details for obtaining the results of sections \ref{sec:1and2LoopUnitarity} and \ref{subsec:high-spin-unitarity}.

\subsection{Stress-Tensor gluon form factors \label{app:bcfw}}
In this section we discuss the derivation of eq.~\eqref{2loopcut1} using BCFW method \cite{Britto:2005fq, Britto:2004ap}. We start by writing the form factor $\langle p_1^gp_2^g|T^{\mu\nu}(p)|0\rangle$ in four-dimensions using spinor-helicity variables,
\begin{equation}
    \langle p_1^{-}p_2^{+}|T^{\mu\nu}(p)|0\rangle^{4d}=\delta^{ab}\frac{\langle 1^{\dot{\alpha}}\langle 1^{\dot{\beta}}\langle 1^{\dot{\gamma}} p^{\alpha}_{\dot{\gamma}} \sigma^{\mu}_{\alpha\dot{\alpha}}\langle 1^{\dot{\rho}}p^{\beta}_{\dot{\rho}} \sigma^{\nu}_{\beta\dot{\beta}}}{\langle 12\rangle^2},
\end{equation}
where $\alpha,\beta,\gamma$ and $\rho$ (and their dotted) version indices are $SU(2)$ indices. We can then obtain $\langle p_1^{+}p_2^{+}p_3^{-}|T^{\mu\nu}(p)|0\rangle$ by shifting $p_3$ and $p_2$ as follows,
\begin{equation}
|\hat{2}]=|\hat2]\quad |\hat{3}]=|3]+z|2] \quad |\hat{2}\rangle=|2\rangle-z|3\rangle \quad |\hat{3}\rangle=|3\rangle,
\end{equation}
The on-shell form factor is then given as:
\begin{equation}
\label{eq:Tggg-4d-v1}
    \langle p_1^{+}p_2^{+}p_3^{-}|T^{\mu\nu}(p)|0\rangle^{4d}=\langle \hat{P}^{+}_{12}\hat{p}_3^{-}|T^{\mu\nu}(p)|0\rangle\frac{1}{P_{12}^2}M_3(p^+_1,\hat{p}^2_2,-\hat{P}^{-}_{12})
\end{equation}
where the 3-gluon on-shell form factor can be written as,
\begin{equation}
    M_3(p^+_1,\hat{p}^2_2,-\hat{P}^{-}_{12})=g_sf^{bcd}\frac{[1\hat{2}]^3}{[1\hat{P}_{12}][\hat{2}\hat{P}_{12}]}.
\end{equation}
With the little bit manipulation eq.~\eqref{eq:Tggg-4d-v1} can be written as,
\begin{equation}
     \langle p_1^{+}p_2^{+}p_3^{-}|T^{\mu\nu}(p)|0\rangle^{4d}=2g_sf^{bcd}\frac{\langle 3\langle 3\langle 3 p\sigma^{\mu}\langle 3p\sigma^{\nu}}{\langle 12\rangle\langle 23\rangle\langle 31\rangle}.
\end{equation}
Here we omitted the $SU(2)$ indices. To go to three dimensions, we use the relation between 3d and 4d polarization, i.e., $\epsilon^{3d}=\frac{\epsilon^++\epsilon^-}{2}$. This yields the result in eq.~\eqref{2loopcut1}.

\subsection{Phase space integrals}
\label{app:2loopcut}

In this section we we discuss the phase space integral yielding the non-analytic part of the two-loop results in sections \ref{sec:1and2LoopUnitarity} and ~\ref{subsec:high-spin-unitarity}. As discussed in the main text in section \ref{sec:1and2LoopUnitarity}, the only cut diagram contributing to non-analytic two loop results is the most right diagram in figure~\ref{fig:2-loop-cut}. So the on-shell form factors needed for two loop calculations are $\langle p_1^gp_2^gp_3^g|O|0\rangle$.

We can calculate the discontinuity by gluing sides of the diagram  in \ref{fig:2-loop-cut} together using,
\begin{align}
    \text{Disc}{}\left(\langle OO'\rangle\right)=&(2f^{abc}g_{s})^2\frac{-i}{3!}\int\frac{d^2p_1}{(2\pi)^22E_1}\frac{d^2p_2}{(2\pi)^22E_2}\frac{d^2p_3}{(2\pi)^22E_3}
    \nn\\&\qquad\times
    (2\pi)^3\delta^3(p-p_1-p_2-p_3)
    \langle 0|O(p)|p_1^gp_2^gp_3^g\rangle\langle p_1^gp_2^gp_3^g|O'(p)|0\rangle.
\end{align}
We can then do the projection to different spin at this level to obtain the integrands which are scalar functions of $p_1,p_2$ and $p_3$,
\begin{equation}
      \text{Disc}{}A_j^{(1)}=-ig_s^2C_A\frac{16}{3 \pi^3}\int\frac{d^2p_1}{E_1}\frac{d^2p_2}{E_2}\frac{d^2p_3}{E_3}\delta^3(p-p_1-p_2-p_3)I_{j}(p_1,p_2,p_3).
\end{equation}
To do the integral, we go to the rest frame of $p$ and define the usual parameters for 3-body phase space calculation,
\begin{equation}
    p=(p^0,0,0), 
    \qquad 
    x_i=\frac{p_i\cdot p}{p^2},  
    \qquad 
    x_1+x_2+x_3=1.
\end{equation}
Now we can write $I_{j}(p_1,p_2,p_3)$ in terms of $x_i$s. Further, using spatial $\delta$-function we can integrate $x_3$ trivially and write the remaining integrals as, 
\begin{equation}
      \text{Disc}{}A_j^{(1)}=ig_s^2C_A\frac{16}{3 \pi^3}\int\frac{d^2x_1}{x_1^2}\frac{d^2x_2}{x_2^2}\delta(\theta-\theta_*)\frac{I_j(x_1,x_2)}{\sin\theta_*},
\end{equation}
where $\cos\theta_*=(1/2+x_1x_2-x_2-x_1)/x_1x_2$. Now the angular integrals can be done and we are left with the two one-dimensional integrals:
\begin{equation}
     \text{Disc}{}A^{(1)}_{j}=ig_s^2C_A\frac{16}{3 \pi^2}\int_{\frac{1}{2}-x_1}^\frac{1}{2} dx_2\int^{\frac{1}{2}}_0 dx_1\frac{x_1^2x_2^2I_j(x_1,x_2)}{(\frac{1}{2}-x_1)(\frac{1}{2}-x_2)(x_1+x_2-\frac{1}{2})}.
\end{equation}
These integrals can then be simply calculated to obtain the non-analytic two loop results quoted in the paper.

\bibliographystyle{JHEP}
\bibliography{3DYM.bib}

\end{document}